\documentclass[onecolumn]{aa}
\usepackage{natbib}
\usepackage{graphicx}
\usepackage{amssymb}
\usepackage{rotating}
\usepackage{txfonts}
\usepackage{ulem}
\usepackage{multirow}
\begin{document}

   \title{A holistic approach to carbon-enhanced metal-poor stars }
   \author{T. Masseron$^{1,2}$, J. A. Johnson$^{1}$, B. Plez$^{2,3}$, S. Van Eck$^{4}$, F. Primas$^{5}$, S. Goriely$^{4}$ and A. Jorissen$^{4}$}

   \offprints{\email{masseron@astronomy.ohio-state.edu}}

   \institute{$^1$ The Ohio State University, Columbus, OH, USA \\
              $^2$ GRAAL, Universit\'e Montpellier II, 34095 Montpellier Cedex 05, France\\
              $^3$ Department of Physics and Astronomy, Uppsala Astronomical Observatory, Box 515, 751 20 Uppsala, Sweden\\
	          $^4$ Institut d'Astronomie et d'Astrophysique, Universit\'e Libre de Bruxelles, Belgium\\
              $^5$ European Southern Observatory (ESO), Karl-Schwarzschild-Str. 2, 85749 Garching b. M\"unchen, Germany }

   \date{Received ...; accepted ...}

   \abstract
    {Carbon-Enhanced Metal-Poor
   (CEMP) stars are known to be the direct witnesses of the
   nucleosynthesis of the first low- and intermediate-mass
   stars, because most have been polluted by a now-extinct AGB star.}
   {By considering the various CEMP subclasses separately, we try to derive, from the specific signatures imprinted on the abundances, parameters (such as metallicity, mass, temperature, and
   neutron source) characterizing
      AGB nucleosynthesis 
 from the specific signatures imprinted on the
   abundances, and separate them from the impact of thermohaline mixing, first dredge-up, and dilution associated with the mass transfer from the companion.}
   {To put CEMP stars in a broad context, we collect abundances for
     about 180 stars of various metallicities (from solar down
   to [Fe/H]=-4), luminosity classes (dwarfs and giants), and
   abundance patterns (C-rich and poor, Ba-rich and poor, etc), from our own sample and from literature.} 
   {First, we show that there are CEMP stars which share the properties of CEMP-s stars and CEMP-no stars (which we call CEMP-low-s stars). We also show that there is a strong correlation
   between Ba and C abundances in the s-only CEMP stars. This strongly
   points at the operation of the $\rm^{13}$C
   neutron source in {\it low-mass AGB stars}. For the CEMP-rs stars
   (seemingly enriched with elements from both the s- and r-processes), 
   the correlation of the N abundances with abundances of heavy elements from the
   2$\rm^{nd}$ and 3$\rm^{rd}$ s-process peaks bears instead the
   signature of the $\rm^{22}$Ne neutron source. Adding the fact
   that CEMP-rs stars exhibit O and Mg enhancements, we conclude that extremely hot conditions prevailed during the thermal pulses of the contaminating AGB stars.  We also note that abundances are not affected by the evolution of the CEMP-rs star itself (especially by the first dredge-up). 
  This implies that mixing must have occurred while the star was on the main sequence and that a large amount of matter must have been accreted so as to trigger thermohaline mixing. 
Finally, we argue that most CEMP-no
   stars (with no overabundances for the
   neutron-capture elements) are likely the extremely
   metal-poor counterparts of CEMP neutron-capture-rich stars. We
   also show that the C enhancement in CEMP-no stars declines with metallicity at extremely low
   metallicity ([Fe/H]~$< -3.2$). This trend is not predicted by any
   of the
   current AGB models.}  
   {}

   \titlerunning{A holistic approach to carbon-enhanced metal-poor stars}
   \authorrunning{T. Masseron et al.}
   \maketitle


\section{Introduction}\label{sec:intro}
The largest wide-field spectroscopic surveys for metal-poor stars to
date, the HK survey \citep{Beers1992} and the HES survey
\citep{Christlieb2001}, have provided a tremendous wealth of
information for the study of the early chemical evolution of our
Galaxy. One of the most interesting and surprising result is the large
frequency of carbon-enhanced stars ([C/Fe] $>$ 1.0, hereafter CEMP
stars) among metal-poor stars. The results from the HK and HES surveys
indicate that they account for 20-30\% of stars with [Fe/H] $<$
-2.5 \citep{Lucatello2005bin,Lucatello2006}. This finding has prompted a number of high-resolution, high
signal-to-noise studies aimed at understanding the origin of the
abundance anomalies in these objects. The carbon-enhancement
phenomenon appears in stars that exhibit four different heavy-element
abundance patterns:

(i) The most numerous class is characterized by
enrichments of neutron-capture elements. From radial-velocity
variations, \citet{Preston2000} and \citet{Lucatello2005bin}
demonstrate that these stars are members of binary
systems. Moreover, some of them (hereafter
CEMP-s) show an abundance pattern for
neutron-capture elements compatible with the operation of the
s-process in Asymptotic Giant Branch (AGB) stars. Thus it is now established that these CEMP-s stars - as well as
classical CH stars - are members of wide binary systems where the former
primary star transferred material during its AGB phase onto the presently
observable companion  \citep{McClure1990}. Actually, we demonstrate
in \citet[][ hereafter Paper~I]{masseron2009I} that CEMP and CH
stars belong to the same class of stars.

(ii) Another class of CEMP stars (hereafter
CEMP-rs), showing large overabundances of elements resulting from the s-process and of elements traditionally assigned to the r-process, has been
discovered by \citet{Barbuy1997} and \citet{Hill2000}. A number of
these stars show radial-velocity variations \citep[e.g. Paper~I;
][]{Sivarani2004,Barbuy2005}. Indeed, there is no doubt that these
stars are binaries (they might even be triple systems) and that the
companion(s?) is(/are) responsible for the peculiar abundance
pattern. Nevertheless, this category of CEMP stars is very puzzling:
they show a large Ba enhancement, which is representative of the s-process, and at the same time a very large Eu enhancement, which is representative of the r-process.
 Most of the scenarios invoked to explain the
 peculiar rs abundance pattern include a double phase: a r-process
 pollution (from a type-II supernovae) followed by a s-process
 pollution (from an AGB star) or vice versa \citep[see hypotheses III,
   IV and VI of][ and references therein]{Jonsell2006}. These
 scenarios are supported by the predictions of \citet{Bisterzo2006}
 which satisfactorily reproduce the general features of CEMP-rs
 neutron-capture patterns by setting in their model an initial high
 r-process enrichment before slow neutron capture begins.  On the
 contrary, \citet{Johnson2004} and \citet{MasseronPhD} do not find a
 satisfactory combination of r- and s-process to reproduce the
 neutron-capture element pattern in CEMP-rs stars, and call for a
 modified neutron-capture process. In addition, the large number of
 CEMP-rs stars observed at low metallicities casts doubt on the
 likelihood of two-phase scenarios. From the present analysis, we suggest instead that the CEMP-rs stars are produced by intermediate-mass AGB stars where both $\rm^{13}$C($\alpha$,n)$\rm^{16}$O and $\rm^{22}$Ne($\alpha$,n)$\rm^{16}$O neutron sources operate.

(iii) Some CEMP stars  with
no enhancements in their neutron-capture-element abundances have been identified
\citep[hereafter
CEMP-no;][]{Aoki2002c}. Unfortunately, not many of these stars currently have
enough radial-velocity measurements to constrain their
binary properties. Consequently, a mass-transfer scenario comparable
to that operating in CEMP-s (and possibly in CEMP-rs) stars is not
firmly established. Nevertheless, the origin of this category is of great
interest as the two most Fe-poor stars known to date
\citep[HE~0107-5240 and HE~1327-2326;][]{Christlieb2002,Frebel2005}
belong to this class. 
The existence of a very large C content in extremely
low-metallicity stars may be explained from 
nucleosynthesis processes operating in basically all mass ranges, including notably
hypernovae \citep[e.g.][]{Heger2002}, type-II supernova models
\citep[e.g.][]{Woosley1995,Limongi2003}, massive stars with rotation 
accounting for the N enhancement
\citep[e.g.][]{Meynet2002,Hirschi2007}, fast-rotating AGB models
\citep[e.g][]{Siess2004}, early-AGB or Red-Giant-Branch (RGB) stars
with modified helium-core flashes 
at very low
metallicity \citep[e.g.][]{Fujimoto2000,Suda2004}. The intriguing
possibility that the abundance pattern of the two most metal-poor stars known to date is
actually shaped by chemical segregation rather than by nucleosynthesis
has been raised by \citet{Venn2008}. 

(iv) Finally, a single case of
a highly r-process-enhanced CEMP star has been noted
\citep[CS22892-052, ][]{Sneden2003}. 

CEMP stars constitute an extremely interesting
probe of the initial mass function (IMF) of the early Galaxy. Those CEMP stars which have been polluted by AGB stars probe
the IMF of the intermediate-mass ($\rm \sim 1-8\; M_\odot$) stars,
provided of course that a reliable mass may be assigned to the companion AGB star. Theoretical models
have indeed shown several ways along which the mass of the AGB star may
influence nucleosynthesis.  For example, N is strongly enhanced by
Hot-Bottom Burning (HBB) at the expense of C in intermediate-mass
($\rm \sim 3-8\;M_\odot$) AGB stars.  On the other hand, F is only produced in low-mass AGB stars \citep[see other examples in][]{Karakas2007}. For the production of the heavy elements through the s-process, the mass of the
AGB star is of importance as well. It has been suggested that low-mass ($\rm \sim 1-3\; M_\odot$) AGB
stars produce neutrons mainly by the
$\rm^{13}$C($\alpha$,n)$\rm^{16}$O reaction operating in radiative
conditions \citep{Straniero1995}, whereas in intermediate-mass AGB
stars, a weak s-process is driven by $\rm^{22}$Ne($\alpha$,n)$\rm^{25}$Mg
operating in convective conditions \citep{Goriely2004,Goriely2005}. However, no
AGB model predicting the formation of the
necessary $\rm^{13}$C pocket in a self-consistent way is available yet, and the
$\rm^{22}Ne(\alpha,n)^{25}Mg$ reaction rate still suffers from large
uncertainties. 

Although yields are not yet available for a wide range of elements,
stellar masses, and
metallicities, some attempts have been made to compare
AGB yields for different masses  with CEMP star
abundances. \citet{Bisterzo2008} daringly derive the mass of the former AGB companion of 74 CEMP-s stars by
fitting abundance patterns with their predictions. They found that all
abundance anomalies in CEMP-s stars originate from AGB stars with $
M\rm<1.4\;M_\odot$. Based on \citet{Fujimoto2000} calculations,
\citet{Komiya2007} argue that CEMP-s stars had an AGB companion with
$\rm 0.8\;M_\odot<$$M$$\rm<3.5\;M_\odot$, whereas CEMP-no stars have an
intermediate-mass AGB companion with $M\rm>3.5\;M_\odot$. 
However, none of these models provide an explanation holding for the 4 CEMP subclasses simultaneously.

In this paper, we review the CEMP phenomenon to shed light on the origin of these intriguing stars based on an unprecedented compilation of abundances of all CEMP classes. After classifying the sample in different categories according to their observed abundances, we first give additional proof that Ba stars are formed by mass-transfer from metal-rich AGB stars (Sect.~\ref{sec:Bastars}). Thus, Ba stars constitute a metal-rich sample which may serve as a comparison for CEMP stars. We also discuss the need for the introduction of CEMP-low-s stars, which show low s-process element abundances (Sect.~\ref{sec:CEMP-low-s}). Our approach consists in looking at element correlations between elemental abundances in each CEMP class. That manner, we bring new insights to the nucleosynthesis in the AGB companion of CEMP-s stars (Sect.~\ref{sec:CEMP-s}). Assuming that AGB stars are also responsible for the CEMP-rs stars peculiar composition, we try to identify their nucleosynthesis processes (Sect.~\ref{sec:CEMP-rs}). In parallel, we qualitatively compare the mean trends between CEMP classes and with non C-rich stars. Finally, we discuss the nature of the companion of CEMP-no stars and evaluate the role of AGB stars at extremely low-metallicity ([Fe/H]$<$-3.0) (Sect.~\ref{sec:CEMP-no}).
This holistic view of CEMP abundances finally attempts to draw a coherent picture of AGB nucleosynthesis at low metallicity (Sect.~\ref{sec:Summary}).

\section{The extended sample}\label{sec:sample}
We have compiled abundances from analyses of high-resolution spectra
($R > 40000$) of CEMP stars (Tables \ref{tab:Cstars_light} and
\ref{tab:Cstars_heavy}), Ba stars, and non-carbon-enhanced metal-poor
stars (Tables \ref{tab:nonCstars_light} and
\ref{tab:nonCstars_heavy}). C-rich stars are defined as stars with
    [C/Fe]~$> 0.9$\footnote{Since not all authors have adopted the same solar abundances for C and Fe, our CEMP criterion ([C/Fe]~$> 0.9$) has been made slightly different from the one used by  \citet{Rossi1999} ([C/Fe]=1.0), in order to keep  
stars like the unique r-process-rich star  CS~22892-052 in the CEMP family with the  \citet{Asplund2005} solar abundances adopted in the present paper.}. All the plots in this paper are exclusively made out
    of data from these tables which include our own data from
    Paper~I. All these data are renormalized by the
    \citet{Asplund2005} solar abundances.

\begin{figure}[!h]
\begin{center}
\includegraphics[width=7cm,angle=-90]{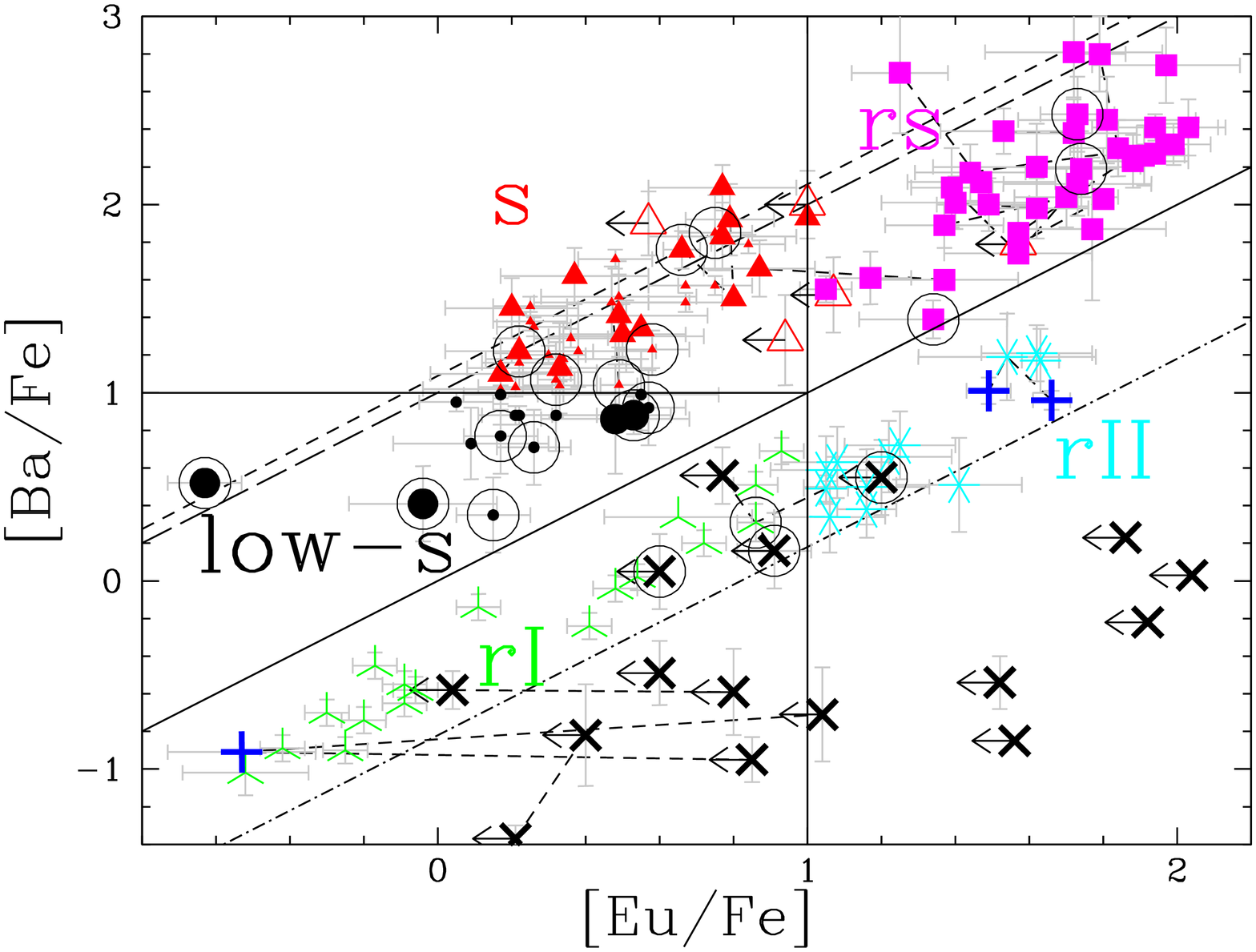}
\includegraphics[width=7cm,angle=-90]{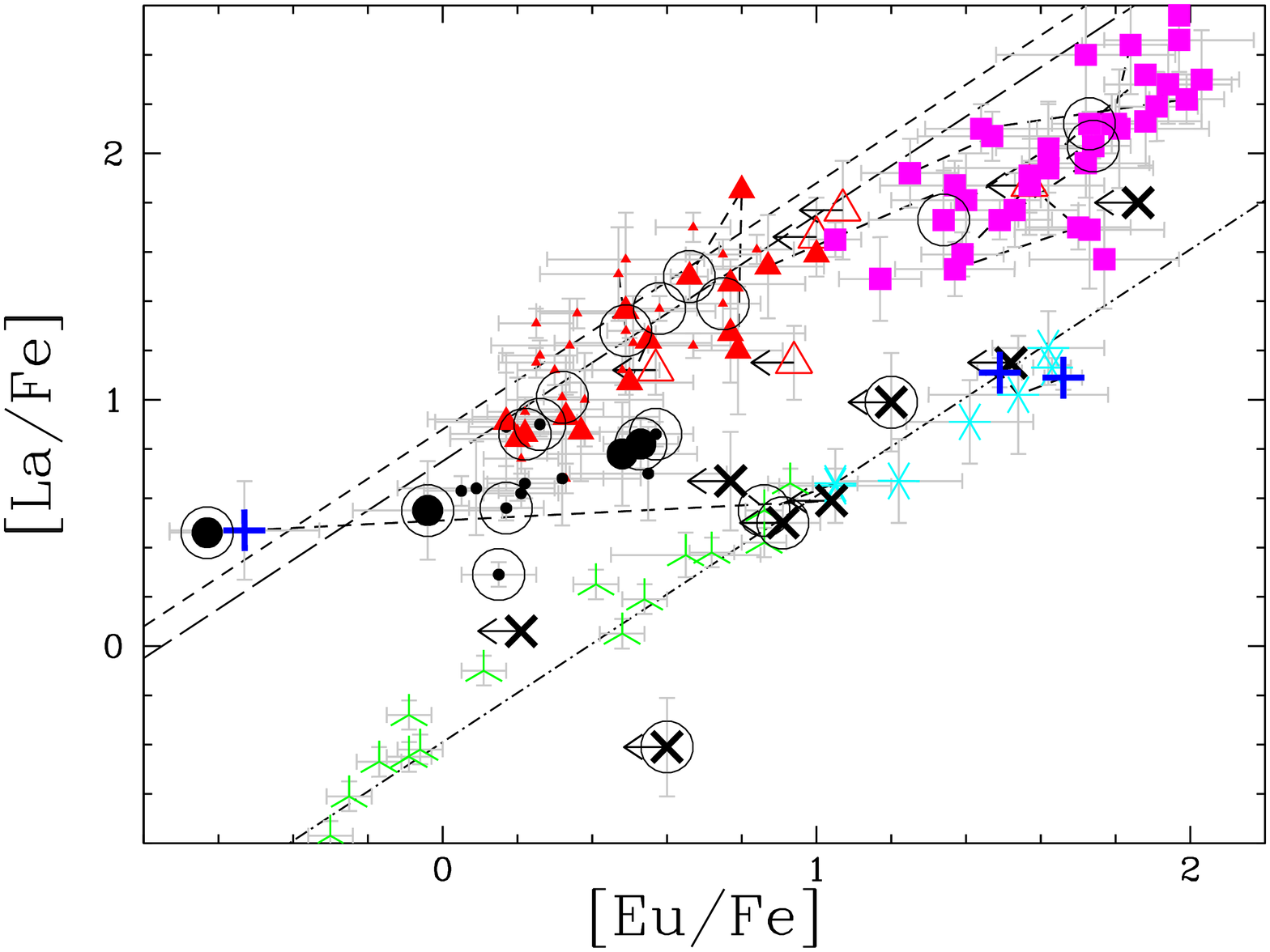}
  \caption{ [Ba/Fe] vs [Eu/Fe] and [La/Fe] vs [Eu/Fe] abundances in sample stars. The (red) triangles stand for CEMP-s stars, the filled (black) circles code the CEMP-low-s stars, the (black) crosses code the CEMP-no stars and the (magenta) squares represent the CEMP-rs stars. We also represent classical Ba stars with tiny black dots and tiny red triangles. Large open circles identify stars studied in Paper~I. The cyan solid lines separate the different classes (see Fig.~\ref{fig:legend} for their explicit definition). The black lines correspond to pure s-process nucleosynthesis predictions for a 0.8 M$\rm_\odot$ \citep[short dash;][]{Masseron2006} and a 3 M$\rm_\odot$ \citep[long dash;][]{Goriely2005} metal-poor AGB star and to pure solar r-process \citep[dash-dot;][]{Goriely1999}. Although La is an excellent s-process tracer, there are fewer abundances available for this element in the literature. Note that the star HE~2356-041 has a typical s-process La/Eu ratio (lower panel), despite the fact that the Ba/Eu ratio qualifies it as a CEMP-r star (upper panel; see also Sect.~\ref{sec:CEMP-low-s})}\label{fig:BaFevsEuFe}
\end{center}
\end{figure}

\begin{figure}[!h]
\begin{center}
\includegraphics[width=6cm,angle=-90]{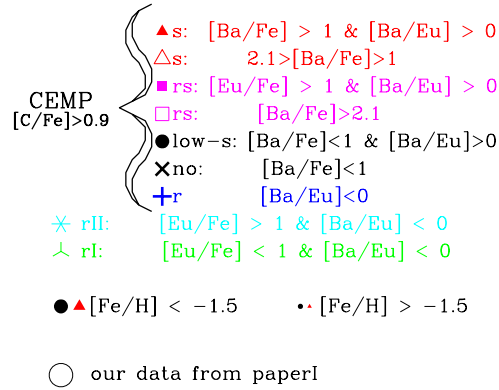}
  \caption{The adopted classification results from Fig.~\ref{fig:BaFevsEuFe}. The
  classification and the corresponding color-shape code adopted in
  this figure are used in all the figures of this paper. Large symbols and small symbols correspond respectively to [Fe/H]$<$-1.5 and $>$-1.5. Open symbols
  are used whenever there is no Eu abundance available. In
  this case, the classification is solely based on the Ba abundance displayed in Fig.~\ref{fig:BaFevsEuFe}.}\label{fig:legend}
\end{center}
\end{figure}

Figure \ref{fig:BaFevsEuFe} shows the distribution of the stars from
Tables \ref{tab:Cstars_heavy} and \ref{tab:nonCstars_heavy} in the
([Ba/Fe], [Eu/Fe]) diagram, which involves two neutron-capture elements. This figure reveals that different families must be distinguished (see Fig.~\ref{fig:legend} for details). Our definitions closely match the ones of \citet{Jonsell2006} for CEMP-s and CEMP-rs stars, and of \citet{Beers2005} for CEMP-no, rI and rII stars (see below for a definition of rI and rII stars). However, these studies do not consider the stars with either no Eu abundance available or with only an upper limit. Thus, to classify those, we rely on the Ba abundance alone. In Fig.~\ref{fig:BaFevsEuFe} we have labeled four stars as CEMP-low-s stars, three of them being based on an Eu measurement from Paper~I (see Sec.~\ref{sec:CEMP-low-s} and \ref{sec:CEMP-no}). At this stage, based on the consideration of Fig.~\ref{fig:BaFevsEuFe} only, the necessity of distinguishing CEMP-low-s stars from CEMP-no stars is not at all obvious, since it may appear  as simply resulting from  the absence of a firm Eu detection in CEMP-no stars. This question will be addressed in more details in Sect.~\ref{sec:CEMP_relation}.

The stars denoted rI and rII by \citet{Beers2005} represent the majority of metal-poor stars (CEMP stars being $\sim 20$\% of them). Because they apparently do not belong to binary systems, we assume that their abundances are representative of the composition of the interstellar medium from which these metal-poor stars formed, and hence of the initial composition of CEMP stars.


In the whole sample, we identify 47 CEMP-s, 44 CEMP-rs and 42 CEMP-no stars, including 32 multiple measurements, summing up to a total of 101 different CEMP stars. Despite the fact that all come from only 2 surveys of metal-poor stars (HK and HES), we do not attempt to extract accurate frequencies from these numbers because each author has potentially biased the subsample of stars with abundance analysis by using specific selection criteria.
There are also 2 CEMP stars which show a pure r-process Ba/Eu ratio (CS~22892-052 and HE~2356-0410, represented by blue 'plus' symbols. It is possible that these stars have received some C-rich material from an AGB star, but the high r-process initial composition overwhelms the low s-process elemental abundances, as suggested by \citet{Aoki2002c}. We also note that the latter star (aka CS 22957-027) has a La/Eu ratio compatible with a pure s-process (lower panel of Fig.~\ref{fig:BaFevsEuFe}), making it difficult to identify the origin of the neutron-capture elements. We also stress that, among three high resolution spectroscopic analyses of this star, only one study was able to derive the Eu abundance at a very low level, and with Eu lines falling in a forest of strong CN lines. Therefore, the accuracy of this measurement should perhaps be carefully be reexamined. 

The error bars shown in all the plots are generally the random errors
published by the authors (when available) and are very often in the range $\approx$ 0.1-0.2 dex). We do not include the
systematic errors, because we consider that they are best represented
by the dispersion in the abundances resulting from the different studies of a given object
(connected by long-dashed lines in all the figures). In fact, in
Paper~I, we show that systematic errors due to different assumptions
about model-atmosphere parameters may lead to large abundance
discrepancies, especially for abundance determinations based on a
single line. As an example,
the [Pb/Fe] ratio derived in different studies of HE~2158-0348 ranges from
2.77 to 3.42 while the [Ba/Fe] values span the much narrower range
1.60 -- 1.66. 

Because of the various oxygen-abundance diagnostics, we choose
to plot O abundances uncorrected for NLTE and 3D effects. However,
we give in column 11 of Table \ref{tab:Cstars_light} NLTE corrections
for O~I measurements using \citet{Takeda2003} formula. The corrections
for O triplet measurements are of the order of 0.2 dex, in agreement
with \citet{GarciaPerez2006}. Notice that for some of the stars,
especially the CEMP-rs, the O-line equivalent widths are above the
applicability limit of the formula.  \citet{GarciaPerez2006} also
found 3D corrections for OH of the order of -0.1 dex. \\

Despite the fact that the production of elements like C, N and s-process elements in the AGB star largely overwhelm the abundance already present in the CEMP star, abundance ratios such as [X/Fe] or [X/H] in CEMP stars are not necessarily identical to the initial yields of the AGB star, especially when not much matter is accreted and/or is heavily diluted in the CEMP star envelope. Various mass-transfer efficiencies are indeed expected. Because of the large variety of orbital parameters, various accretion and dilution factors contribute to the abundance scatter observed for C, N or s-process elements. Unfortunately, there 
are currently only very few orbits available for CEMP stars. Therefore, it is 
difficult to constrain the initial abundances from observations of
[X/Fe] or [X/H] ratios. Nevertheless, the reader should keep in mind that using abundance ratios of enhanced s-process elements (e.g. [Ba/Eu], [Pb/Ba]) represents a way to partially cancel the uncertainty produced by the various mass transfer efficiencies and to get a handle on the AGB nucleosynthesis.
\begin{table}[h!]
{\small
\begin{tabular}{ l c c c c c c c c c c c c c c }
\hline
Object       & T$\rm_{eff}$ $\log$ g & [Fe/H] & $\sigma$ & [C/Fe] &  $\sigma$ & $\rm^{12}C/^{13}C$ &  $\sigma$ & [N/Fe] & $\sigma$ & [O/Fe]  & $\sigma$ & Non-LTE & [Mg/Fe] &  $\sigma$     \\           
   (1)       &   (2)             & (3)    & (4)      & (5)    &  (6)      & (7)             &  (8)      & (9)    &  (10)    & (11)    & (12)     &  (13)   &   (14)  &  (15)       \\        
\hline																																																		       
  HE 0007-1832      &   6515 3.8 &   -2.72 &   0.19 &     2.45 &   ...  &      ... &  ...   &     1.67 &   ...    &      ...  &   ...  &   ...  &      0.79 &   0.35\\
  HE 0012-1441      &  5730 3.50 &   -2.59 &   0.16 &     1.70 &   ...  &      ... &  ...   &     0.59 &   ...    &      ...  &   ...  &   ...  &      0.94 &   0.16\\
  HE 0024-2523      &   6625 4.3 &   -2.70 &   0.12 &     2.62 &   0.10 &        6 &   1    &     2.12 &   0.10   & (2)  0.65 &   ...  &   NaN  &      0.60 &   0.06\\
  HE 0039-2635      &  4900 1.50 &   -2.91 &   0.27 &     2.72 &   0.20 &      ... &  ...   &     2.63 &   0.53   &      ...  &   ...  &   ...  &      1.31 &   0.12\\
  HE 0107-5240      &   5100 2.2 &   -5.39 &   ...  &     3.85 &   0.15 &   $>$ 30 &  ...   &     2.39 &   0.15   &      ...  &   ...  &   ...  &      0.24 &   ... \\
  HE 0131-3953      &  5928 3.83 &   -2.71 &   0.17 &     2.32 &   0.27 &      ... &  ...   &     ...  &   ...    &      ...  &   ...  &   ...  &      0.30 &   0.24\\
  HE 0143-0441      &   6240 3.7 &   -2.38 &   0.18 &     2.08 &   ...  &      ... &  ...   &     1.68 &   ...    &      ...  &   ...  &   ...  &      0.66 &   0.08\\
                    &   6370 4.4 &   -2.23 &   0.21 &     1.56 &   ...  &      ... &  ...   &    -0.22 &   ...    &      ...  &   ...  &   ...  &      0.50 &   0.17\\
  HE 0202-2204      &  5280 1.65 &   -1.98 &   0.19 &     1.03 &   0.28 &      ... &  ...   &     ...  &   ...    &      ...  &   ...  &   ...  &     -0.01 &   0.23\\
  HE 0206-1916      &  5200 2.70 &   -2.09 &   0.20 &     2.10 &   0.19 &       15 &   5    &     1.61 &   0.33   &      ...  &   ...  &   ...  &      0.52 &   0.15\\
  HE 0212-0557      &  5075 2.15 &   -2.34 &   0.26 &     1.84 &   ...  &        4 &   1    &     1.04 &   ...    &      ...  &   ...  &   ...  &      0.07 &   ... \\
  HE 0231-4016      &  5972 3.59 &   -2.08 &   0.18 &     1.23 &   0.27 &      ... &  ...   &     ...  &   ...    &      ...  &   ...  &   ...  &      0.22 &   0.24\\
  HE 0336+0113      &   5947 3.7 &   -2.39 &   0.09 &     2.72 &   0.10 &   $>$ 60 &  ...   &     1.72 &   0.20   &      ...  &   ...  &   ...  &      0.32 &   0.20\\
                    &  5700 3.50 &   -2.75 &   0.22 &     2.35 &   ...  &   $>$  7 &   1    &     1.55 &   ...    &      ...  &   ...  &   ...  &      1.07 &   0.18\\
\hline	
\end{tabular}
}																	      
\caption{Abundances of light elements in carbon-enhanced metal-poor stars from literature. Columns (4), (6), (8), (10), (12), (15) list random errors (when available). Column (3) lists metallicities preferentially from Fe I. Column (5) list C abundances from preferencially CH determinations. (7) N abundances preferentially from CN. (9) For oxygen abundances  (2) stands for measurements from O~I triplet and (3) from OH lines. By default, the oxygen measurement comes from the [OI] line at 6300 \AA. (11) Non-LTE corrections of O~I triplet measurement according to \citet{Takeda2003} (when applicable).\textit{The entire table will be published in A\&A}}\label{tab:Cstars_light}  
\end{table}

\begin{table}[h!]
{\small
\begin{tabular}{ l c c c c c c c c c c |c c l}
\hline
Object              & [Ba/Fe]       &  $\sigma$ &  [La/Fe]  &  $\sigma$ & [Ce/Fe] & $\sigma$ &   [Eu/Fe]   & $\sigma$ &  [Pb/Fe]   & $\sigma$ & Class &    solar ref &  references        \\
                    &  (16)         &  (17)     &  (18)     &  (19)     &   (20)  &  (21)    &  (22)       &  (23)    & (24)       &  (25)    & (26) &  (27)  & (28)      \\
\hline
  HE 0007-1832      &          0.23 &   ...  &    $<$  1.80 &   ...  &     ...  &   ...  &       $<$  1.86 &   ...  &    $<$  3.21 &   ...  &    no &   An89 & (1)                                                                                                                   \\                                                                                            
  HE 0012-1441      &          1.22 &   0.27 &         ...  &   ...  &     ...  &   ...  &            ...  &   ...  &    $<$  2.09 &   ...  &     s &   An89 & (2)                                  							       			     \\
  HE 0024-2523      &          1.52 &   0.20 &         1.77 &   0.20 &     ...  &   ...  &       $<$  1.07 &   0.10 &         3.32 &   0.10 &     s &   An89 & (3)                                  							       			     \\
  HE 0039-2635      &          2.23 &   0.13 &         ...  &   ...  &     ...  &   ...  &            ...  &   ...  &         ...  &   ...  &    rs &   As05 & (4)                                  							       			     \\
  HE 0107-5240      &     $<$  0.82 &   ...  &         ...  &   ...  &     ...  &   ...  &       $<$  2.80 &   ...  &         ...  &   ...  &    no &   Gr98 & (7,8)              							       			     \\
  HE 0131-3953      &          2.20 &   0.23 &         1.94 &   0.27 &     1.93 &   0.29 &            1.62 &   0.28 &         ...  &   ...  &    rs &   Gr98 & (5)                                  							       			     \\
  HE 0143-0441      &          2.39 &   0.12 &         1.77 &   0.15 &     2.00 &   0.21 &            1.53 &   0.17 &         3.28 &   ...  &    rs &   An89 & (2)                                  							       			     \\
                    &          2.38 &   0.18 &         1.96 &   0.14 &     2.20 &   0.21 &            1.72 &   0.17 &         3.67 &   ...  &    rs &   An89 & (1)                                  							       			     \\
  HE 0202-2204      &          1.41 &   0.22 &         1.36 &   0.34 &     1.30 &   0.26 &            0.49 &   0.24 &         ...  &   ...  &     s &   Gr98 & (5)                                  							       			     \\
  HE 0206-1916      &          1.97 &   0.16 &         ...  &   ...  &     ...  &   ...  &            ...  &   ...  &         ...  &   ...  &     s &   As05 & (4)                                  							       			     \\
  HE 0212-0557      &          2.25 &   0.06 &         2.27 &   0.22 &     2.21 &   ...  &            ...  &   ...  &         ...  &   ...  &    rs &   An89 & (2)                                  							       			     \\
  HE 0231-4016      &          1.47 &   0.23 &         1.22 &   0.28 &     1.53 &   0.27 &            ...  &   ...  &         ...  &   ...  &     s &   Gr98 & (5)                                  							       			     \\
  HE 0336+0113      &          2.12 &   ...  &         2.07 &   0.10 &     1.87 &   0.19 &            1.47 &   0.20 &         2.82 &   0.20 &    rs &   An89 & (6)                       									       			     \\
                    &          2.70 &   0.32 &         1.92 &   0.14 &     2.37 &   0.18 &            1.25 &   0.13 &    $<$  2.45 &   ...  &    rs &   An89 & (2)                                  							       			     \\
\hline
\end{tabular}
}	
\caption{Abundances of neutron-capture elements in carbon enhanced metal-poor stars from literature. Columns (17), (19), (21), (23), (25) list random errors (when available). (27) Solar abundances reference adopted in the original reference source. \textit{The entire table will be published in A\&A}
references: (1) \citet{Cohen2004} 2004ApJ...612.1107C 
, (2) \citet{Cohen2006} 2006AJ....132..137C   
, (3) \citet{Lucatello2003} 2003AJ....125..875L   
, (4) \citet{Aoki2007} 2007ApJ...655..492A  
, (5) \citet{Barklem2005}  2005A\&A...439..129B   
, (6) \citet{LucatelloPhD}   
, (7) \citet{Christlieb2002} 2002Natur.419..904C   
, (8) \citet{Christlieb2004} 2004ApJ...603..708C   
, (9) \citet{Jonsell2006} 2006A\&A...451..651J   
, (10) \citet{Norris2007} 2007ApJ...670..774N   
, (11) \citet{Cohen2008} 2008ApJ...672..320C   
, (12) \citet{Goswami2006} 2006MNRAS.372..343G  
, (13) \citet{Frebel2007} 2007ApJ...658..534F   
, (14) \citet{Frebel2005} 2005Natur.434..871F   
, (15) \citet{Aoki2006}  2006ApJ...639..897A   
, (16) \citet{Frebel2006} 2006ApJ...638L..17F   
, (17) \citet{Cohen2003} 2003ApJ...588.1082C   
, (18) \citet{Johnson2002} 2002ApJ...579L..87J  
, (19) \citet{TsangaridesPhD}    
, (20) \citet{Aoki2002c} 2002ApJ...567.1166A  
, (21) \citet{Aoki2002b} 2002ApJ...580.1149A   
, (22) \citet{Aoki2002d} 2002PASJ...54..933A   
, (23) \citet{Preston2001} 2001AJ....122.1545P  
, (24) \citet{Sneden2003_CS22892} 2003ApJ...591..936S   
, (25) \citet{Cayrel2004} 2004A\&A...416.1117C   
, (26) \citet{Spite2005} 2005A\&A...430..655S   
, (27) \citet{Spite2006} 2006A\&A...455..291S   
, (28) \citet{Barbuy2005} 2005A\&A...429.1031B   
, (29) \citet{Francois2007} 2007A\&A...476..935F  
, (30) \citet{Depagne2002} 2002A\&A...390..187D   
, (31) \citet{Sneden2003} 2003ApJ...592..504S  
, (32) \citet{Norris1997b} 1997ApJ...489L.169N   
, (33) \citet{Bonifacio1998} 1998A\&A...332..672B   
, (34) \citet{Aoki2002a} 2002ApJ...576L.141A  
, (35) \citet{Sivarani2006} 2006A\&A...459..125S   
, (36) \citet{Sivarani2004} 2004A\&A...413.1073S  
, (37) \citet{Ivans2005} 2005ApJ...627L.145I   
, (38) \citet{Aoki2004} 2004ApJ...608..971A   
, (39) \citet{Johnson2004} 2004ApJ...605..462J   
, (40) \citet{Zacs1998} 1998A\&A...337..216Z  
, (41) \citet{Aoki2001} 2001ApJ...561..346A   
, (43) \citet{Plez2005} 2005A\&A...434.1117P   
, (44) \citet{Deroo2005} 2005A\&A...438..987D   
, (45) \citet{Honda2004} ApJ...607..474H   
, (46) \citet{Allen2006a} 2006A\&A...454..895A   
, (47) \citet{Christlieb2004} 2004A\&A...428.1027C   
, (48) \citet{Hill2002} 2002A\&A...387..560H   
, (49) \citet{Plez2004} 2004A\&A...428L...9P   
, (50) \citet{Honda2006} 2006ApJ...643.1180H   
, (51) \citet{Roederer2008} 2008ApJ...679.1549R
, (52) Paper~I }\label{tab:Cstars_heavy}
\end{table}

\begin{table}[h!]
{\small
\begin{tabular}{ l c c c c c c c c c c c c c c }
\hline
Object       & T$\rm_{eff}$ $\log$ g & [Fe/H] & $\sigma$ & [C/Fe] &  $\sigma$ & $^{12}C/^{13}C$ &  $\sigma$ & [N/Fe] & $\sigma$ & [O/Fe]  & $\sigma$ & Non-LTE & [Mg/Fe] &  $\sigma$     \\           
   (1)       &   (2)             & (3)    & (4)      & (5)    &  (6)      & (7)             &  (8)      & (9)    &  (10)    & (11)    & (12)     &  (13)   &   (14)  &  (15)       \\        
\hline	
\multicolumn{15}{c}{Ba stars}																																										                                                \\
\hline													       
  BD +18:5215        & 6300 4.20 &   -0.44 &   0.04 &     0.59 &   0.08 &    ... &   ...   &     ...  &   0.13   &      0.43 &   0.10 &    ... &      0.17 &   0.05    \\
  HR 107            & 6650 4.00 &   -0.34 &   0.04 &     0.23 &   0.08 &    ... &   ...   &     ...  &   0.13   &     -0.07 &   0.10 &    ... &      0.08 &   0.05    \\
  HD 749            & 4580 2.30 &   -0.06 &   0.18 &     0.20 &   0.20 &    ... &   ...   &    -0.04 &   0.18   &      0.21 &   0.24 &    ... &     -0.12 &   0.20    \\
 HD 5424            & 4600 2.30 &   -0.21 &   0.14 &     0.12 &   0.10 &     8  &   2     &   0.43    &   0.10 &       0.00 &   0.20 & ??     &   0.34    &   0.34 \\
                    & 4700 1.80 &   -0.51 &   0.18 &     0.39 &   0.20 &    ... &   ...   &     0.41 &   0.18   &      0.18 &   0.24 &    ... &      0.20 &   0.20    \\
  HD 8270           & 6070 4.20 &   -0.44 &   0.04 &     0.31 &   0.08 &    ... &   ...   &     ...  &   0.13   &      0.08 &   0.10 &    ... &     -0.04 &   0.05    \\
  HD 12392          & 5000 3.20 &   -0.06 &   0.18 &     0.40 &   0.20 &    ... &   ...   &     0.57 &   0.18   &      0.18 &   0.24 &    ... &      0.05 &   0.20    \\
  HD 13551          & 6050 3.70 &   -0.44 &   0.04 &     0.24 &   0.08 &    ... &   ...   &     ...  &   0.13   & $<$  0.43 &   0.10 &    ... &      0.25 &   0.05    \\
  HD 22589          & 5630 3.30 &   -0.12 &   0.04 &     0.30 &   0.08 &    ... &   ...   &     0.19 &   0.13   & $<$  0.03 &   0.10 &    ... &      0.21 &   0.05    \\
 HD 24035           & 4500 2.00 &  -0.14 &   0.18 &   0.15 &   0.10 &    1    20 &     5 &   0.46 &   0.10 &    1  -0.12 &   0.20 & ?? &  -0.24 &   0.14 \\
\hline	
\end{tabular}
}																			      
\caption{Same as Table \ref{tab:Cstars_light} for stars with [C/Fe]$<$0.9.\textit{The entire table will be published in A\&A}}	\label{tab:nonCstars_light} 
\end{table}												 																																				      
																																																		      
\begin{table}[h!]
{\small
\begin{tabular}{ l c c c c c c c c c c c l}
\hline
Object              & [Ba/Fe]       &  $\sigma$ &  [La/Fe]  &  $\sigma$ & [Ce/Fe] & $\sigma$ &   [Eu/Fe]   & $\sigma$ &  [Pb/Fe]   & $\sigma$ & solar ref &  references        \\
                    &  (16)         &  (17)     &  (18)     &  (19)     &   (20)  &  (21)    &  (22)       &  (23)    & (24)       &  (25)    & (26)      &  (27)       \\																																													      
\hline	
\multicolumn{13}{c}{Ba stars}														            \\
\hline																																																		       
  BD +18:5215        	&          1.46 &   0.05 &         1.15 &   0.06 &     1.23 &   0.05 &            0.25 &   0.10 &         0.45 &   0.19 &    Gr98 & (46)     \\       																																																	      
  HR 107            	&          0.95 &   0.05 &         0.63 &   0.06 &     0.53 &   0.05 &            0.05 &   0.10 &         0.90 &   0.19 &    Gr98 & (46)     \\   																																																	      
  HD 749            	&          1.18 &   0.19 &         1.22 &   0.19 &     1.62 &   0.19 &            0.34 &   0.21 &         0.38 &   0.29 &    Gr98 & (46)     \\   																																																	       
HD 5424                 &   1.04 &   0.20 &       1.28 &   0.05 &       1.66 &   0.16 &       0.49 &   0.05 &       0.91 &   0.10 & As05 & (52) \\
                    	&          1.48 &   0.19 &         1.51 &   0.19 &     1.98 &   0.19 &            0.47 &   0.21 &         1.10 &   0.29 &    Gr98 & (46)     \\   
  HD 8270           	&          1.11 &   0.05 &         0.96 &   0.06 &     0.95 &   0.05 &            0.33 &   0.10 &         0.50 &   0.19 &    Gr98 & (46)     \\   
  HD 12392          	&          1.51 &   0.19 &         1.57 &   0.19 &     1.79 &   0.19 &            0.49 &   0.21 &         1.15 &   0.29 &    Gr98 & (46)     \\   
  HD 13551          	&          1.16 &   0.05 &         0.95 &   0.06 &     1.03 &   0.05 &            0.22 &   0.10 &         0.50 &   0.19 &    Gr98 & (46)     \\   
  HD 22589          	&          0.88 &   0.05 &         0.66 &   0.06 &     0.57 &   0.05 &            0.22 &   0.10 &        -0.15 &   0.19 &    Gr98 & (46)     \\   
HD 24035                &   1.07 &   0.20 &       1.01 &   0.05 &      1.63 &   0.13 &       0.32 &   0.05 &       0.94 &   0.10 & As05 & (52) \\
\hline
\end{tabular}
}	
\caption{Same as Table \ref{tab:Cstars_heavy} for stars with [C/Fe]$<$0.9.\textit{The entire table will be published in A\&A}}	\label{tab:nonCstars_heavy} 
\end{table}														
\newpage

\section{Ba stars and CEMP-s stars}\label{sec:Bastars}
\begin{figure}[!h]
\begin{center}
\includegraphics[width=5cm,angle=-90]{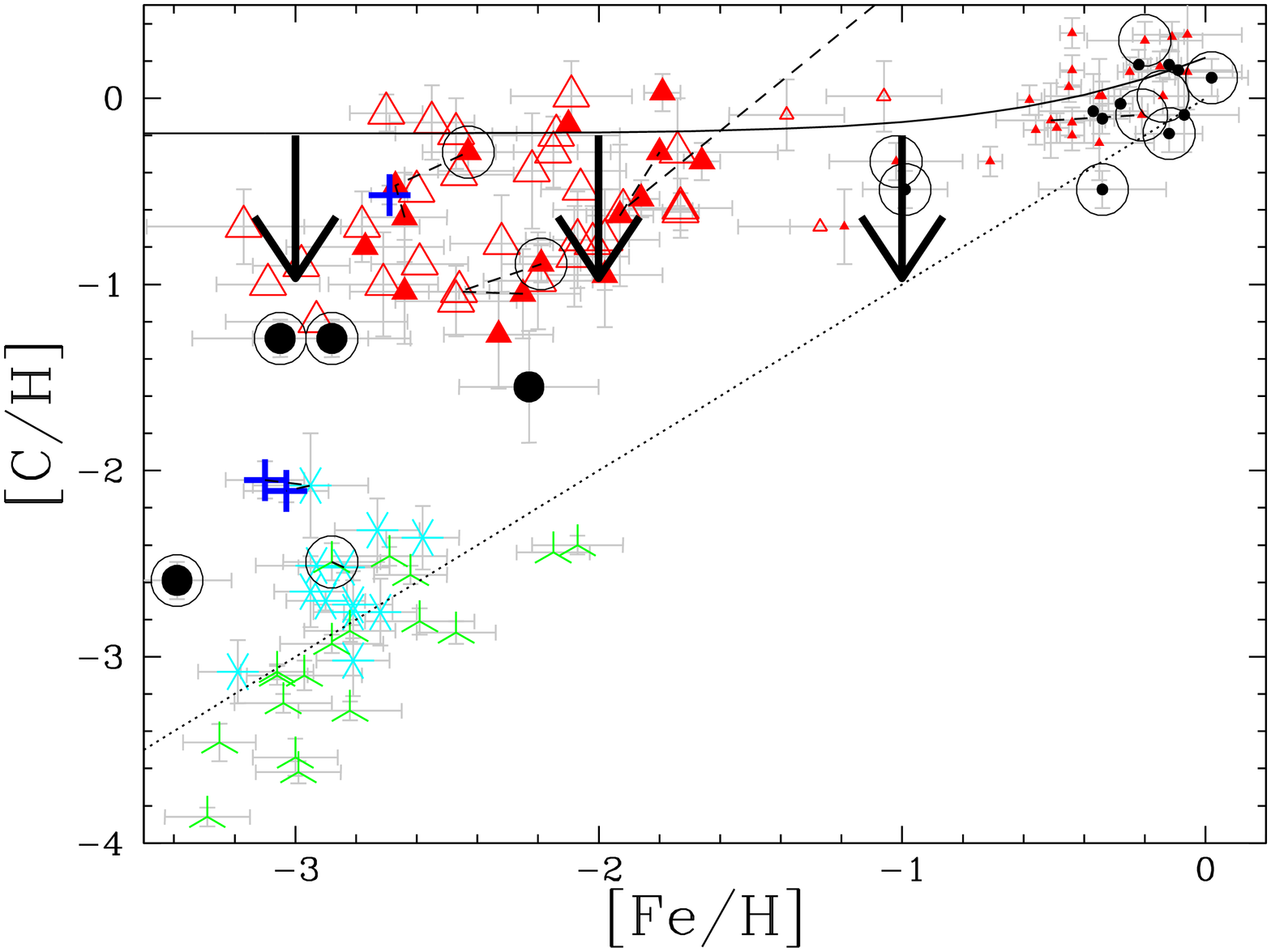}
\includegraphics[width=5cm,angle=-90]{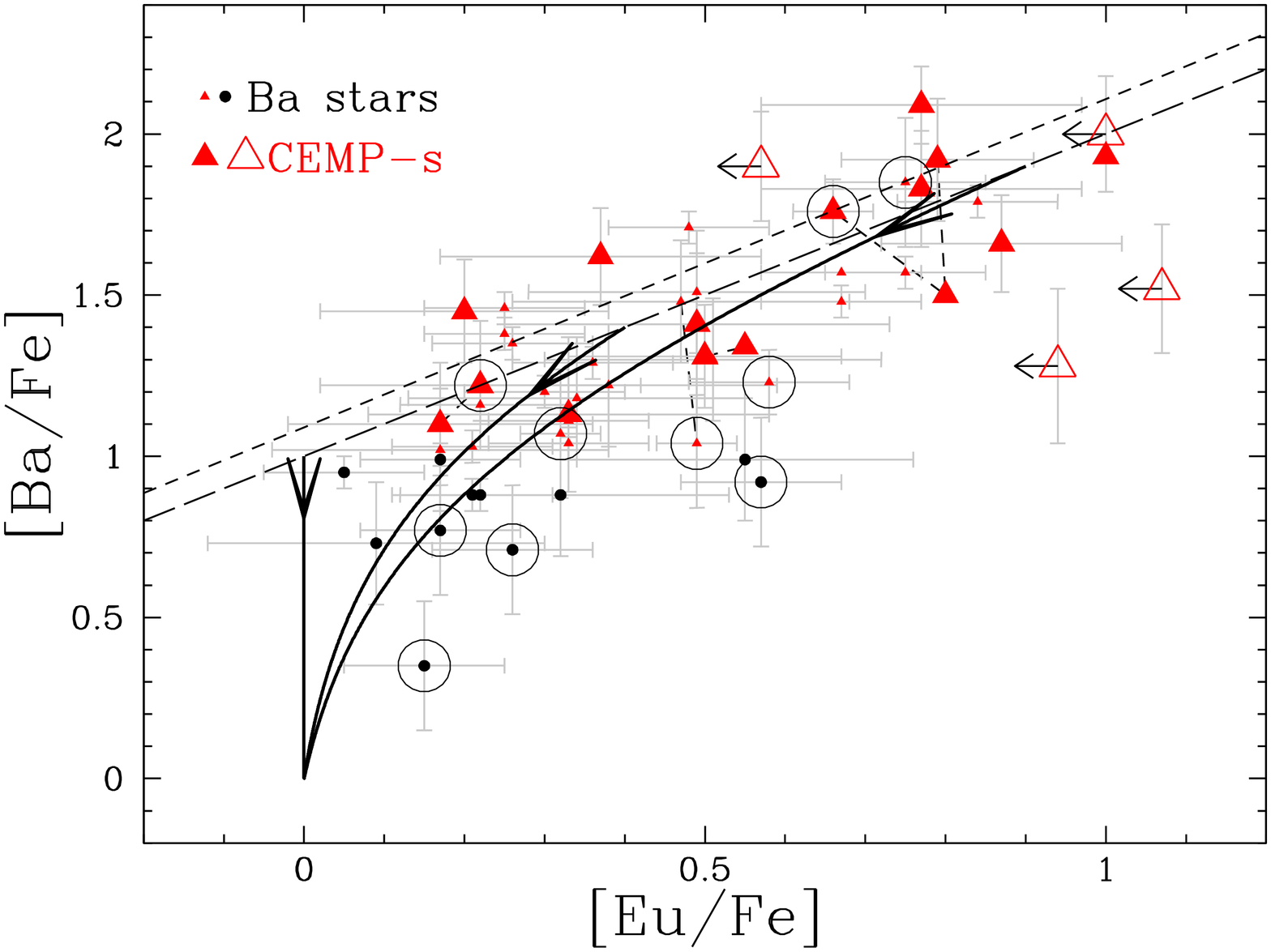}
  \caption{ (left panel) The C content in CEMP-s stars and in Ba stars (see
    Fig.~\ref{fig:legend} for a description of symbols). The
    dotted line represents the Galactic average C content
    ([C/Fe]~$\approx0$) and the  solid line stands for a constant
    amount of C ([C/H]~$=-0.2$) added to the initial Galactic average
    content. The black arrows represent a arbitrary dilution factor of the accreted C. This assumption of a constant C abundance in the accreted material is consistent with a
    primary C production in AGB stars. This simple calculation shows
    that Ba stars are indeed the analogues of C stars at high
    metallicities. (right panel) [Ba/Fe] as a function of [Eu/Fe] for Ba stars and CEMP-s/low-s stars. The black arrows represent the track followed by the abundance ratios when increasing the dilution of s-process enriched material in a solar-composition material. This demonstrates that a varying dilution of s-process-rich material into solar-composition material can explain the global trend of [Ba/Fe] and [Eu/Fe] ratios in Ba stars. }
\label{fig:C_BaEu_Bastars}
\end{center}
\end{figure}

 \citet{Jorissen2000} demonstrate that Ba stars are just part of a
binary evolutionary sequence which also involves MS, S and C stars
without lines from the unstable element Tc (see their Figure 1), and
\citet{Allen2006b} conclude that Ba stars have the same s-process
signature as AGB stars. Figure~\ref{fig:BaFevsEuFe} shows that they exhibit  [Ba/Fe] and [Eu/Fe] ratios identical to CEMP-s stars. Following \citet{Cohen2006}, we thus suggest that CEMP-s stars and Ba stars belong to the same category of mass-transfer binaries from a former AGB companion, and differ only on the ground of their metallicity. Ba stars are not as carbon-enhanced as CEMP-s stars because their composition prior to the mass transfer from the AGB star had already high C and O content, and, in such 
circumstances, the C present in the accreted material is not sufficient to bring the resulting C/O ratio above unity. 
Hence, even after the transfer of C-rich material from the AGB
companion, the [C/H] ratio remains close to the Galactic average
(dotted line in Fig.~\ref{fig:C_BaEu_Bastars}). Thus, the C/O ratio remains below 1, and CH
or C$\rm_2$ lines are less intense in the spectra of Ba stars than in their
more metal-poor counterparts. Therefore, we argue that the same nucleosynthesis processes are responsible for the C and s-element production in Ba stars and CEMP-s stars, which just
differ by metallicity. Note that in the following plots we
include Ba stars to give a broader view of AGB nucleosynthesis, thus making it
possible to identify the impact of metallicity. In Fig.~\ref{fig:C_BaEu_Bastars}, we also highlight the effect of dilution (either in the AGB envelope or when the material transferred from the AGB star is mixed with material in the companion's envelope). We calculate the dilution tracks of the neutron-capture elements as follows: 
\begin{eqnarray}\label{formula_dilu}
Eu & = & (1-d) \times Eu_{s} + d \times Eu_{init}\\
Ba & = & \frac{Ba}{Eu}\bigg|_s \times (1-d) \times Eu_{s} + \frac{Ba}{Eu}\bigg|_{init} \times d \times Eu_{init}
\end{eqnarray}
with $d$ being the dilution factor (ranging from 0 to 1) and $Eu$ and $Ba$ being the resulting Ba and Eu abundances after dilution. We chose $\frac{Ba}{Eu}\big|_s$ so that [Ba/Eu]$_s=1$ as observed in CEMP-s stars and  $\frac{Ba}{Eu}\big|_{init}$  and $Eu_{init}$ so that [Ba/Eu]$_{init}=0$ and [Eu/Fe]$=0$ as it is expected in matter of solar metallicity. We apply this formula to 3 values of s-process Eu ($Eu_{s}$) so that [Eu/Fe]$_s=0.0,0.4,0.9$, matching the observed range. 
This simple calculation demonstrates that the scatter observed for the neutron-capture elements in Ba stars and for C in CEMP-s stars may at least be partly ascribed to dilution.  

\section{CEMP-low-s stars: the low s-process counterparts of CEMP-s stars}\label{sec:CEMP-low-s}
The discovery of CEMP stars that have low Ba abundances (black crosses in Fig.~\ref{fig:C_BaEu_Bastars}) was exciting,
because, as discussed in Sect.~\ref{sec:intro}, it suggested that the
carbon enrichment seen in these stars could be due to pollution by a star different from an AGB. The identification of the origin of
these stars with low Ba abundances represents a special challenge, because very few
abundance data are available for the neutron-capture elements. This
difficulty may be ascribed either to a true absence of large
overabundances, or to the difficulty of detecting the spectral lines
when the metallicity is very low, even in the presence of
overabundances. This situation is illustrated by Fig.~\ref{fig:BaFevsEuFe}, where stars
currently classified as CEMP-no stars (large crosses) only have an
upper limit on their Eu abundances. In Paper~I and in \citet{Masseron2006}, we derived the Eu abundances for three CEMP-no stars (HE~1419-1324, HE~1001-0243 and CS~30322-023, represented by large
circled black dots). Their low Ba abundances would classify them as CEMP-no stars according to \citet{Beers2005}.

\begin{figure}[!h]
\begin{center}
\includegraphics[width=8cm,angle=-90]{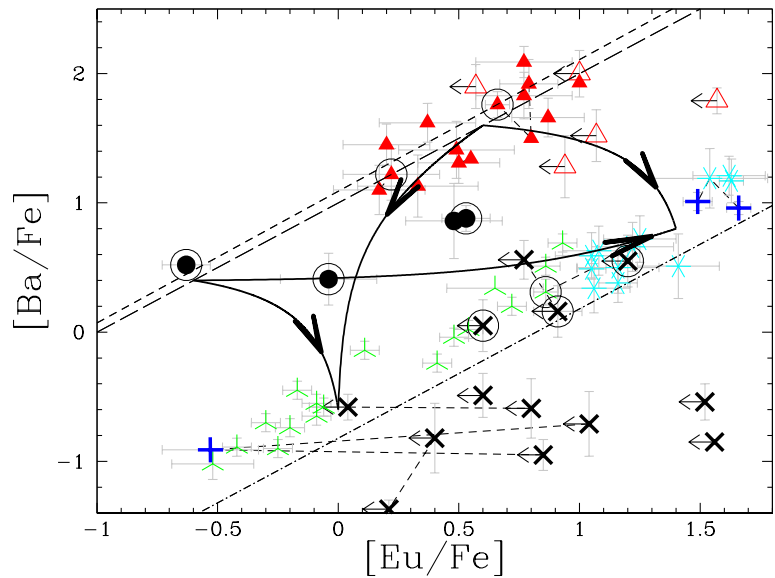}
\caption{ Ba/Fe and Eu/Fe in metal-poor stars (see Fig.~\ref{fig:legend} for a description of symbols). The black arrows represent the track followed by the abundance ratios when increasing the dilution of s-process enriched material in a pure r-process material. Because the initial pure s-process composition and the pure r-process composition are unknown, this calculation as been made for 2 sets of s-process Ba and Eu abundances representative of what is observed in CEMP-s and CEMP-low-s stars, and 2 sets of r-process Ba and Eu abundances typical of what is observed in rI and rII stars.}
\label{fig:BaFevsEuFe_diluCEMP}
\end{center}
\end{figure}

There is yet another star, namely HKII~17435-00532, with properties similar to the three just described (large circled black dots in Fig.~\ref{fig:BaFevsEuFe_diluCEMP}). So, in addition to high C, they are different from ``normal'' field stars in their heavy-element abundance. These stars fall between the s- and r-process lines, thus may be accounted for by diluting pure s-process material in pure r-process matter. We have therefore identified stars with the low Ba enhancements of CEMP-no stars, but the [Ba/Eu] ratios showing contamination by s-process material. We labeled these as CEMP-low-s stars.

We stress that, because the dilution lines (arrows in Fig.~\ref{fig:BaFevsEuFe_diluCEMP}) cross the region between the s-process and r-process lines, CEMP-low-s stars could erroneously be classified as either CEMP-rs or CEMP-no stars. So far, not many CEMP-low-s stars are known, but we suggest in Sect.~\ref{sec:CEMP_relation} that a fair fraction of CEMP-no stars could actually turn out to be CEMP-low-s stars after the missing Eu abundances become available.

The situation is a bit different for CS~30322-023. Although its low [Ba/Fe] ratio formally locates it among CEMP-low-s stars, it falls along the pure s-process line, so that there is no need to invoke the above argument involving dilution. CS~30322-023 is also special in being a genuine low-metallicity AGB star, with a very low initial Eu content \citep[Paper~I and][]{Masseron2006}! Thus, there has been no dilution associated with mass-transfer from an AGB companion for this star, and its envelope contains pure s-processed matter brought there by the third dredge-up. All the Eu initially present has been overwhelmed by the s-process Eu brought by the dredged-up
matter. 

In the remainder of this paper, the four stars forming the CEMP-low-s category will be included with the CEMP-s and CEMP-no stars in figures and discussions.

\newpage
\section{The nature of the companion of CEMP-s stars}\label{sec:CEMP-s}
\begin{figure}[!h]
\begin{center}
\includegraphics[width=5cm,angle=-90]{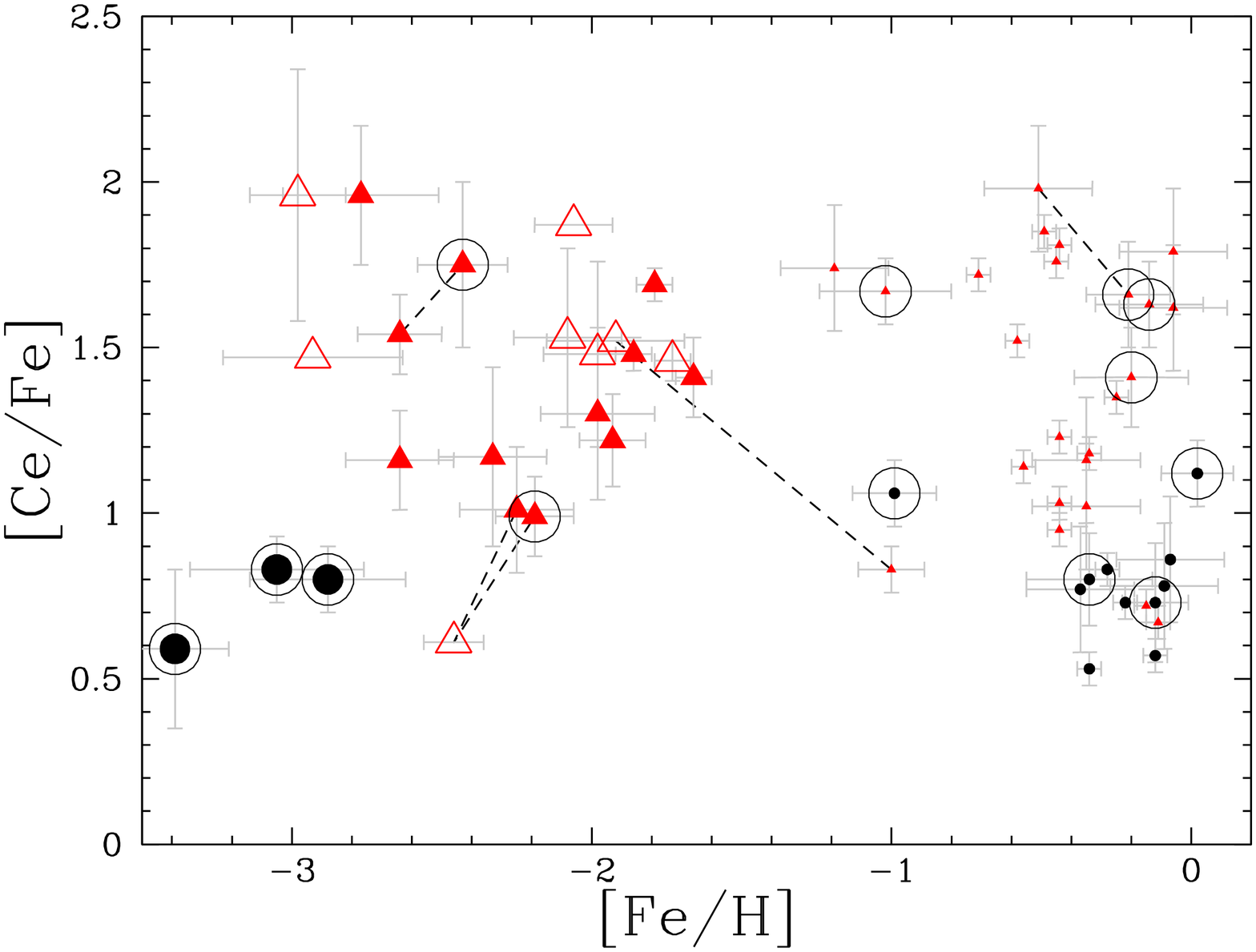}
\includegraphics[width=5cm,angle=-90]{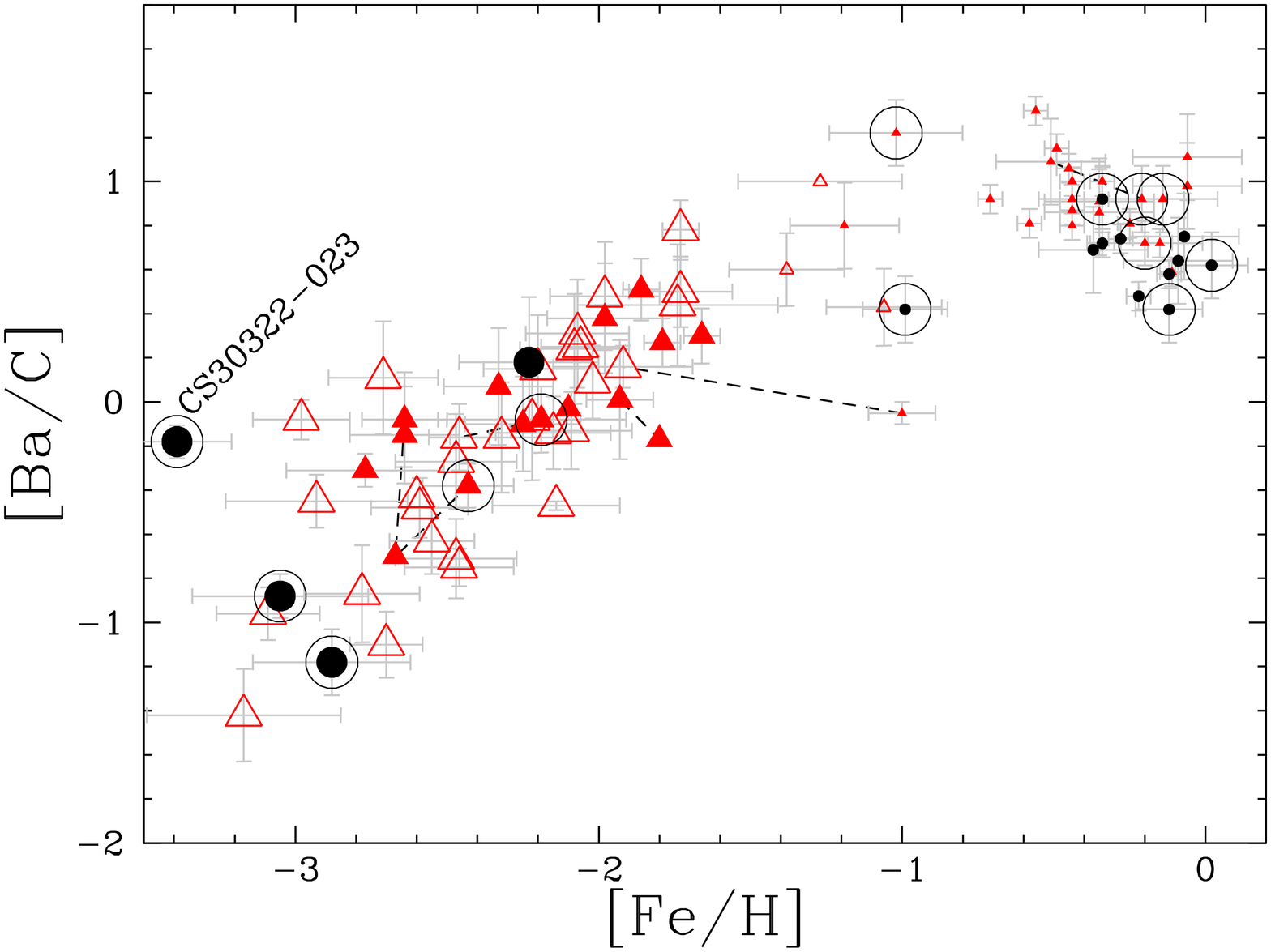}
  \caption{[Ce/Fe] (left panel) and [Ba/C] (right panel) as a function of metallicity for CEMP-s and Ba stars (see Fig.~\ref{fig:legend} for a
  description of symbols). Ce as a representative s-process element is very scattered at all metallicities. Note that the right panel is equivalent to
  [Ba/Fe] vs [C/H] \citep[e.g. Fig.~8 of][]{Aoki2002c} with the
  advantage of canceling dilution and mass-transfer effects. There is
  a strong correlation between Ba and C synthesis.  A maximum is
  obtained for [Fe/H] between -1.0 and -0.6 as expected by
  \citet{Goriely2000} and \citet{Busso2001}.}\label{fig:BaCvsFe_s}
\end{center}
\end{figure}

 The good agreement between  the predicted and observed [Ba/Eu] and [La/Eu] ratios in CEMP-s and CEMP-low-s stars (Fig.~\ref{fig:BaFevsEuFe}) supports the standard model for the operation of the s-process in AGB stars through the $^{13}$C($\alpha$,n)$^{16}$O neutron source \citep{Straniero1995,Goriely2000}. Consequently, we expect the Ba/C ratio to depend on metallicity \citep{Clayton1988}. This is because C is of primary origin (independent of metallicity; due to the triple-$\alpha$ reaction in the He-burning shell), whereas the production of Ba is secondary (it depends on the availability of Fe seed nuclei). The trend of Ba with metallicity predicted by the operation of the $^{13}$C($\alpha$,n)$^{16}$O neutron source is, however, a complicated one. First, one should remark that the Ba abundance at the surface of AGB stars where $^{13}$C($\alpha$,n)$^{16}$O operates is affected by three different factors:

(i) the low number of available Fe seed nuclei is the major limiting factor at the lowest metallicities. Hence, the [Ba/C] increases up to [Fe/H] $= -1$ (Fig.\ref{fig:BaCvsFe_s}).

\begin{figure}[!h]
\begin{center}
\includegraphics[width=5cm,angle=-90]{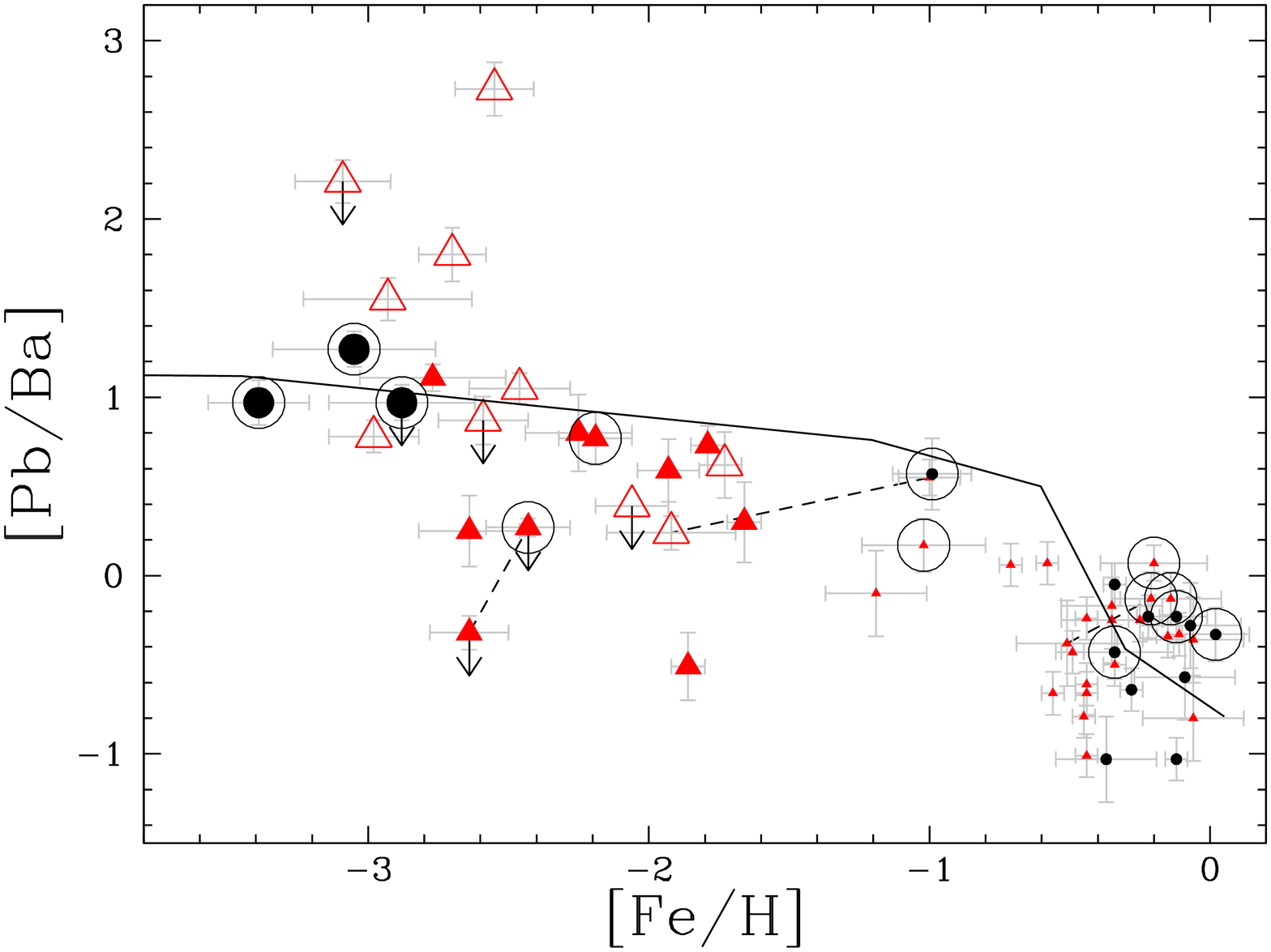}
\includegraphics[width=5cm,angle=-90]{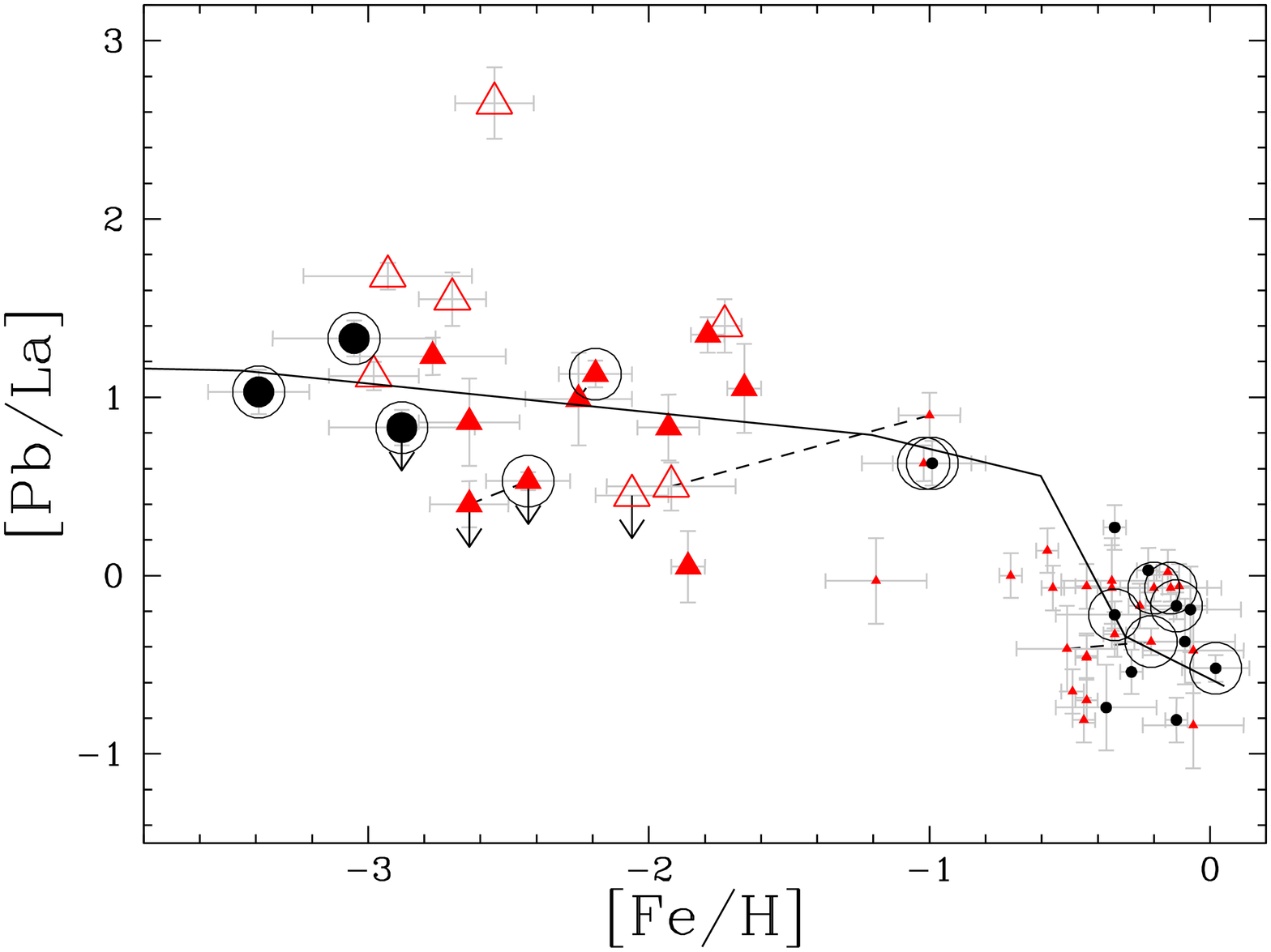}
 \includegraphics[width=5cm,angle=-90]{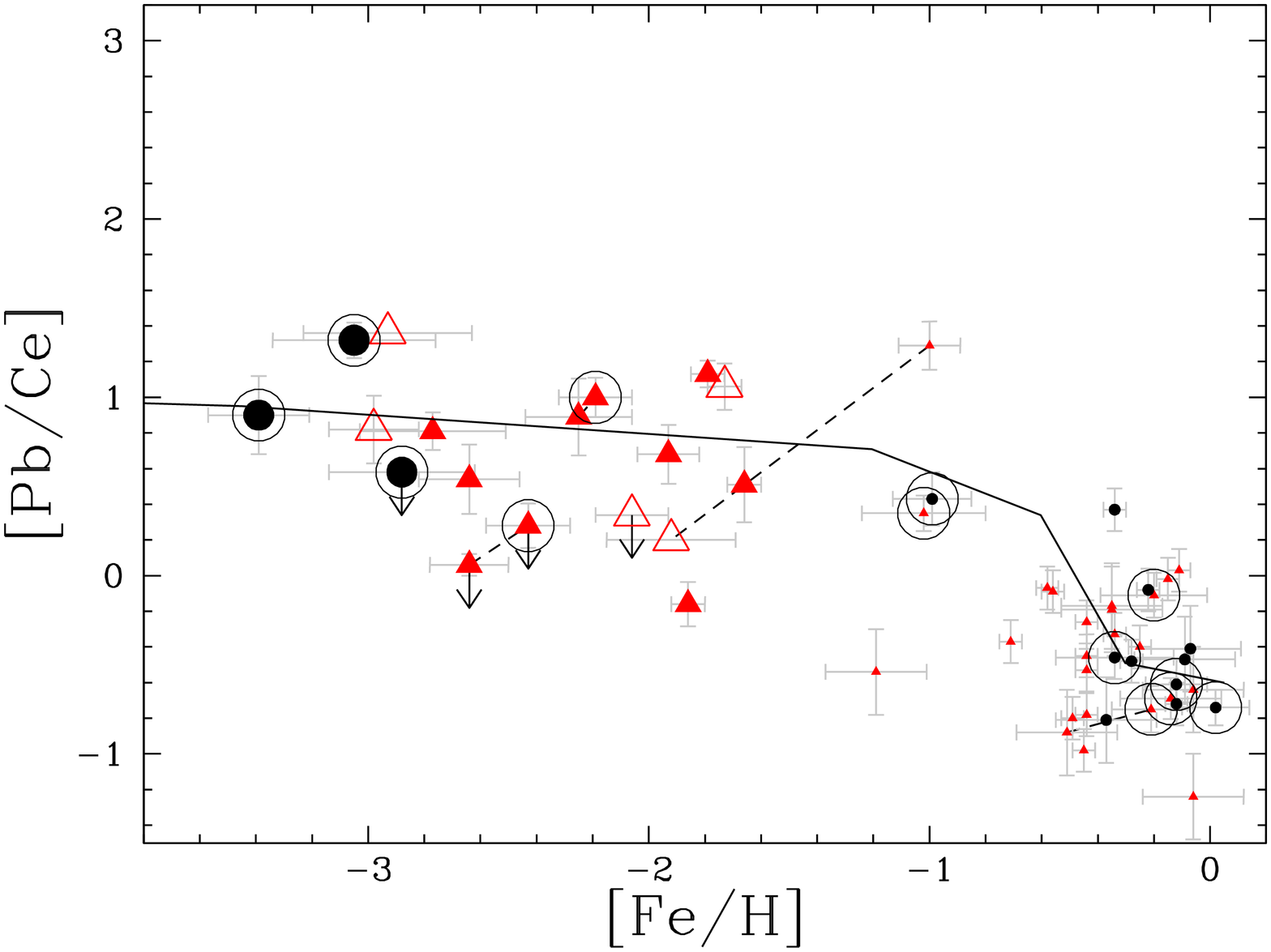}
 \caption{Third-peak to second-peak s-process element ratios for
   CEMP-s and Ba stars (see Fig.~\ref{fig:legend}  for a
   description of symbols). We also plot here our Pb measurements of
   CEMP-low-s stars as these are the only ones in the literature for this
   category. The solid line connects the predictions for different
   metallicities from \citet{Goriely2000} (after 10 dredge-ups), \citet{Goriely2001} and \citet{Masseron2006}. }\label{fig:Pb2ndpicvsFe_s}
\end{center}
\end{figure}

(ii) the s-process abundance pattern (i.e., the ratio [vhs/hs] where 'vhs' stands for third-peak s-process elements like Pb and 'hs' stands for second-peak s-process elements like Ba, La or Ce) varies as well with metallicity, since the number of neutrons captured per seed nuclei increases with decreasing metallicity. This is because the  $^{13}$C($\alpha$,n)$^{16}$O neutron source involves primary fuels, namely $^{12}$C and protons, through  $^{12}$C(p,$\gamma$)$^{13}$N($\beta^+)^{13}$C. Hence, the number of available neutrons remains the same at all metallicities (see , however, item (iii) below). But since the available Fe seed nuclei decreases with metallicity, at low metallicities the number of neutrons captured per Fe seed nuclei is large, and heavy s-process elements like Pb are produced (see Fig.~\ref{fig:Pb2ndpicvsFe_s}).
At intermediate metallicities, though, the number of neutrons captured is just enough to synthesize second-peak elements like Ba. The Ba abundance should thus reach a maximum at intermediate metallicities.

(iii) the number of neutrons available will depend on the size of the proton pocket mixed in the carbon zone, which is currently not constrained by the models, since the physical mechanism responsible for the proton diffusion in the C-rich shell remains unknown. The Ba enrichment predicted by the models is affected by this uncertainty, but the C enrichment is not. Hence, the [Ba/C] ratio is not totally independent of mixing: although not affected by the third dredge-up, it is dependent on the size of the proton pocket, the variation of  which may account for the scatter in s-process-element abundances such as Ce (left panel of Fig.~\ref{fig:BaCvsFe_s}). 

Figure~\ref{fig:BaCvsFe_s} shows that the [Ba/C] ratio in CEMP-s stars qualitatively  follows the expected trend, being maximum around [Fe/H] $= -1$ \footnote{As [C/N]$\approx$0 in CEMP-s stars (see Sect.\ref{sec:CandN}), identical values are found for [Ba/C+N]}.  The small scatter (0.28 dex) on the [Ba/C] vs. [Fe/H] relationship is indeed quite remarkable.  It is comparable to the root-mean-square of the uncertainties on the measurements ($\sim0.2$~dex). Therefore, the Ba/C ratio resulting from the s-process nucleosynthesis is not expected to vary much from star to star at a given metallicity. Although current AGB models explain well the [Ba/C] behavior as a function of metallicity, none of them reproduce it quantitatively. Hence, our results set strict constraints on the proton-diffusion mechanism producing $^{13}$C as this correlation links C production in the AGB with the s-process.

Concerning the s-process abundance pattern, recent improvements in the accuracy of the
abundances have revealed that not all the elements in a
given s-process peak behave similarly. Therefore, we choose to show
single-element ratios in Fig.~\ref{fig:Pb2ndpicvsFe_s}. This figure shows that the
Pb/Ba ratio is increasing as metallicity decreases, as expected from
the models \citep{Gallino1998,Goriely2000,Busso2001}. We stress in particular that three stars formerly classified as CEMP-no for which we derived the Pb abundance (Paper~I) fall along the expected trend; hence, they were reclassified as CEMP-low-s in Sect.~\ref{sec:CEMP-low-s}.

However, these figures show as well a significant scatter around these 
broad trends. We remind the reader that Pb and Ba might be affected by observational uncertainties. Lead is
very challenging to measure as it generally relies on one single line
(405.77~nm) blended by a regular CH line (405.78~nm) and a broad
CH-predissociation line (405.58~nm) \citep{Plez2008}, and is very sensitive to stellar
parameters. The large spread in the Pb abundances derived by different
authors  in a given
star illustrates these difficulties. For example, CS~22942-019 has been
analyzed in this work as well as by \citet{Aoki2002c}, and there is a
large discrepancy between the upper limits on the Pb abundance derived
by these two studies
($\approx$ 0.6 dex). Ba might as well be affected by large measurement
errors as illustrated by the different [Ba/Fe] ratios in CS~22942-19,
as most of the Ba lines used in spectroscopic studies
are resonance lines sensitive to non-LTE effects and usually quite 
strong.

There are also
theoretical uncertainties. In the proton-mixing scenario (Goriely \& Mowlavi 2000), 
the [vhs/hs] ratio is essentially controlled by the metallicity. As
shown by Van Eck et al. (2003), uncertainties originating from unknown dilution factors or from the proton mixing profile has an impact of $\pm 0.2$ dex on the Pb/Ba ratio. The scatter observed in Fig.~\ref{fig:Pb2ndpicvsFe_s} 
clearly indicates that additional parameters need to be considered. In 
particular, it has been suggested \citep{Goriely2004} that the 
ingestion of protons in hot AGB stars can modify the neutron 
irradiation and lead to noticeably different s-abundance 
distributions. A deep hot third dredge-up tends to reduce 
the s-process efficiency, the protons injected into the C-rich layers 
producing a $^{14}$N-rich region rather than a $^{13}$C-rich pocket.

\newpage
\section{The possible nature of the companions to CEMP-rs stars}\label{sec:CEMP-rs}

The puzzling abundance pattern in CEMP-rs stars, characterized by large overabundances of neutron-capture elements (as illustrated in Fig.~\ref{fig:BaFevsEuFe}), 
has given rise to various hypotheses \citep[see][for a detailed review]{Jonsell2006}. The binarity of these stars has now been confirmed \citep[e.g. Paper~I, and ][]{Barbuy2005}.
Because of the relatively old age of these low-metallicity halo stars, an initial pollution of their original gas by C, N and neutron-capture elements requires a very fast evolving object. But current models of massive stars predict a pure r-process pattern \citep{Woosley1995} or a weak s-process \citep{Pignatari2008}, but none support a rs pattern (and in particular the large Pb enhancement). 

In addition, \citet{Jonsell2006} noticed that the large number of CEMP-rs stars observed at low metallicities casts doubt on the probability of a two-source pollution scenario including at least one massive star. However, neither the IMF nor the multiple system frequency are known at low metallicity. Actually, the estimates from \citet{Tumlinson2007} support the fact that the IMF should be pushed toward high masses at low metallicities, and the simulation of \citet{Vanhala1998} predicts that the explosion of the first generations of massive stars would favour the formation of binaries, thus would privilege the formation of binary systems with the imprints of a massive star. Alternatively, \citet{Cohen2003} hypothesized that CEMP-rs stars are first polluted by the s-process from an AGB primary companion, which subsequently turns into a white dwarf. This white dwarf later accretes material back from the secondary companion, and if the white dwarf is an O-Ne-Mg dwarf, accretion-induced collapse of the white dwarf ensues and leads to a neutron star \citep{Nomoto1991,Justham2009}. A neutrino-driven wind from the forming neutron star enriches the secondary star in the r-process elements, leading to the final abundances in the CEMP-rs star. We stress that this scenario involves an O-Ne-Mg white dwarf, and this requirement strongly reduces the frequency  of occurrence of the scenario. 

Nevertheless, the calculations of \citet{Aoki2006} established that {\textit“s-process and r-process behave are almost independent contributors to the final yields”}. Following this argument, \citet{Johnson2004} failed to reproduced in detail the extensive abundance pattern observed in the CEMP-rs star CS~31062-50 by adding an s-process pattern to an r-process pattern. Furthermore, \citet{MasseronPhD} demonstrated that the addition of Ba and Eu abundances as observed in CEMP-s stars (representing the contribution of a low-metallicity AGB star) to the Ba and Eu abundance as observed in rII stars (representing the contribution of a low-metallicity massive star) falls below the amount of Ba and Eu observed in CEMP-rs stars. Hence, we argue that the double enhancement scenario does not hold.  Thus, it appears that a non standard s-process is the best candidate to explain the CEMP-rs phenomenon and most certainly from a unique companion (likely an AGB star).

\subsection{Evidence for the operation of the 
$\rm^{22}Ne(\alpha,n)^{25}Mg$ neutron source in CEMP-rs companions}

\begin{figure}[!h]
\begin{center}
\includegraphics[width=8cm,angle=-90]{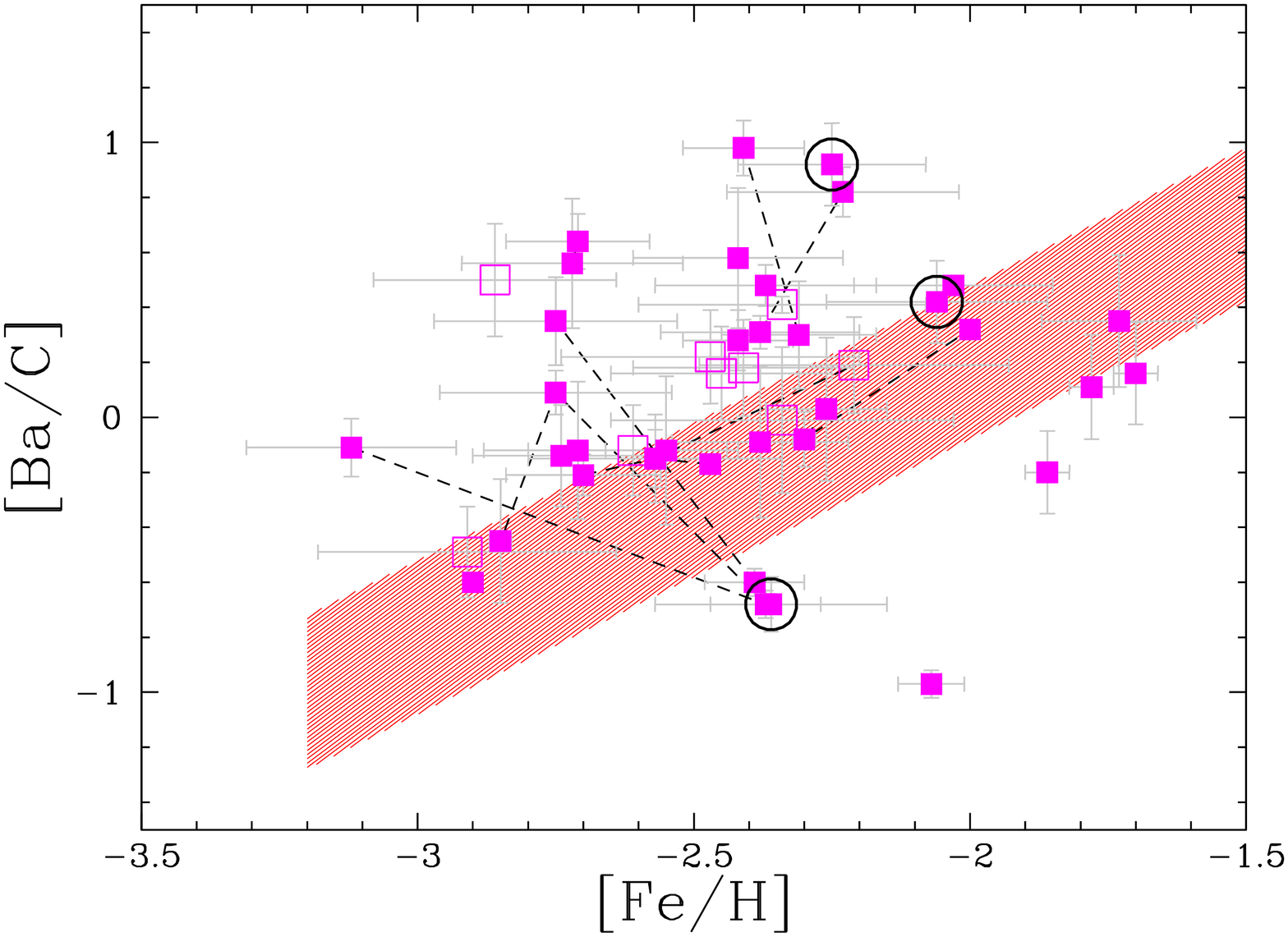}
\includegraphics[width=8cm,angle=-90]{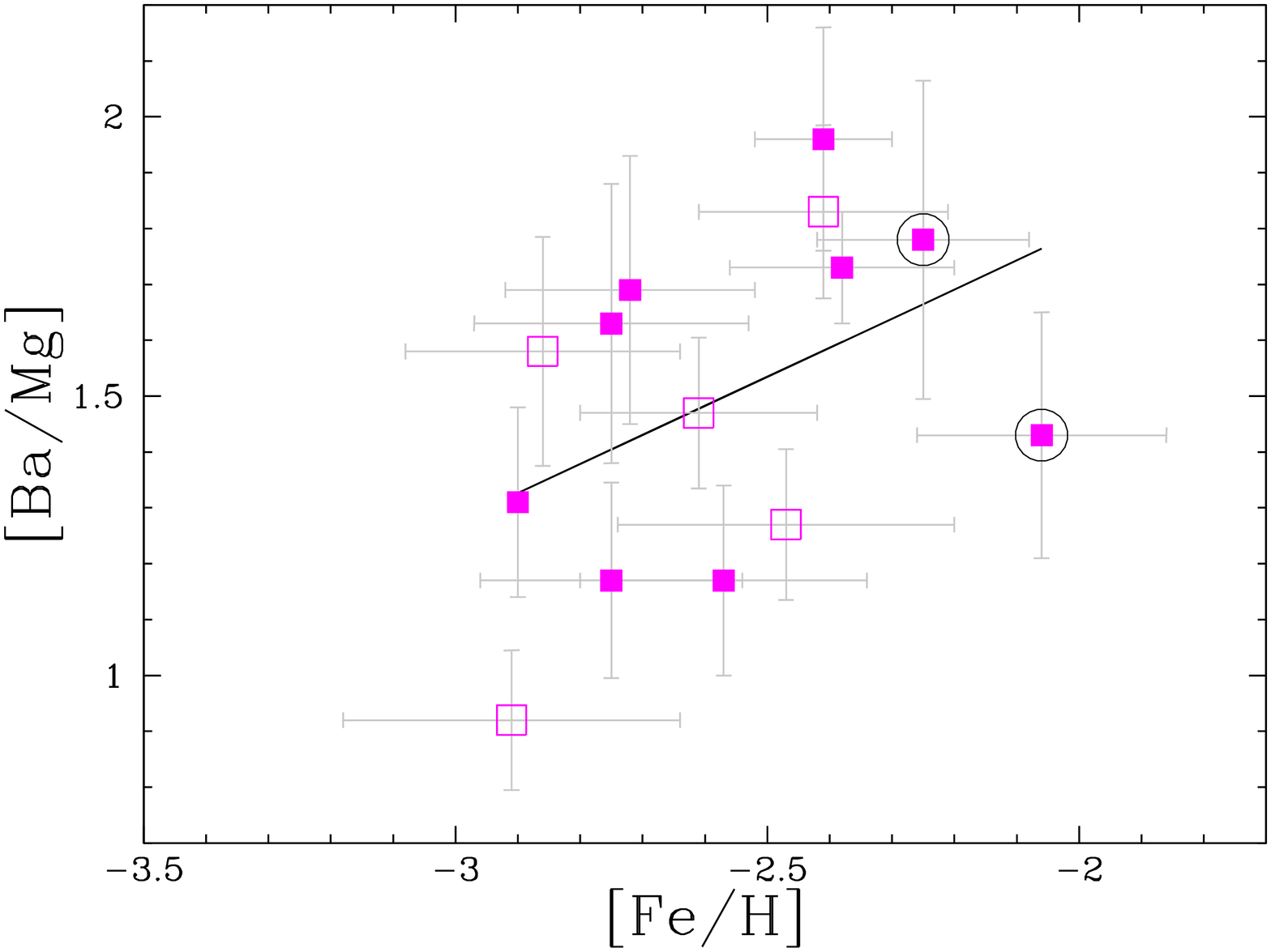}
  \caption{Upper panel: [Ba/C] in CEMP-rs stars as  a function of metallicity (see
    Fig.~\ref{fig:legend} 
for a description of symbols). No correlation is observed, in contrast to CEMP-s stars (shaded area from Fig.~\ref{fig:BaCvsFe_s}). Lower panel: [Ba/Mg] in CEMP-rs stars with [Mg/Fe]~$> 0.6$, as  a function of metallicity. There is an apparent trend (the correlation coefficient of the least-square fit (solid line) is 0.55)}\label{fig:BaCvsFe_rs}
\end{center}
\end{figure}

\begin{figure}[!h]
\begin{center}
\includegraphics[width=5cm,angle=-90]{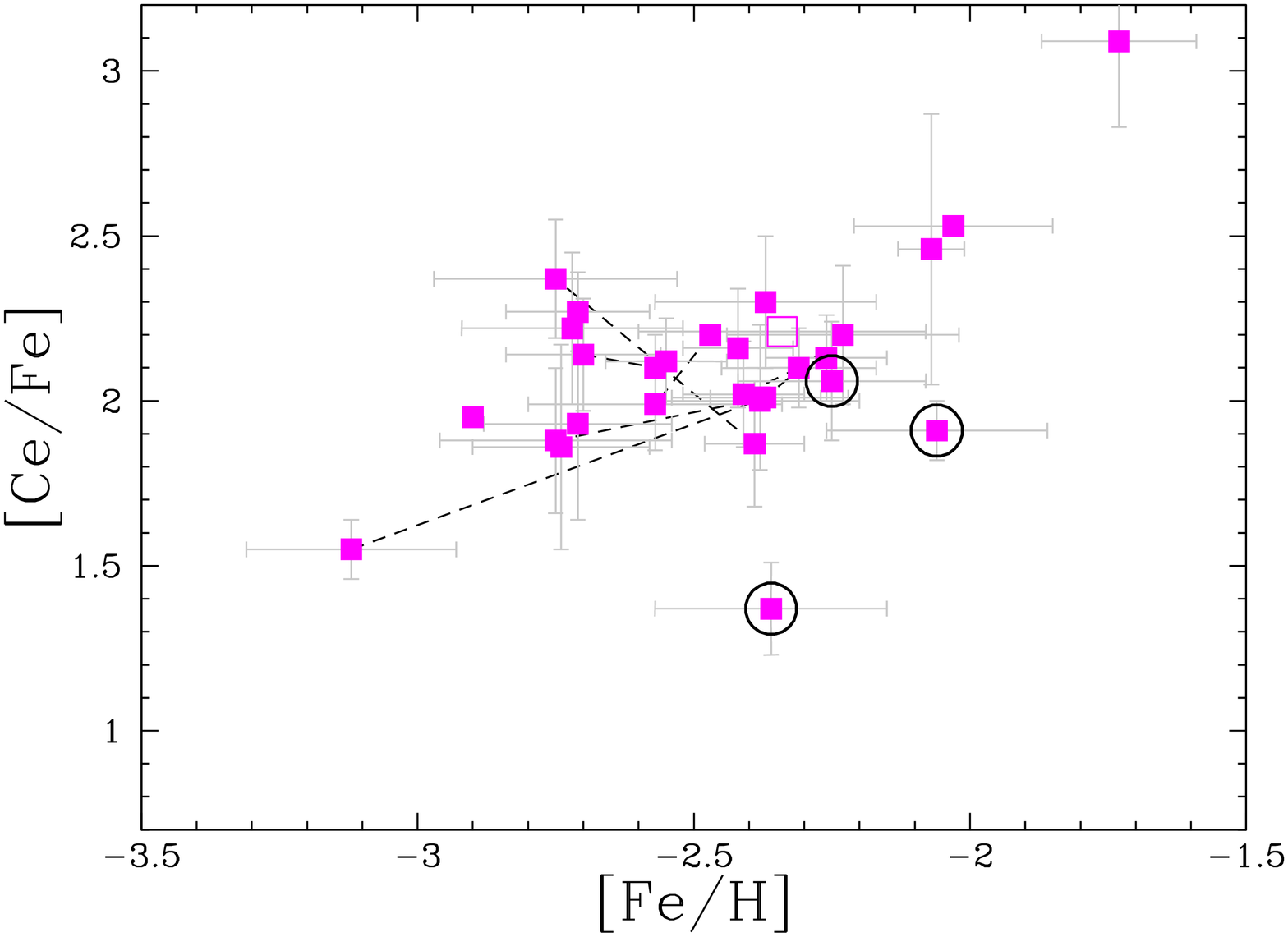}
\includegraphics[width=5cm,angle=-90]{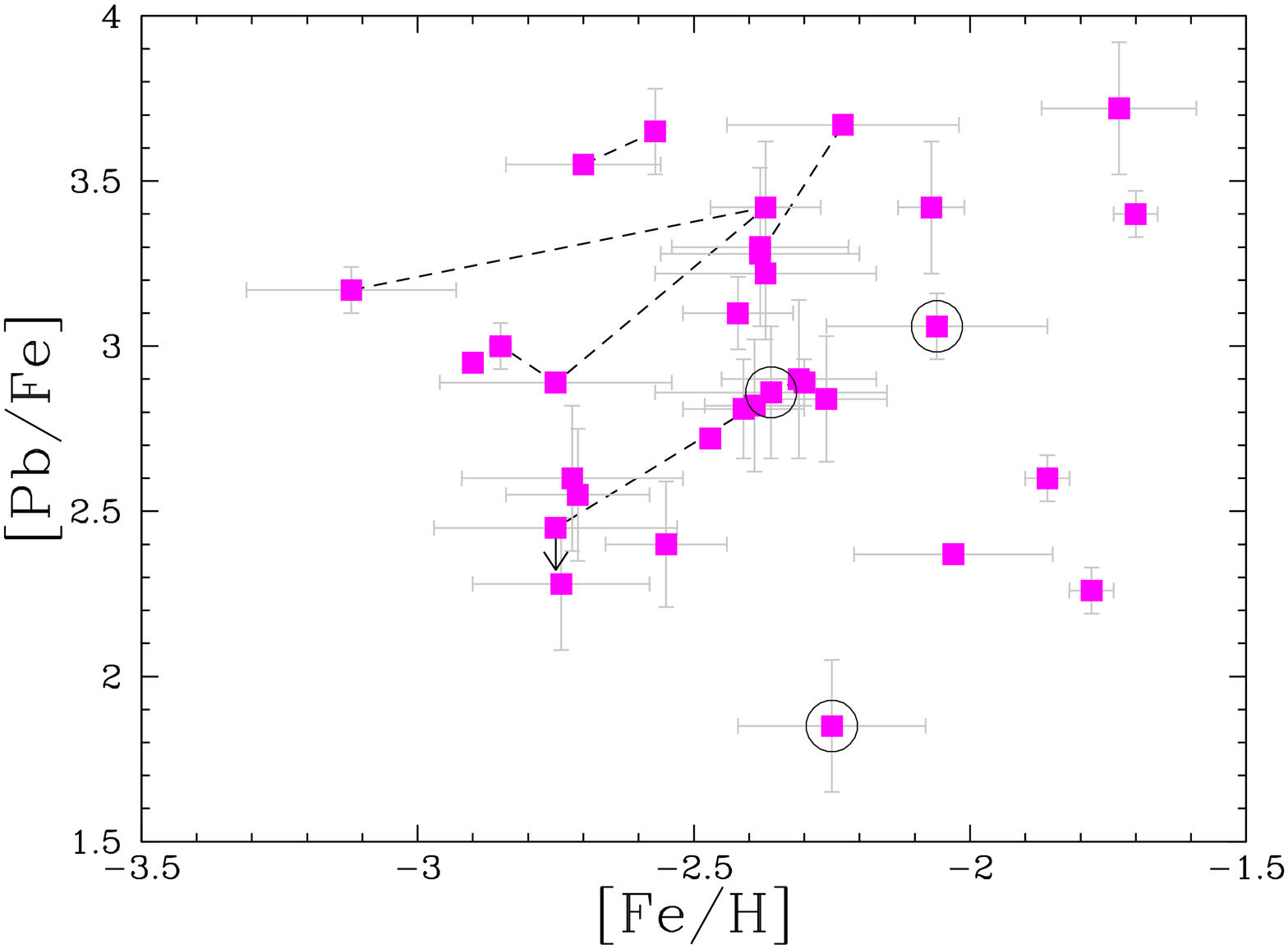}
  \caption{Ce and Pb enhancements as a function of
    metallicity in CEMP-rs stars (see Fig.~\ref{fig:legend} for a
    description of symbols). The narrow scatter in Ce/Fe
    highlights a strong dependence between Ce production
    with metallicity in contrast to Pb, also produced in large quantities but
    with a large range of abundances. If dilution was causing the scatter in [Pb/Fe], we would expect a similar scatter in [Ce/Fe].}\label{fig:CeFevsFe_rs}
\end{center}
\end{figure}

\begin{figure}[!h]
\begin{center}
\includegraphics[width=5cm,angle=-90]{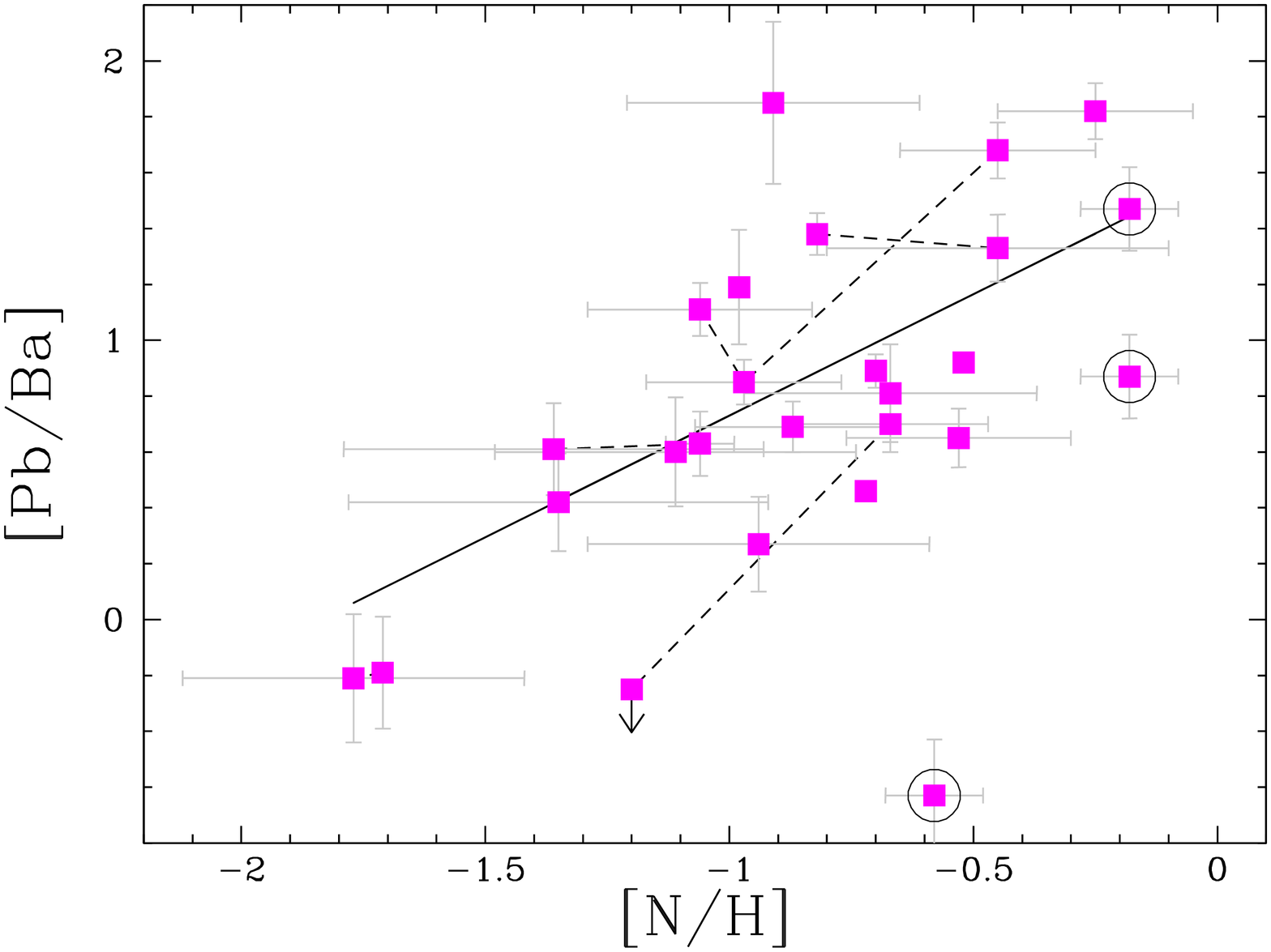}
\includegraphics[width=5cm,angle=-90]{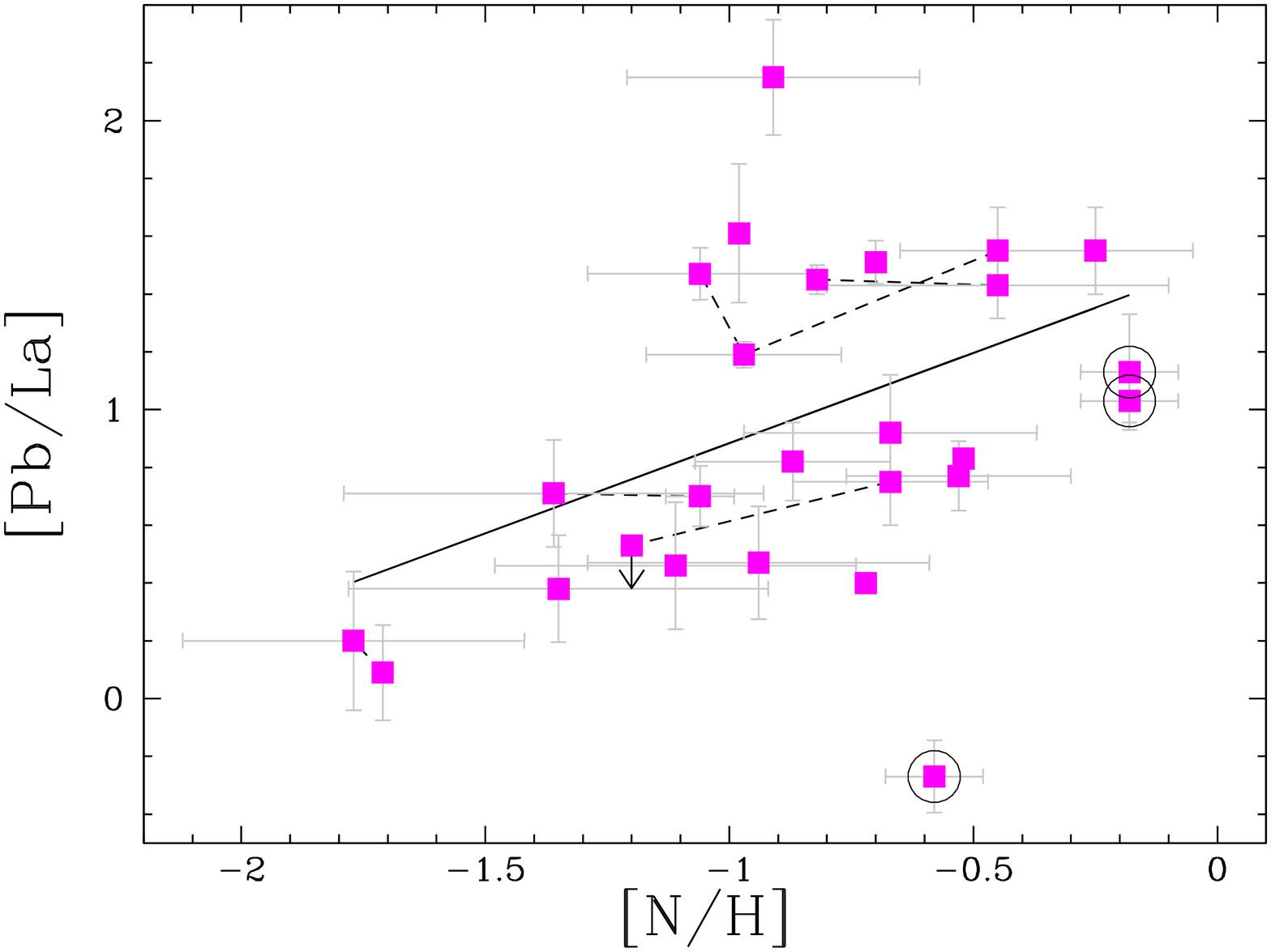}
\includegraphics[width=5cm,angle=-90]{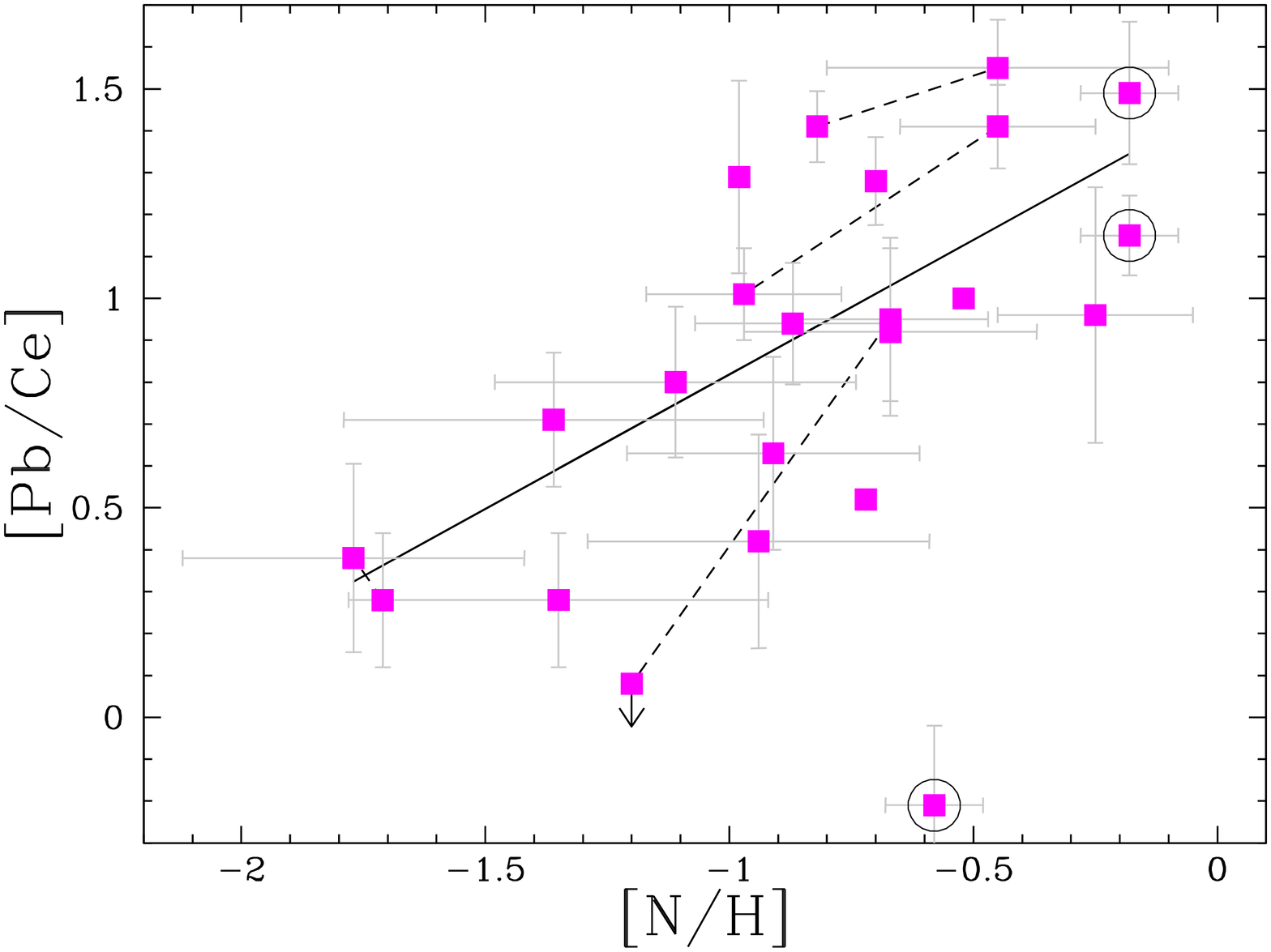}
\includegraphics[width=5cm,angle=-90]{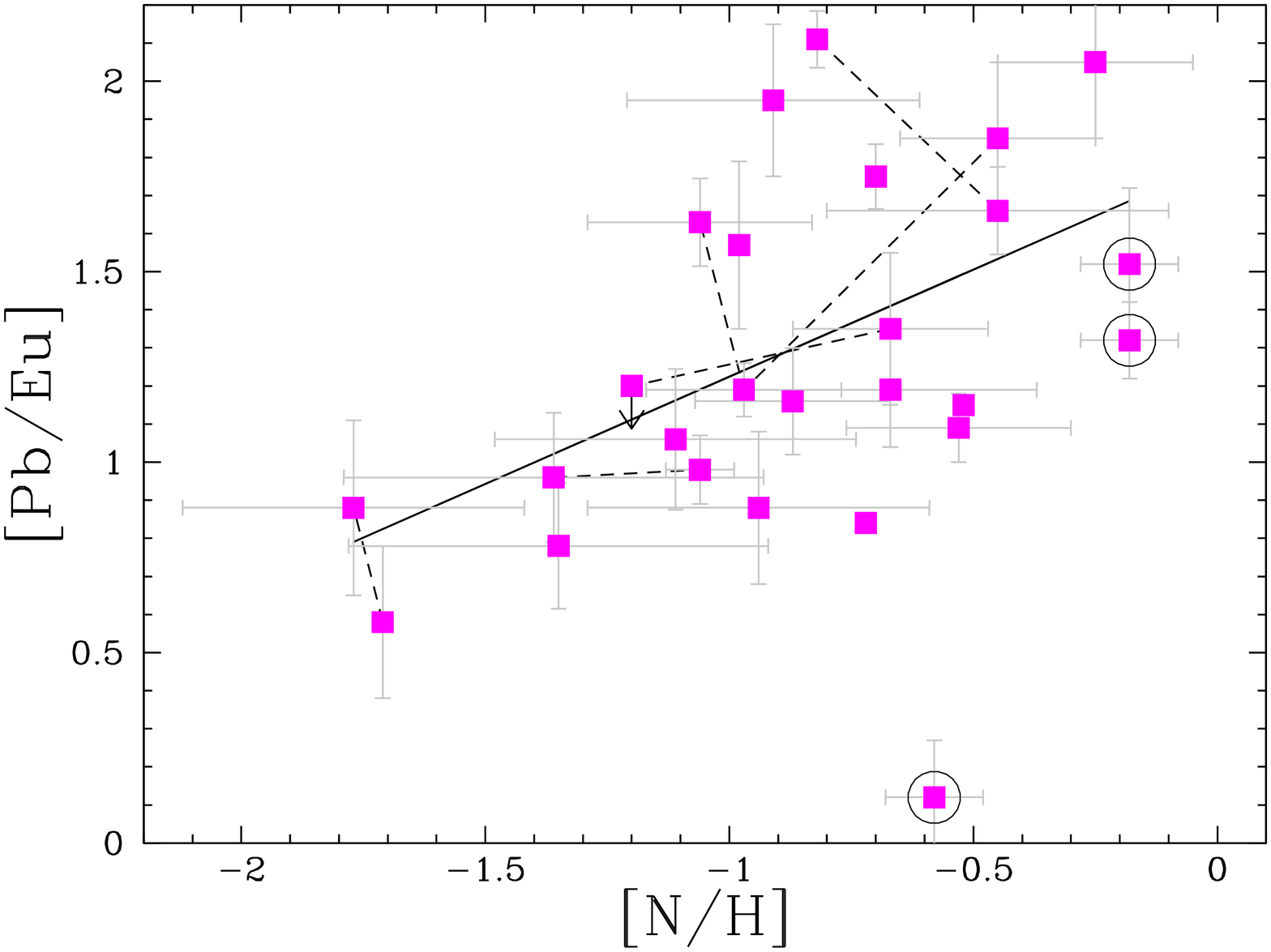}
  \caption{Third- to second-peak abundance ratios as a function of N
    in CEMP-rs stars (see Fig.~\ref{fig:legend} for a description
    of symbols). Solid lines are least-square fits. The corresponding
    correlation coefficients from top-left to down-right panels are
    0.65, 0.51, 0.68 and 0.58. Clear correlations are observed between
    third- and second-peak abundance ratios, especially in the case of Ce.}\label{fig:Pb2ndpicvsN_s}
\end{center}
\end{figure}

In Sect.~\ref{sec:CEMP-s}, we established that in CEMP-s
stars, both the s-process element overabundances (reflected in Ba/C) and the
s-process efficiency (reflected in the third-peak to second-peak
abundance ratios) depended, at least to some extent, 
on the metallicity of the AGB star. 
One striking fact in CEMP-rs stars is that the Ba/C ratio
does not show any correlation with metallicity
(Fig.~\ref{fig:BaCvsFe_rs}), 
despite the strong correlation between the production of Ce and metallicity highlighted by the small scatter in [Ce/Fe] values in CEMP-rs stars (Fig.~\ref{fig:CeFevsFe_rs}). Similarly, the s-process efficiency does not
show any correlation with metallicity but rather with N (Fig.~\ref{fig:Pb2ndpicvsN_s})! 

\begin{figure}[!h]
\begin{center}
\includegraphics[width=5cm,angle=-90]{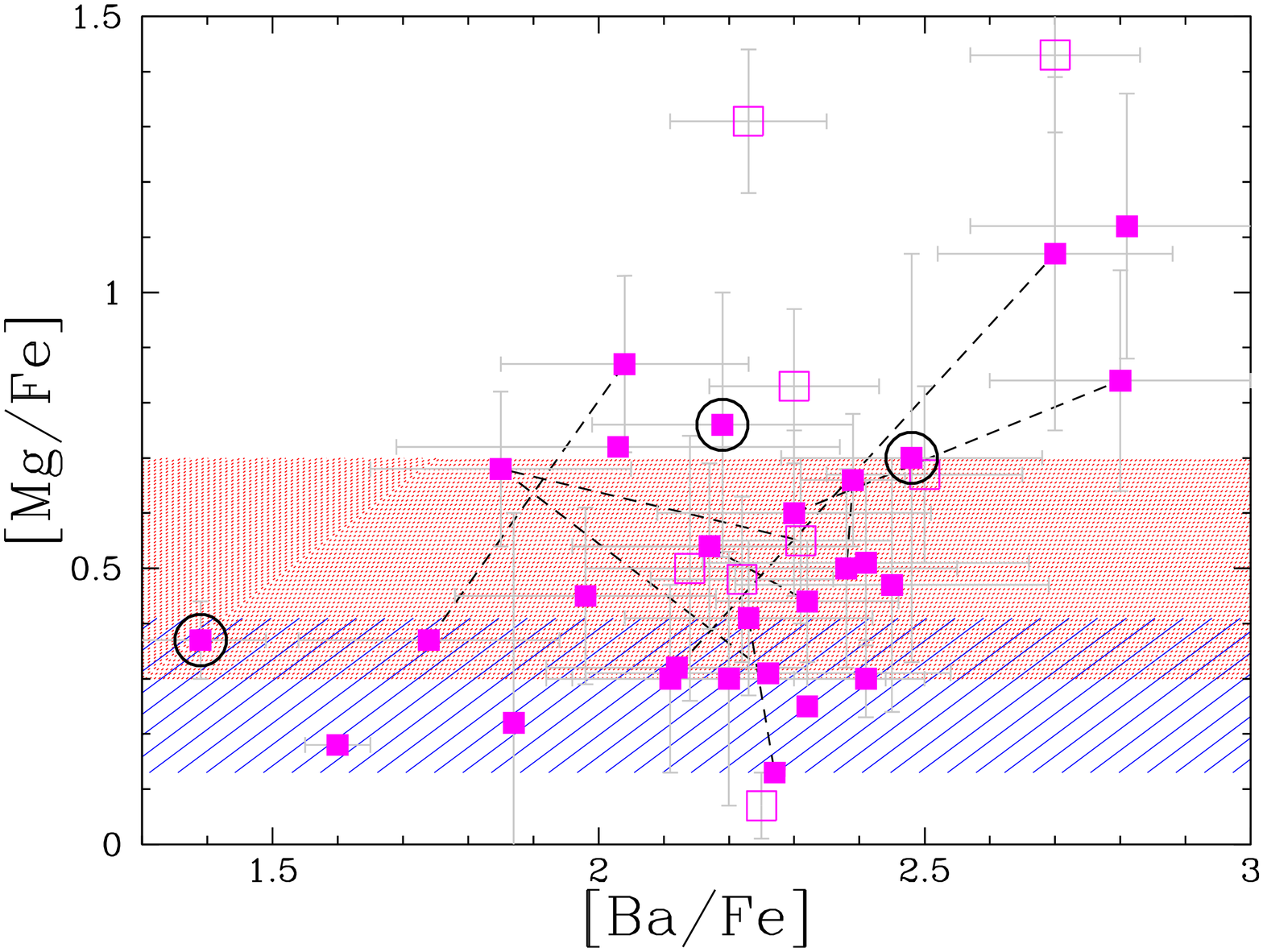}
  \caption{Mg enhancement vs Ba enhancement in CEMP-rs stars (see
    Fig.~\ref{fig:legend} for a description of symbols). The red
    shaded area represents the average [Mg/Fe] $\pm 1\sigma$ for
    CEMP-s stars while the blue hatched area represent  the average
    [Mg/Fe] $\pm 1\sigma$ for rI and rII stars reflecting the galactic
    mean value. Mg and Ba are  simultaneously enhanced in some CEMP-rs
    stars compared to the galactic mean value, reinforcing the idea that the $\rm^{22}Ne(\alpha,n)^{25}Mg$ reaction occurred intensively in these stars. }\label{fig:MgFevsBaFe_rs}
\end{center}
\end{figure}

In fact, according to \citet{Goriely2005}, the correlation between
[Pb/hs] (with hs = Ba, La or Ce) and N is evidence for a convective
s-process driven by the $\rm^{22}Ne(\alpha,n)^{25}Mg$ neutron source operating in thermal pulses of AGB stars. When
a thermal pulse occurs, N (left over from the former hydrogen-burning) is fully burnt through the
$\rm^{14}N(\alpha,\gamma)^{18}F(\beta^+,\nu)^{18}O(\alpha,\gamma)^{22}Ne$
reaction, and neutrons are released by the subsequent
$\rm^{22}Ne(\alpha,n)^{25}Mg$ reaction. When the temperature is high
enough in the convective pulse (i.e., the AGB star must be rather massive),
this reaction is a
very efficient neutron source leading to an efficient production of s-process
elements. 
This is consistent with the large overabundances of Ba, La,
Ce, and Pb observed in CEMP-rs stars. It is also expected that
$\rm^{25}Mg$ and $\rm^{26}Mg$ are significantly produced 
in AGB stars with high
masses \citep{Karakas2003}. This is what is potentially observed in
Fig.~\ref{fig:MgFevsBaFe_rs}, 
where the observed Mg (=
$^{24}$Mg+$^{25}$Mg+$^{26}$Mg) is enhanced in some of the CEMP-rs
stars. \\

 Moreover the high neutron density associated with
 $\rm^{22}Ne(\alpha,n)^{25}Mg$ is predicted to favour the so-called
 sr-nuclei production \citep{Gallino1998,Goriely2000}, notably
 $\rm^{142}Ce$, as well as r-process nuclei like $\rm^{151}Eu$ and
 $\rm^{153}Eu$ \citep{Goriely2005}. It then becomes clear why [Eu/Fe]
 increases at the same time as [Ba/Fe]
 (Fig.~\ref{fig:BaFevsEuFe}). 
 
 If the  $\rm^{22}Ne(\alpha,n)^{25}Mg$ neutron source is responsible for the synthesis of the heavy elements now observed in CEMP-rs stars,  [Ba/Mg]  (rather than [Ba/C], as is the case for CEMP-s stars) should correlate with metallicity.  The right panel of Fig.~\ref{fig:BaCvsFe_rs} hints at such a trend, although the scatter is large. The causes for such a scatter are many: (i) the Mg abundance may be dominated by the isotope $^{24}$Mg, not altered by the operation of $\rm^{22}Ne(\alpha,n)^{25}Mg$; (ii) the efficiency of the $\rm^{22}Ne(\alpha,n)^{25}Mg$ neutron source is very sensitive to the temperature at the base of the thermal pulse, which is in turn a function of mass, metallicity and pulse number. The lifetime of $\rm^{22}Ne$ against $\rm^{22}Ne(\alpha,n)^{25}Mg$ (which controls the s-process efficiency) is moreover difficult to predict with certainty, given the large uncertainties still plaguing that reaction rate. To activate the $\rm^{22}Ne(\alpha,n)^{25}Mg$ reaction, temperatures larger than about 
$ 3.5\times 10^8$~K are needed at the base of the convective
pulse. These temperatures are only expected in stars more massive than about 3$\;M_\odot$.

\begin{figure}[!h]
\begin{center}
\includegraphics[width=7cm,angle=-90]{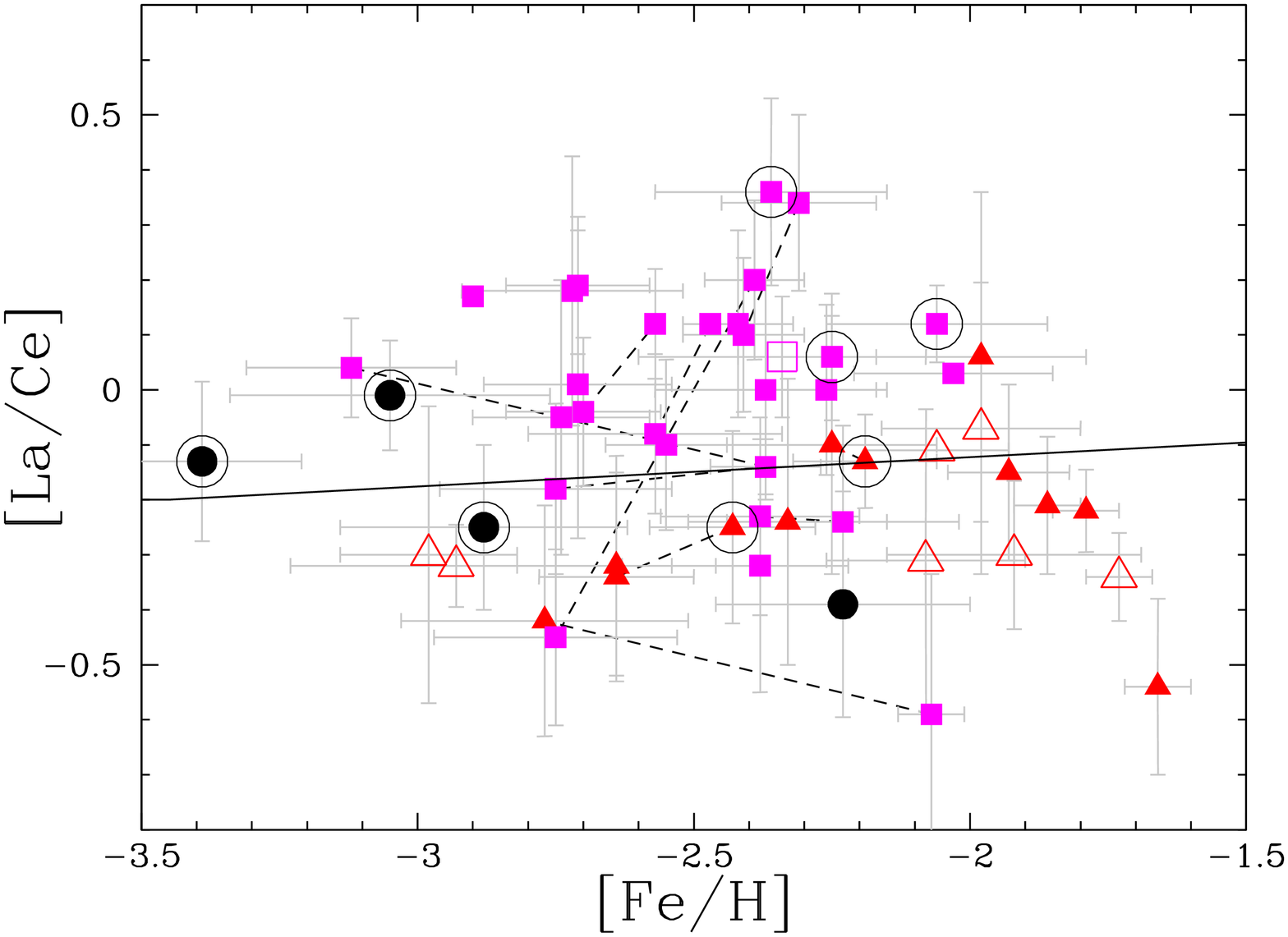}
  \caption{[La/Ce] in CEMP-s stars (red triangles) and in CEMP-rs stars (magenta squares). The solid line connects the predictions for different
   metallicities from \citet{Goriely2000} (after 10 dredge-ups), \citet{Goriely2001} and \citet{Masseron2006}. While CEMP-s stars show negative [La/Ce] ratios in agreement with predictions, CEMP-rs show [La/Ce]$\approx$0 ratios.}\label{fig:LaCevsFe_srs}
\end{center}
\end{figure}

Figure~\ref{fig:LaCevsFe_srs} shows that [La/Ce]~$\approx$0 in CEMP-rs stars while the calculations by \citet{Goriely2005} predict a negative value from the operation of the $\rm^{13}C(\alpha,n)^{16}O$ neutron source. The consistently larger [La/Ce] values observed for CEMP-rs stars are thus a strong indication that  the $\rm^{13}C(\alpha,n)^{16}O$ neutron source does not operate in those stars, as already suggested above from various other arguments. Moreover, it is very meaningful that the  [La/Ce] values observed in CEMP-rs stars are in fact compatible with the values 0.2 - 0.4~dex predicted from the operation of the $\rm^{22}Ne(\alpha,n)^{25}Mg$ neutron source in warm pulses, and after dilution in the AGB envelope \citep{Goriely2005}. The operation of  $\rm^{22}Ne(\alpha,n)^{25}Mg$ alone does not, however, lead to Pb production, and would thus appear inconsistent with the large [Pb/hs] ratios seen on Fig.~\ref{fig:Pb2ndpicvsN_s}. Fortunately, such large Pb overabundances are predicted when $\rm^{13}C(\alpha,n)^{16}O$ and $\rm^{22}Ne(\alpha,n)^{25}Mg$ operate jointly, which should be the case in a limited mass range \citep{Goriely2004,Goriely2005}.
Finally, we emphasize that the last up-to-date models from \citet{Cristallo2009} support a non negligible contribution of the $\rm^{22}Ne(\alpha,n)^{25}Mg$ reaction to the s-process in a 2 $M_\odot$ low-metallicity AGB star.

\subsection{Evidence for mixing in CEMP-rs stars}\label{sec:CEMP-rs_dilu}

It must be stressed that the evolution of the CEMP star itself could also affect the abundance ratios like [X/Fe] or [X/H], and even the [C/N] ratio. 
In general, after stars leave the main sequence and before they ascend the giant branch,
they undergo the first dredge-up. This convective process brings up material that has been processed by the CN cycle during the main sequence, and brings N-rich and C-poor material to the surface. Therefore, in normal stars that have 
homogeneous envelopes, the first dredge-up is responsible for low [C/H] and high
[N/H] above some luminosity threshold ($\log L/L_\odot \sim 1$). 

However, the situation may be different from the above picture
describing the standard first dredge-up, if a star had
transferred AGB material in a thin
layer on its surface while on the main sequence, as CEMP stars did. The dilution resulting from mass transfer is governed by the convective or radiative nature of the accretor
envelope, and by the difference between the molecular weights of the
accretor's initial envelope and of the accreted matter. 
 When the abundance differences are large enough between the initial composition and the accreted material, thermohaline mixing will operate \citep{Proffitt1989,Barbuy1992}, and dilute the accreted material in the star's envelope before the first dredge-up occurs. 
The respective depths of thermohaline mixing and first dredge-up will fix the surface [C/H] and [N/H] ratios.
If thermohaline mixing does not extend very deeply, dilution in the star's envelope by the first dredge-up will dominate and all abundances of accreted elements (e.g. [C/H], [N/H], [Ba/Fe]) will drop 
\citep{Denissenkov2008}. In the situation where thermohaline mixing is deep enough, the first dredge-up may not even leave any observable signature on the C and N abundances. In the extreme case where thermohaline mixing drags the accreted C and N down into the H burning region, C is processed through the CN cycle. The first dredge-up is then expected to bring to the surface depleted C, enhanced N \citep{Stancliffe2007} and leave unchanged the other ratios (e.g. [Ba/Fe], [C+N/H]).

\begin{figure}[!h]
\begin{center}
\includegraphics[width=8cm,angle=-90]{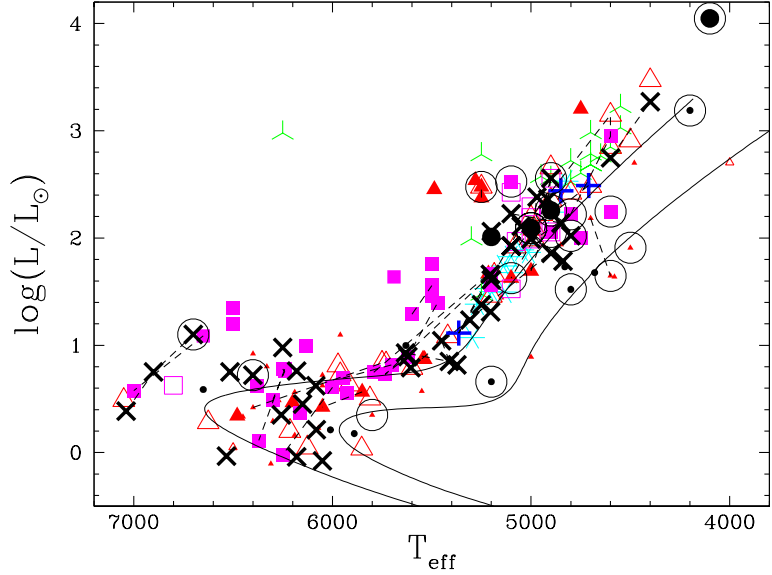}
  \caption{The Hertzsprung-Russell  diagram for the entire sample (see Figure
  \ref{fig:legend} for a description of symbols). The luminosity
  has been calculated with the following formula: $\log L/L_\odot
  = \log (M/M_\odot) + 4 \log(T_{\rm eff}/T_{\rm eff\odot})-\log (g/g_\odot)$
  adopting 0.8~M$_\odot$ for all the stars and taking $T_{\rm eff}$
  and $\log (g)$ from column (2) of Tables \ref{tab:Cstars_light} and
  \ref{tab:nonCstars_light}. The black solid lines are 12 Gyr
  isochrones for [Fe/H]=-1.01 (right curve) and -2.31 (left curve),
  both $\alpha$-enhanced, from \citet{VandenBerg2006}. While most of
  the stars fall on the isochrones, some CEMP-rs stars (magenta
  squares) and a couple of CEMP-no stars (black crosses) are bluer and
  more luminous. }\label{fig:HR}
\end{center}
\end{figure}

On top of  putting constraints on the nucleosynthesis processes in CEMP companions, we can use abundance ratios to study the depth of the dilution in the polluted star by comparing abundance ratios at different evolutionary stages. We emphasize that this approach is only valid when the overabundances in accreted material overwhelm the initial abundances. In this regard, CEMP-rs stars offer the best conditions because they show the largest overabundances. 
We determined the
luminosities of our sample stars from their surface parameters ($T\rm_{eff}$ and $\log g$)
and assumed they all are 0.8~M$\rm_\odot$ stars (see Fig.~\ref{fig:HR}).

\begin{figure}[!h]
\begin{center}
\includegraphics[width=5cm,angle=-90]{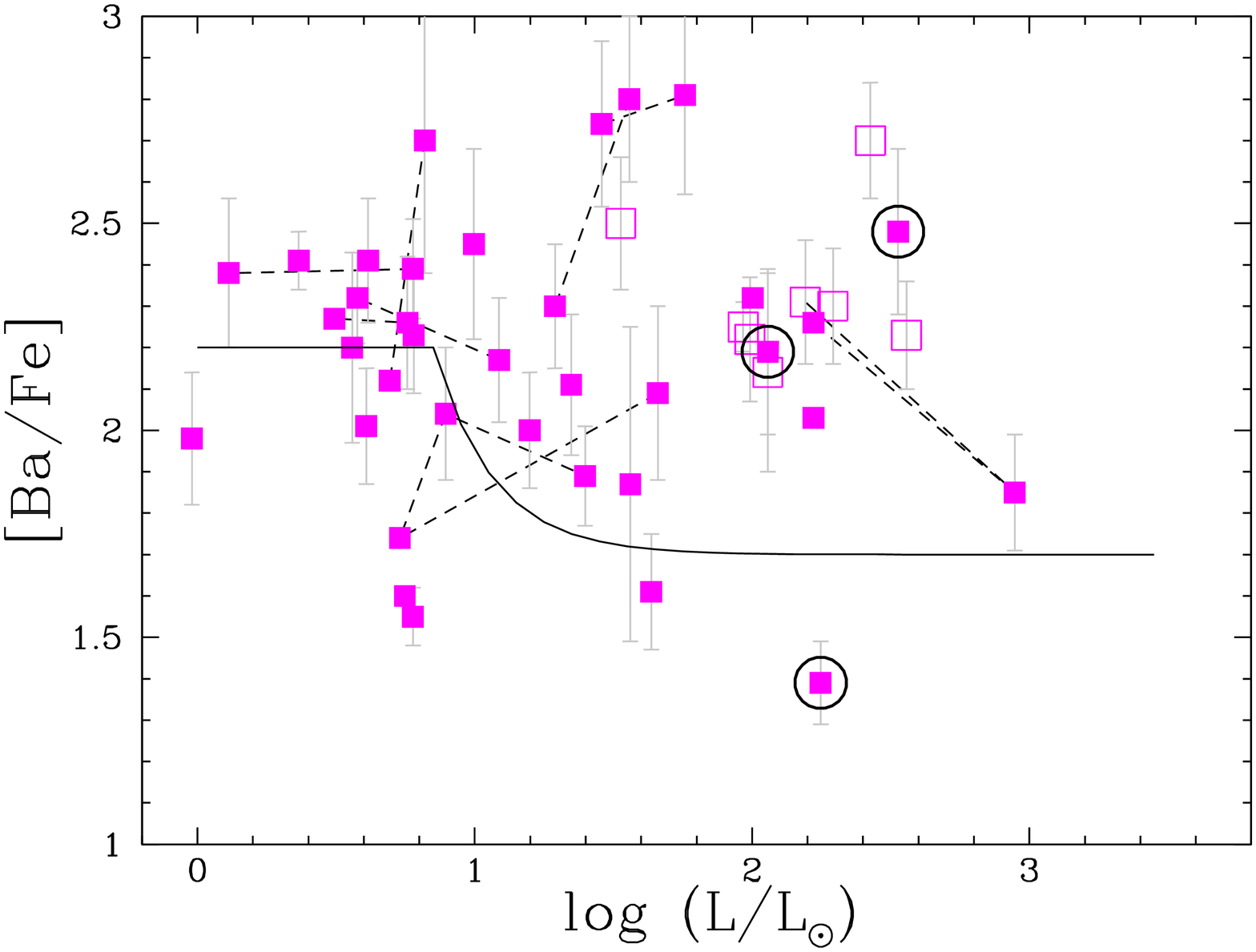}
\includegraphics[width=5cm,angle=-90]{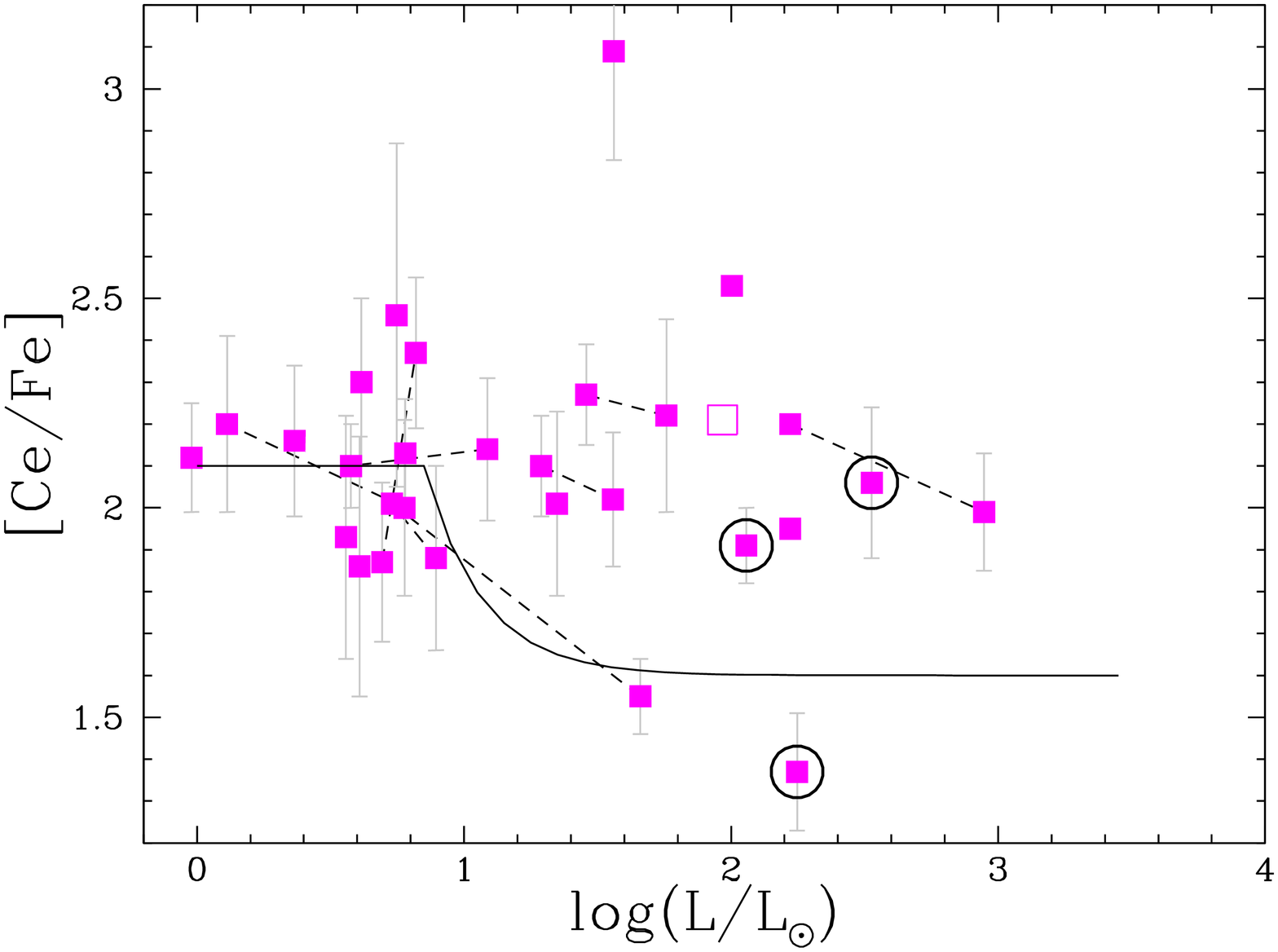}
  \caption{[Ba/Fe] and [Ce/Fe] vs luminosity in CEMP-rs stars (see Fig.~\ref{fig:legend} for a description of symbols). Whereas the
  accreted [Ba/Fe] is expected to be diluted in the star after the first
  dredge-up ($\log(L/L_\odot) \approx 1$) as predicted by
  \citet{Denissenkov2008} (solid line), [Ba/Fe] in CEMP-rs stars does not
  show any such signature of depletion between main sequence 
  and giant stars ($\rm<[Ba/Fe]>_{dwarfs} = 2.23\pm0.39$ and
  $\rm<[Ba/Fe]>_{giants} = 2.21\pm0.34$). The [Ce/Fe] scatter is very small ($\sigma = 0.24$ dex compared to $0.19$ dex random measurement uncertainty) and independent of the star luminosity.}\label{fig:dilution_rs}
\end{center}
\end{figure}

\begin{figure}[!h]
\begin{center}
\includegraphics[width=5cm,angle=-90]{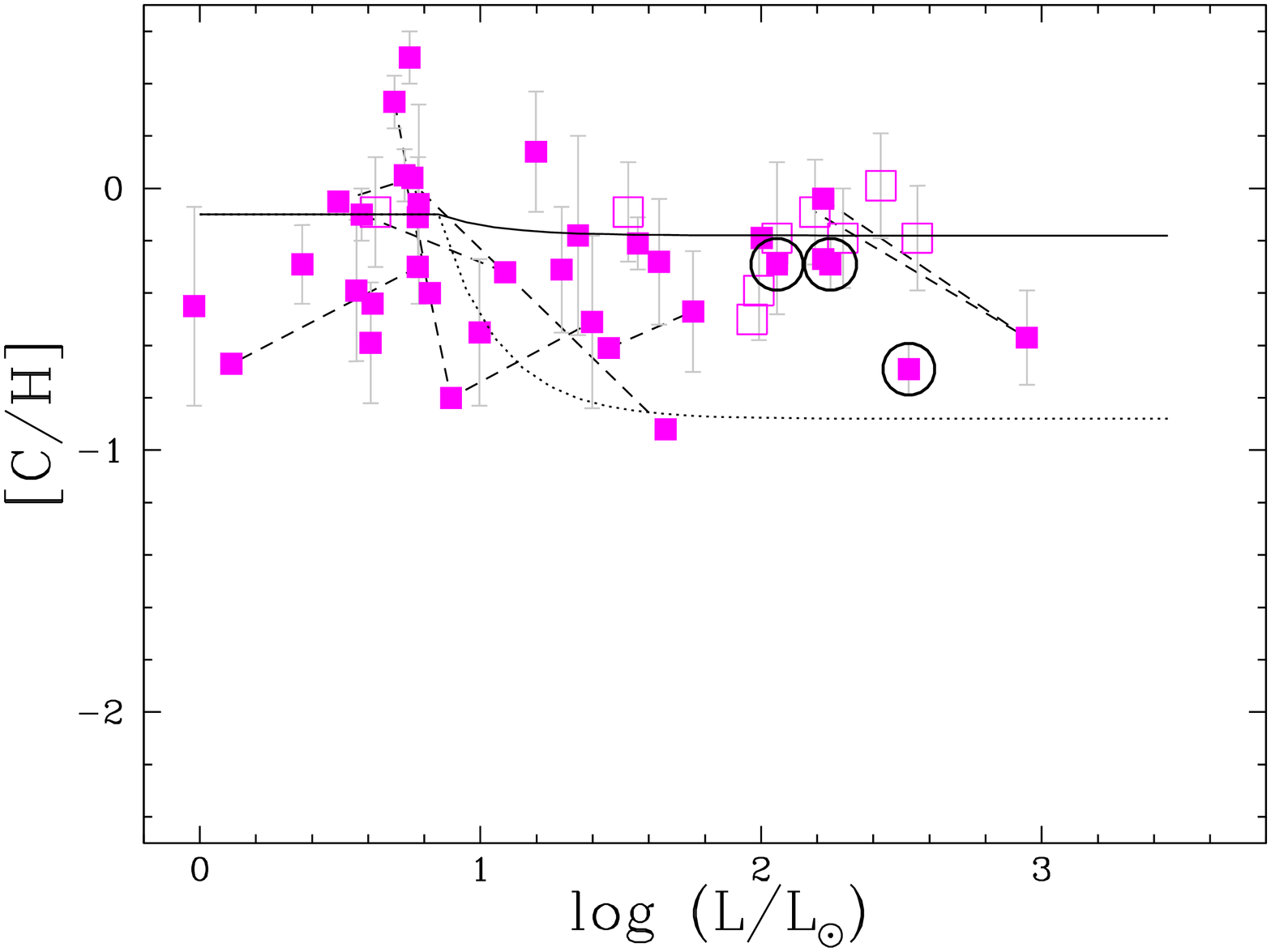}
\includegraphics[width=5cm,angle=-90]{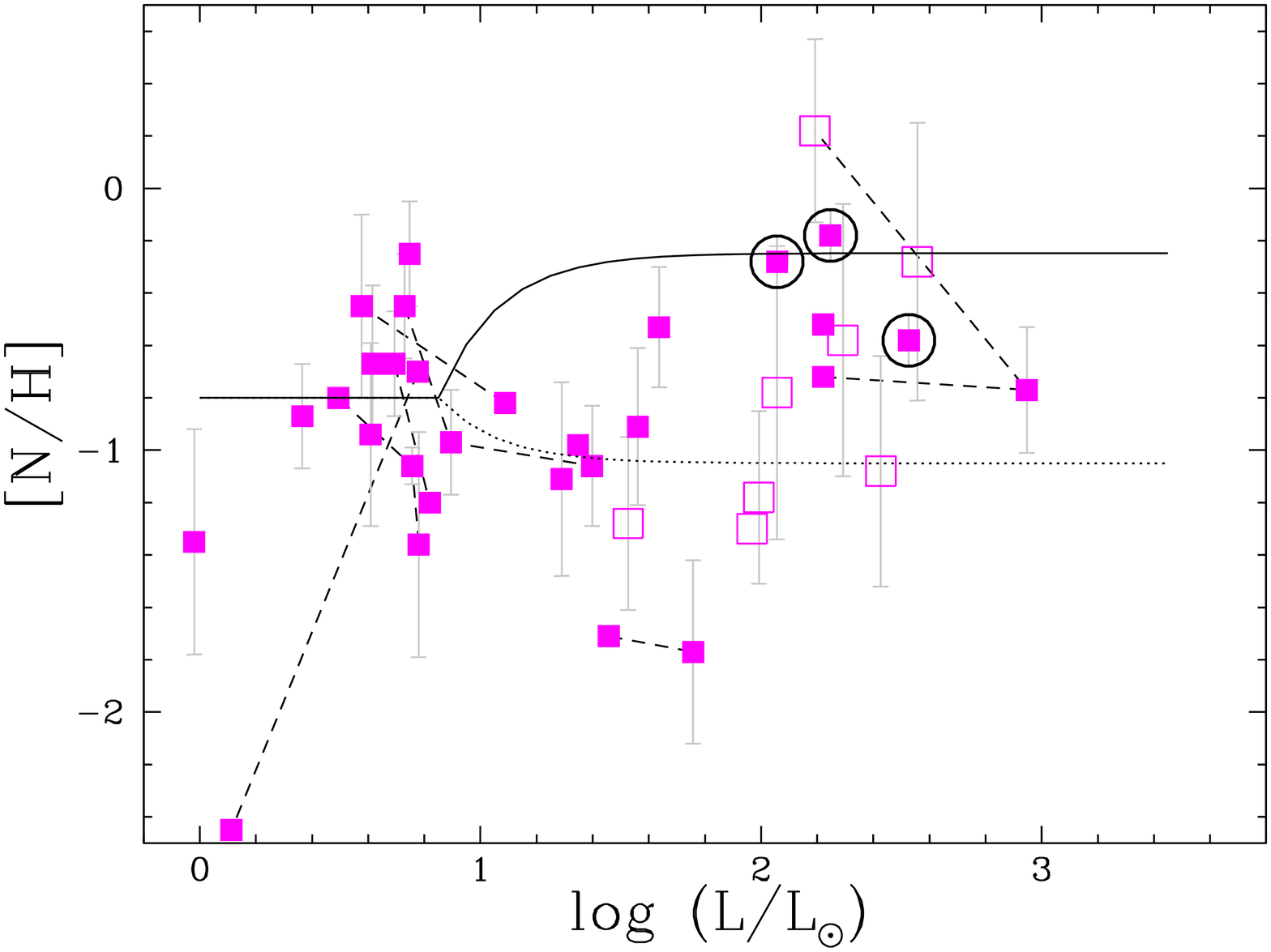}
  \caption{C and N abundances in CEMP-rs stars (see Fig.~\ref{fig:legend} for a description of symbols). The solid line is the predicted abundance trend with thermohaline mixing while the dotted line is without thermohaline mixing but dilution after the first dredge-up from \citet{Stancliffe2007}. C is constant with luminosity ($\rm<[C/H]>_{dwarfs}=-0.23\pm0.33$ and $\rm<[C/H]>_{giants}=-0.30\pm0.24$). But it is not clear whether N is enhanced after the 1$\rm^{st}$ dredge-up. We note as well that the C scatter is much lower than the N scatter. }\label{fig:CNvsL_rs}
\end{center}
\end{figure}

C abundances, [Ba/Fe] and [Ce/Fe] ratios in Figs.~\ref{fig:dilution_rs}
and \ref{fig:CNvsL_rs} clearly show no distinction between dwarfs and
giants. Therefore we see no signature of dilution in CEMP-rs stars, so that the accreted 
material must have been mixed in the star during the main sequence and/or turn off.

The accretion of material, particularly on the main sequence, may affect the position of a star in the HR diagram. It is remarkable in Fig.~\ref{fig:HR} that all stars agree well with a 12~Gyr isochrone, except for some of the main-sequence and turn-off CEMP-rs stars which are significantly bluer. (One should keep in mind, however, that in some studies, isochrones were used to determine gravities). The off-sequence stars occupy a similar region of the HR diagram as the blue stragglers, where mixing has occurred on the main sequence. 
\citet{Stancliffe2007} found in their model that thermohaline mixing of the C-rich material is deep enough to reach the H combustion layers, so that it stimulates H burning and increases the luminosity of the star. \citet{Jonsell2006} already suspected a relation between CEMP-rs stars and blue stragglers (hypothesis VIII).
According to \citet{Stancliffe2007}, the additional luminosity is a direct consequence of the enhanced CN mixed throughout the star, thus boosting the CNO cycle.
Therefore, they  also expect strong N enhancement after the
  1$\rm^{st}$ dredge-up (occurring around $\log(L/L_\odot)\approx 1$).
But Fig.~\ref{fig:CNvsL_rs} shows no clear increase in N for stars with $ \log(L/L_\odot)> 1$. \\
The examination of the abundances as a function of CEMP-rs parameters leads to 3 main conclusions: 
(i) the abundances observed in CEMP-rs stars are not affected by the dilution associated with the first dredge-up in the atmosphere of the CEMP star;

(ii) on the contrary, the accreted material seems to alter the evolution of CEMP-rs stars, like for blue stragglers. This requires a large amount of accreted material, consistent with the fact that the companion had to loose a large amount of mass, thus being relatively massive;

(iii) the variation  of the amount of transferred material from one CEMP-rs star to another is below the measurement uncertainties (right panel of Fig.~\ref{fig:dilution_rs}). Therefore, abundance trends such as that observed for N/H are free from the scatter associated with variable mass-transfer efficiencies. \\

\subsection{Thermohaline mixing and first dredge-up dilution in CEMP-s and CEMP-no stars.}\label{sec:CEMP-s_thermohaline}
\begin{figure}[!h]
\begin{center}
\includegraphics[width=5cm,angle=-90]{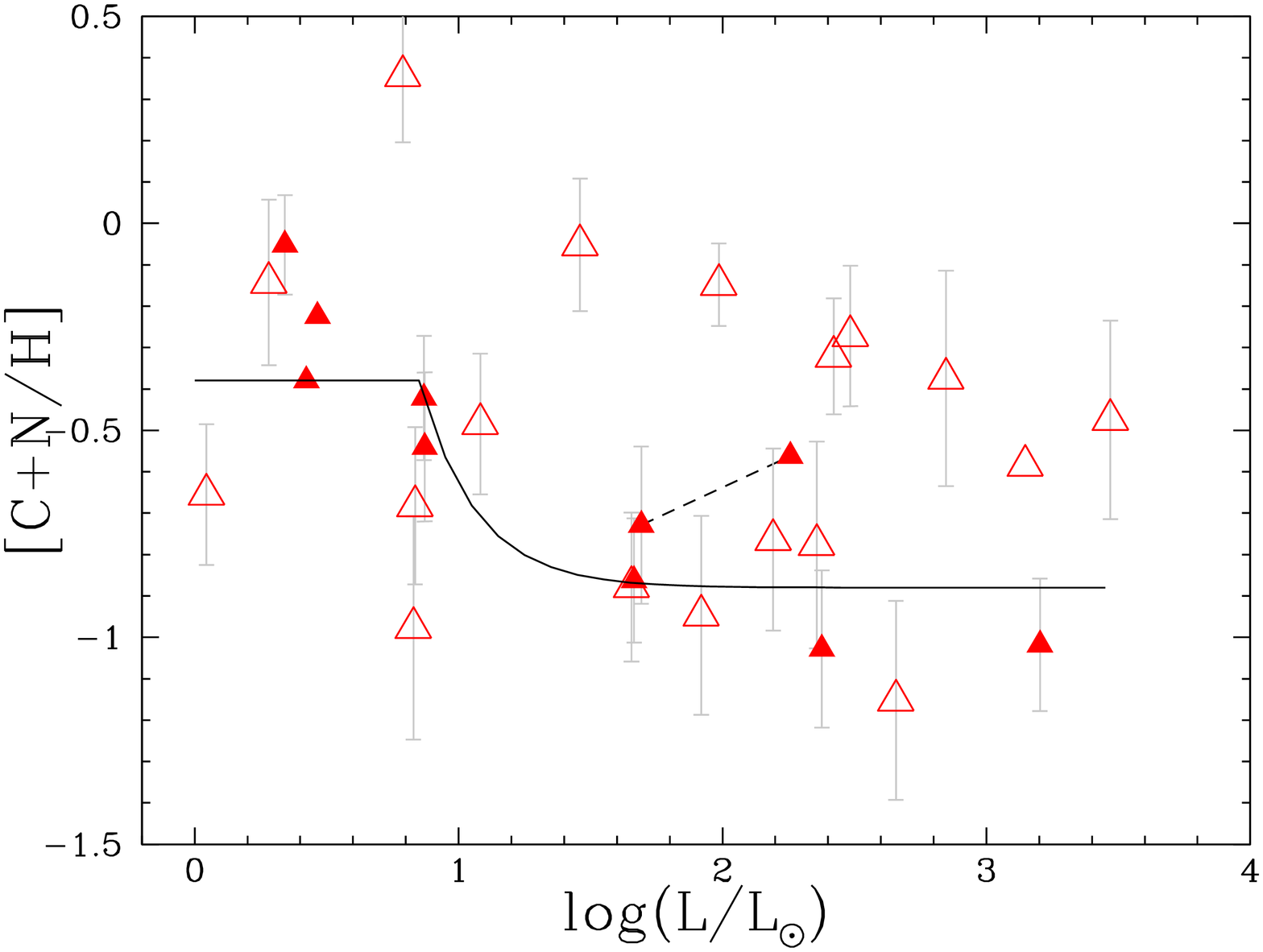}
\includegraphics[width=5cm,angle=-90]{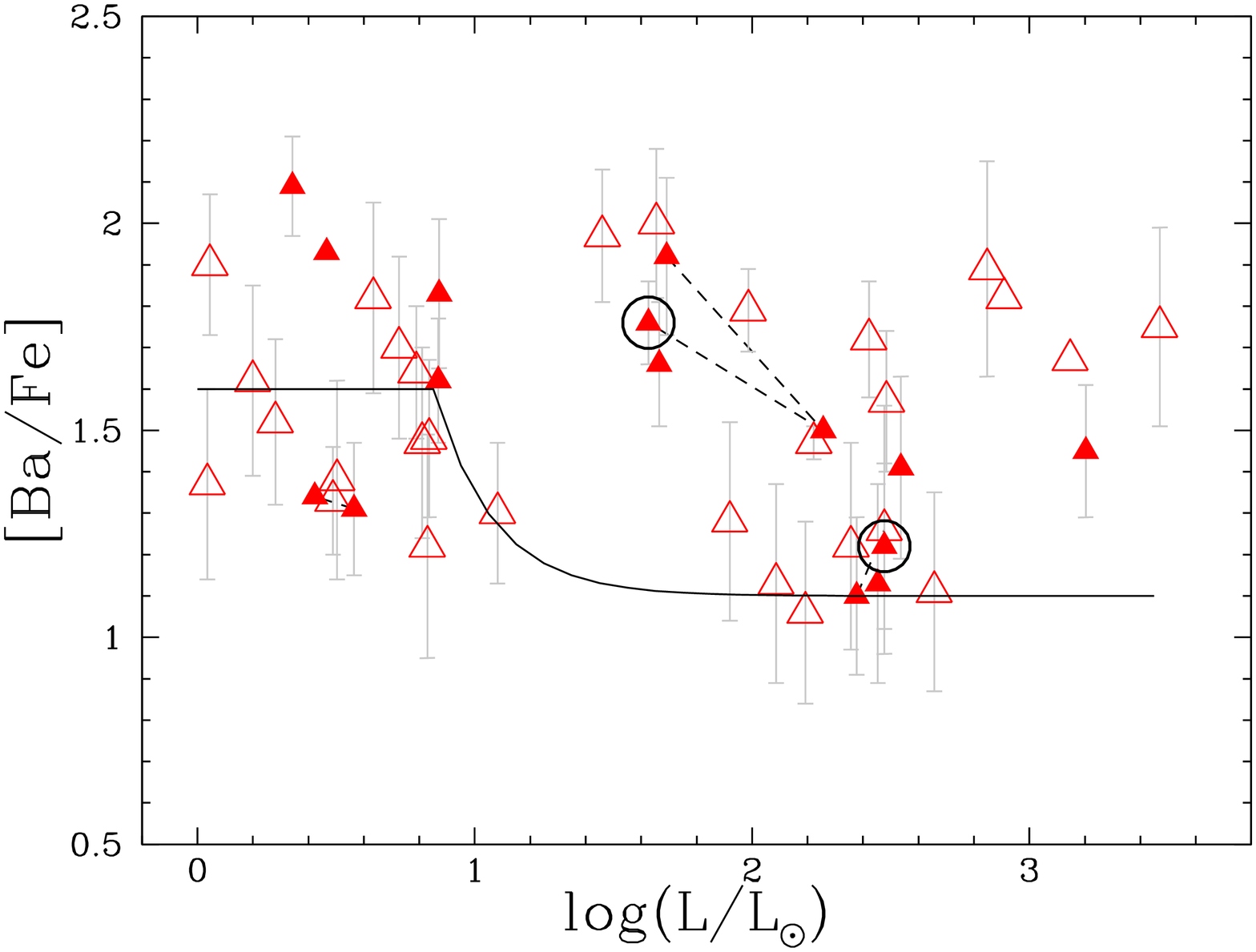}
  \caption{[C+N/H] and [Ba/Fe] in CEMP-s stars as a function of luminosity. While the average [C+N/H] shows a discrepancy between dwarfs and giants ($<$[C+N/H]$>$=-0.38$\pm$0.35 for dwarfs and -0.64$\pm$0.32 for giants) in apparent agreement with the predictions of \citet{Denissenkov2008} (solid line), [Ba/Fe] is very homogeneous between dwarfs and giants ($<$[Ba/Fe]$>$ =1.57$\pm$0.25 for dwarfs and 1.51$\pm$0.30 for giants).}\label{fig:C+N_BaFevsL_s}
\end{center}
\end{figure}

Like CEMP-rs stars, CEMP-s stars show large C enhancements in their atmosphere; therefore thermohaline mixing could be expected to be at work as well. \citet{Aoki2008} point out that their earlier data \citep{Aoki2007} for Ba-enhanced CEMP stars do not support thermohaline mixing. We confirm that in CEMP-s stars, the average [C+N/H] show an apparent discrepancy between giants and dwarfs, but [Ba/Fe] remains remarkably homogeneous (Fig.~\ref{fig:C+N_BaFevsL_s}). The dilution should affect in a similar way all the accreted elements. Since this is not what is observed in CEMP-s stars from the comparison of C+N and Ba data, we conclude that there is no clear signature of dilution of the accreted material after the first dredge-up in CEMP-s stars, in contrast to Aoki et al. (2008)'s conclusions.

For CEMP-no stars, the situation is different. As noted by \citet{Aoki2002c}, CEMP-no stars show mild
C-enrichment on average: CEMP-no stars have [C/H]=-1.42$\pm$0.86 whereas CEMP-rs have [C/H]=-0.28$\pm$0.28 and CEMP-s
stars have [C/H]=-0.62$\pm$-0.34. Because thermohaline mixing depends on the molecular weight gradient induced by the accretion of heavy elements, a less efficient thermohaline mixing than the one in CEMP-rs stars must be expected for CEMP-no stars.  
\citet{Lucatello2006} and \citet{Aoki2007} do find a trend of [C+N/H] with the CEMP luminosity. Hence, these two works conclude that thermohaline mixing was negligible and that what is seen is the dilution of the accreted material by the first dredge-up, as estimated by \citet{Denissenkov2008}. Note however that \citet{Lucatello2006} do not separate CEMP classes, so the trend they found is certainly smoothed between CEMP-no, CEMP-s and CEMP-rs stars.

\begin{figure}[!h]
\begin{center}
\includegraphics[width=5cm,angle=-90]{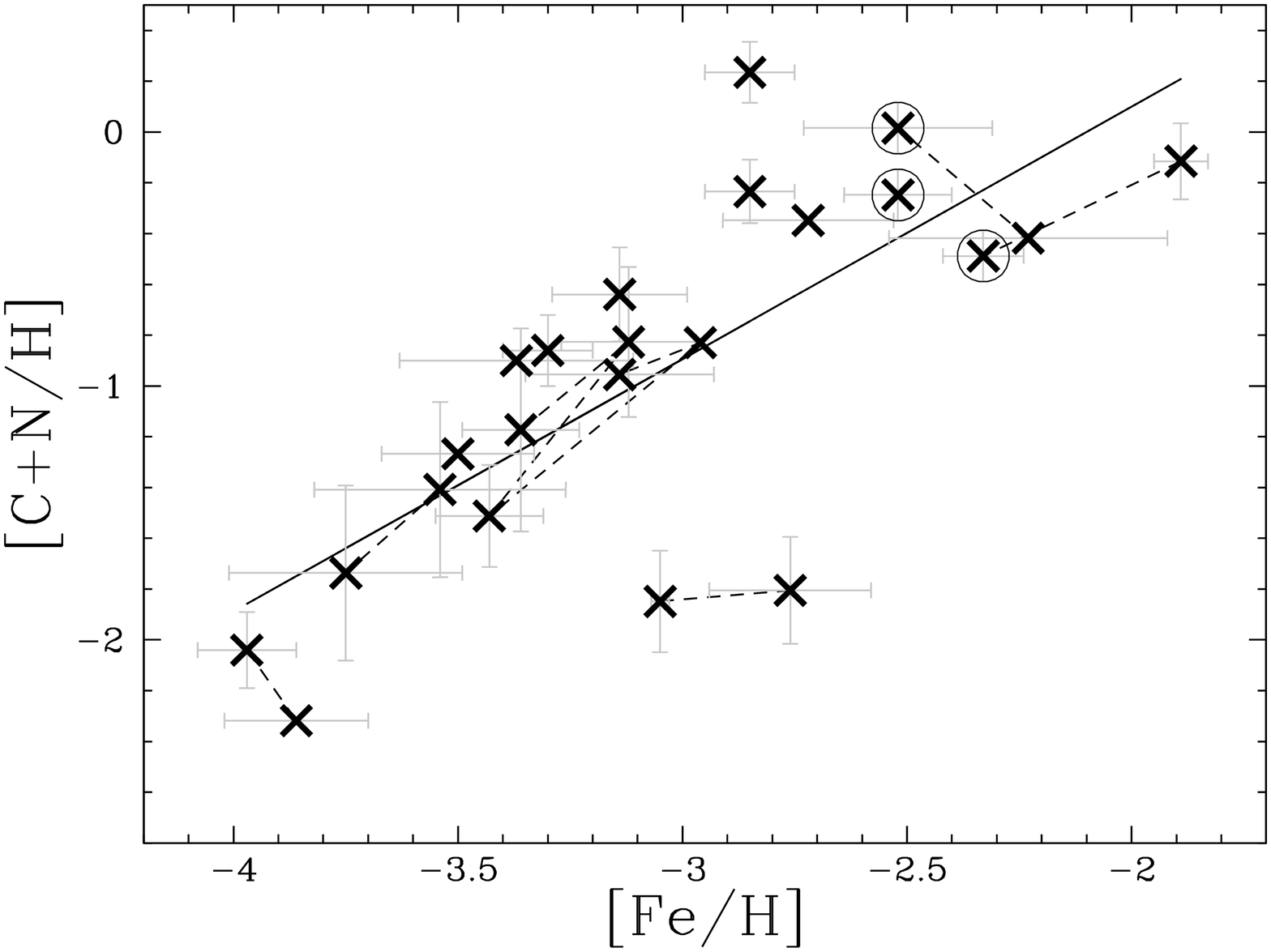}
\includegraphics[width=5cm,angle=-90]{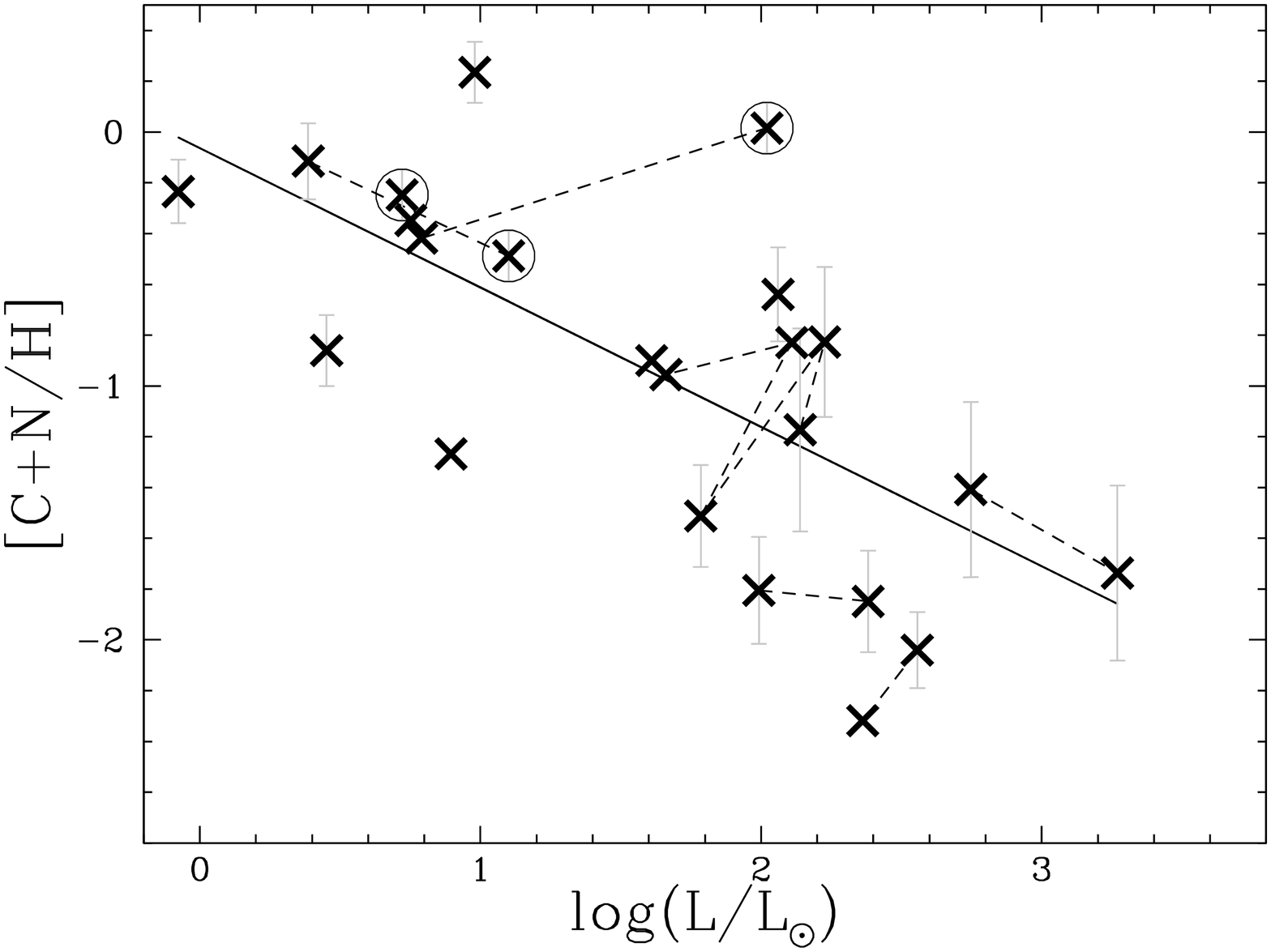}
\includegraphics[width=5cm,angle=-90]{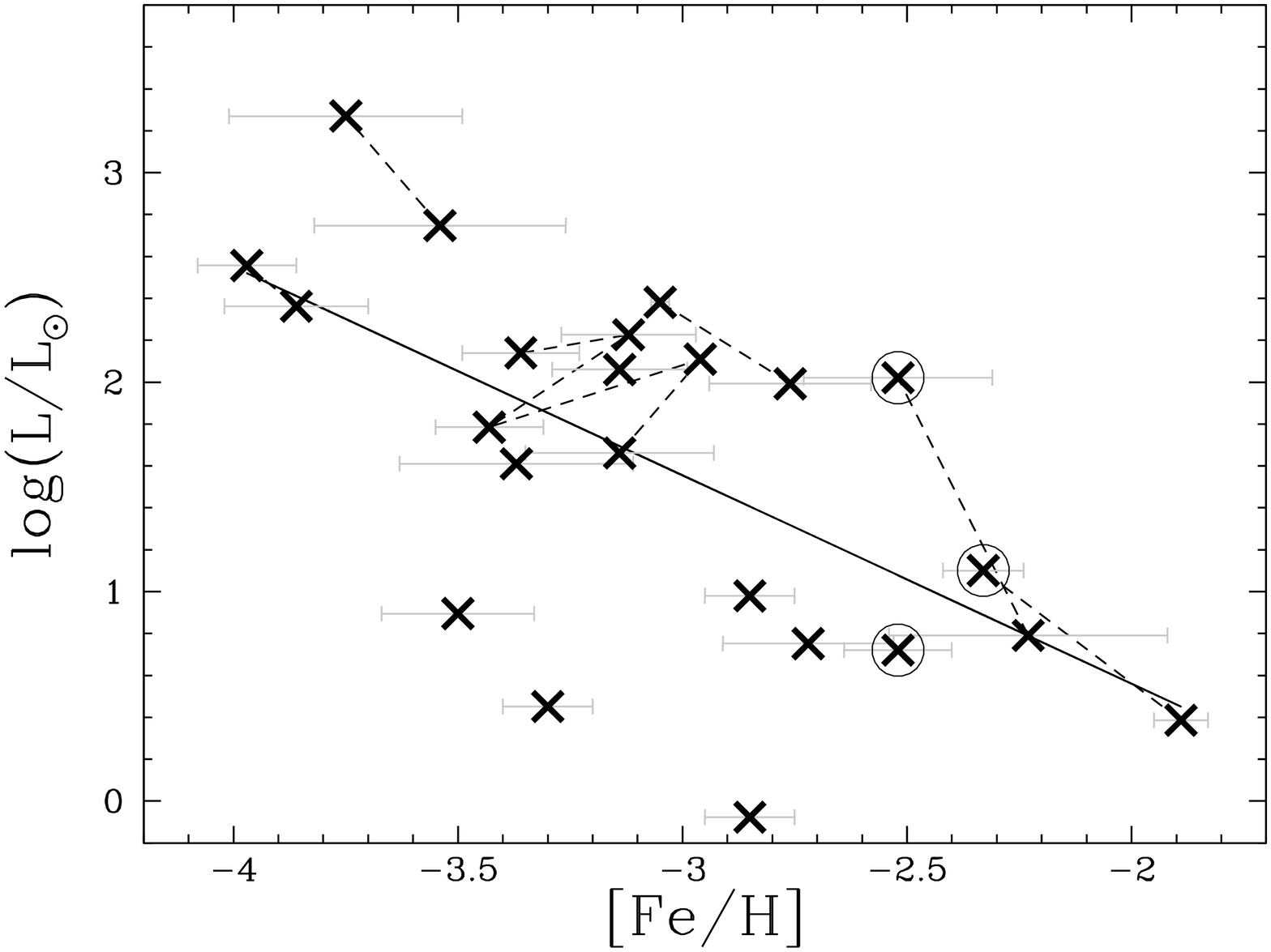}
  \caption{[C+N/H] in CEMP-no stars as function of metallicity (upper left) and luminosity (upper right). There is a correlation in both cases (the solid lines represent the least-square fits), implying that luminosity is anti-correlated with metallicity (lower panel). This highlights an observational bias in CEMP-no stars.}\label{fig:C+NvsFeL_no}
\end{center}
\end{figure}

We show in Fig.~\ref{fig:C+NvsFeL_no} that there is a correlation between metallicity and luminosity in CEMP-no stars. In surveys, more giants than dwarfs are expected to be found as their large luminosity allows us to detect them in a larger volume. Consequently, it would appear natural that the more the metallicity decreases and the fewer metal-poor stars detected \citep[e.g. ][]{Ryan1991}, the more the frequency of giants increases, leading to an apparent correlation between metallicity and luminosity \citep[see ][for more detailed estimates]{Masseron2006}.
Therefore we argue that the conclusions of \citet{Lucatello2006} and \citet{Aoki2007} are observationally biased so that the decrease of [C+N/H] as a function of luminosity is dominated by the metallicity trend. Thus, we conclude that we cannot with the current data measure the impact of thermohaline mixing in CEMP-no stars.

\subsection{Comparison between CEMP-s and CEMP-rs stars}\label{sec:CEMPs-rs_relation} 
After examining the neutron-capture elements in CEMP-rs stars, we compare in this section light-element abundances in CEMP-s and CEMP-rs to compare the properties of their respective companions.

\subsubsection{Carbon and Nitrogen}\label{sec:CandN} 

N is generally thought to be the result of the CN cycle. As a 
consequence of this cycle, $\rm ^{14}$N increases at the expense of 
$\rm^{12}C$, and $\rm^{12}C/^{13}C$ is lowered from its initial value. 
But in low- and intermediate-mass metal-poor stars, $\rm ^{14}$N can be 
produced at many different stages, notably after the first dredge-up, 
during the RGB phase through ``extra mixing'' 
\citep[e.g.][]{Gilroy1989,Boothroyd1999} and during the AGB phase if HBB 
occurs at the bottom of the convective envelope or if there is ``extra 
mixing'' below the convective envelope, of a similar nature than the 
one occurring during the RGB, but called Cool-Bottom Processing (CBP) in 
the framework of the AGB evolution \citep[][]{Abia2001,Nollett2003}. 
For CEMP stars, the situation is thus more complex as both the CEMP star 
and its companion can have undergone CN processing. Based on the 
observation that CEMP stars are also N-rich stars, it is commonly 
admitted that N originates from the same source as C, i.e. the 
companion. Very few CEMP stars in our sample have high enough luminosities ($\rm \log(L/L_\odot) \gtrsim 2.2$) to undergo extra mixing \citep[][]{Spite2005}, and \textit{a fortiori} neither HBB nor CBP. Thus, CEMP stars could not have produced N through these processes. Moreover, we have demonstrated in Sect.\ref{sec:CEMP-rs_dilu} that there is no significant enhancement of N after CEMP stars' first dredge-up. Finally, there are strong arguments in favour of N originating from the former AGB companion, notably the fact that the abundances of the s-process elements in CEMP-rs 
stars are correlated with N  (Fig.~\ref{fig:Pb2ndpicvsN_s}). 
Since it is very unlikely that CEMP stars have synthesized any s-process elements in their interiors, their AGB companions must have 
produced both the heavy elements and all the observed N (as well as $\rm^{12}$C and $\rm^{13}C$).

\begin{figure}[!h] 
\begin{center} 
\includegraphics[width=8cm,angle=-90]{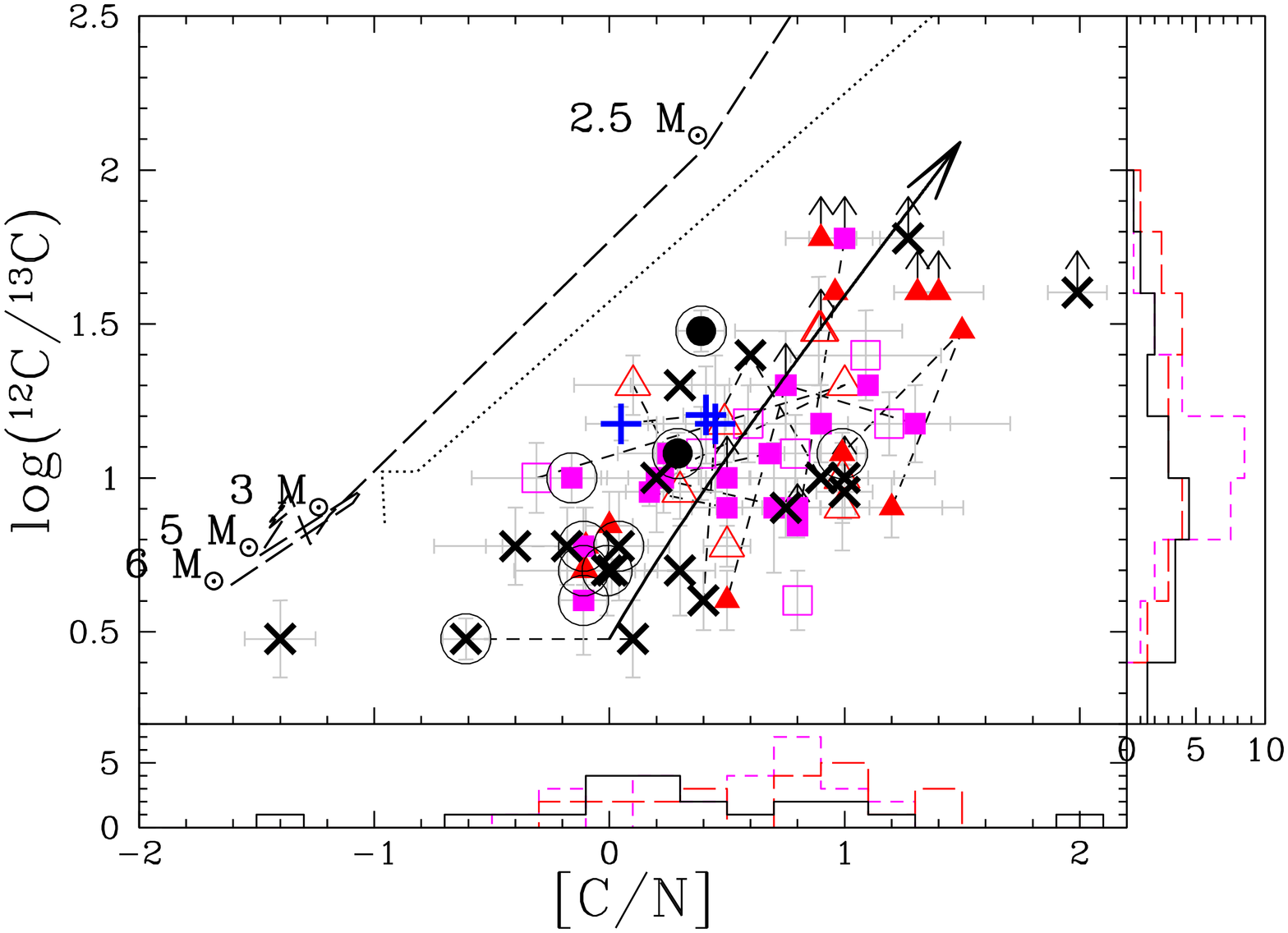} 
\caption{$\rm^{12}C/^{13}C$ as a function of [C/N] ratios for CEMP stars 
(see Fig.~\ref{fig:legend} for symbols). The 
long-dashed line is the prediction for low-metallicity AGB stars 
from \citet{Karakas2007} and the dotted line is the prediction from 
\citet{Herwig2004}. It is remarkable that all CEMP stars fall in the 
same region in this diagram. The black arrow represents the addition of 
an increasing amount of pure $\rm^{12}C$ starting from a 
$\rm^{12}C/^{13}C$ near the CN equilibrium value and [C/N]=0. No current 
models are able to predict the intermediate C/N ratios observed in CEMP 
stars (see text). We note that CEMP-no stars have on average a slightly 
lower [C/N] ratio and a lower $\rm^{12}C/^{13}C$ than CEMP-s and CEMP-rs 
stars.}\label{fig:C13vsCN} 
\end{center} 
\end{figure}

The model predictions displayed in Fig.~\ref{fig:C13vsCN} reveal that 
intermediate-mass AGB stars with HBB should have low [C/N] and low 
$\rm^{12}C/^{13}C$ ratios. In contrast, for low-mass AGB stars, no CN 
processing is expected after the second dredge-up. Consequently, in 
those stars, the cumulative amount of  $\rm^{12}C$  dredged-up during the 
AGB phase results in a high [C/N] ratio and a high $\rm^{12}C/^{13}C$ 
ratio. In contrast, all CEMP stars show relatively low $\rm^{12}C/^{13}C$ ratios. 
This suggests that CN processing may have occurred between 3$\rm^{rd}$ dredge-ups, which are known
to bring up large amount of $\rm^{12}C$ to the surface. But, as already 
noticed by \citet{Johnson2007}, CEMP stars show C/N ratios intermediate 
between the high ratios expected in low-mass stars and the low ratios 
expected in intermediate-mass stars with HBB. Actually, a low 
$\rm^{12}C/^{13}C$ ratio with [C/N] $\approx 0$ cannot be obtained by a 
complete CN cycle (which leads instead to
 $\rm^{12}C/^{13}C \approx 4$  and 
[C/N]$\approx -1.3$, as does HBB). Therefore, as already concluded by 
\citet{Aoki2002c}, C has only been partly processed (possibly in the H-burning shell) before it reaches the quiescent convective envelope. This is indeed a characteristic feature of CBP. 
Moreover, we show in Fig.~\ref{fig:C13vsCN} that there is a correlation 
between $\rm^{12}C/^{13}C$ and C/N in CEMP stars. It is remarkable that 
all observed CEMP stars are located in the same region of this diagram 
and follow the same trend. 
This correlation is well reproduced by varying the amount of pure 
$\rm^{12}C$. Indeed, the observed trend might be ascribed to the 
competition between third dredge-ups and CBP. \\

In brief, although the mechanism responsible for the N production may be 
attributed to CBP, no current AGB models  reproduce the trend observed 
in Fig.~\ref{fig:C13vsCN}. Thus we cannot use N to bring 
additional constraints concerning the mass of the progenitor. Finally, 
as long as the physical mechanism responsible for the CBP remains 
unidentified, the relation between the nitrogen and s-process 
productions in CEMP-rs stars is difficult to interpret because these two 
elements are supposedly produced in two distinct parts of the AGB star. Nevertheless, 
assuming that some common mechanism, such as rotation 
\citep{Decressin2006}, thermohaline mixing \citep{Cantiello2008} or Dual 
Shell Flashes \citep{Campbell2008}, drives the CBP and the s-process, 
this correlation can be used to test these different models.

\subsubsection{Oxygen}
\label{Sect:oxygen}

 Oxygen is also expected to be enhanced
in low-metallicity AGB stars, because of hotter conditions
in the He-burning shell [thus activating
$\rm^{12}C(\alpha,\gamma)^{16}O$] and deeper dredge-ups compared to
solar-metallicity AGB stars \citep{Herwig2004,Karakas2007}. 
\begin{figure}[!h]
\begin{center}
\includegraphics[width=8cm,angle=-90]{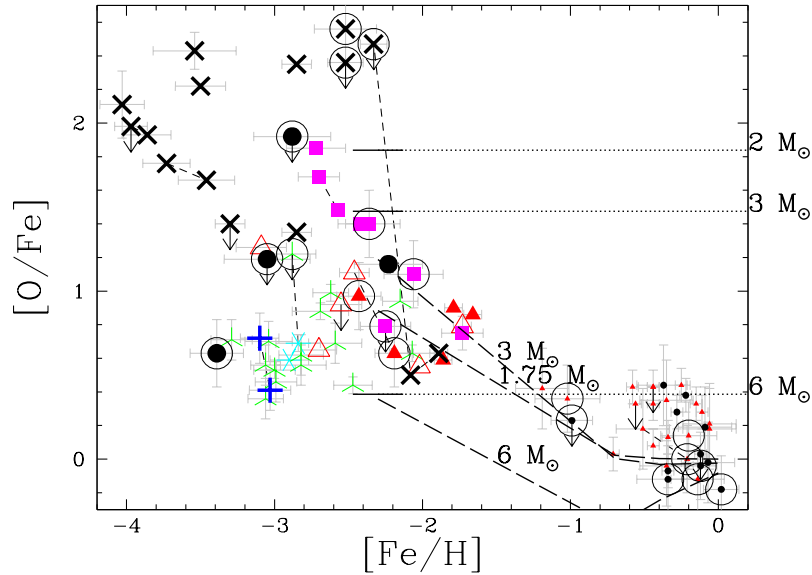}
  \caption{O abundances in various classes of metal-poor stars (see
  Fig.~\ref{fig:legend} for a description of symbols). Dashed
  lines represent O predictions as a function of metallicity for
  different masses from \citet{Karakas2007} and dotted lines point
  toward \citet{Herwig2004} predictions for
  different masses (as indicated on the right-hand scale) at 
  [Fe/H]~$\approx -2.3$. O is more enhanced in CEMP-rs (magenta squares)
  than in CEMP-s stars (red triangles) with the same metallicity.  We also notice that most of
  CEMP-no stars have even larger O enhancements. Note that none of the displayed O abundances have been
  corrected for 3D or non-LTE effects.}\label{fig:OFevsFe}
\end{center}
\end{figure}
Fig.~\ref{fig:OFevsFe} indeed shows that CEMP-rs stars are enhanced
in O. 
In this figure, we do not attempt to correct the O abundance for
systematic effects because they are coming from different indicators
([OI], O triplet and OH lines). Although NLTE and 3D effects are
expected to decrease the O abundance, the corrections  \citep[$\approx
  0.2$~dex; ][]{Takeda2003,GarciaPerez2006} are smaller than the actual O
enhancement in CEMP-rs compared to the normal galactic content
represented by rI and rII stars. In this figure, CEMP-s stars also
show a slight O enhancement but in this case, uncertainties from
non-LTE and 3D effects are not negligible compared to the observed enhancement
so that no firm conclusion can be drawn.\
It must be stressed that O yields predicted by different models do not agree with each other.  \citet{Herwig2004} finds that the dilution in the AGB envelope always dominates over the quantity of dredged-up matter, thus yielding lower O enhancements as mass increases. On the contrary, \citet{Karakas2007} found that there is a competition between dilution and hotter conditions in the He shell boosting the operation of the $\rm^{12}C(\alpha,\gamma)^{16}O$ reaction) as mass increases. Thus, they predict maximum O yields in their models around 3~M$_\odot$. Because of these uncertainties  in the model predictions, from the oxygen data alone, it  is not possible to infer the mass
of the former AGB companion of the CEMP-rs stars. Nevertheless, we have shown evidence for the operation of $\rm^{22}Ne(\alpha,n)^{25}Mg$ in the He-flash-driven convective zone in the former AGB companions of CEMP-rs stars. This reaction requires high temperatures ($3\times10^8$~K) which are also more favorable for the $\rm^{12}C(\alpha,\gamma)^{16}O$ reaction. Therefore, it appears that the enhancement of O and the occurrence of $\rm^{22}Ne(\alpha,n)^{25}Mg$ in the companions of CEMP-rs stars are consistent with the fact that they had a more massive AGB companion than CEMP-s stars.\\
Furthermore, it is established that the ON cycle occurs on a longer timescale than the CN cycle so that O is marginally burnt by HBB. We demonstrated in Sect.~\ref{sec:CandN} that the CN cycle occurring in the companion of the CEMP stars is incomplete and this implies that the ON cycle is certainly far from taking place in the companions of CEMP stars. Therefore we can reasonably conclude that the O enhancements observed  in CEMP stars do not come from H-burning. Most of the CEMP stars  are not evolved enough to process the accreted material themselves, with possibly two exceptions, though: CS~22891-171 and CS~30322-023. CS~22891-171 is not enriched in O, but is very
N-rich and does not follow the [Pb/Ce] vs [N/H] trend (Fig.~\ref{fig:Pb2ndpicvsN_s}). Its luminosity locates it either at the top
of the RGB or in the early-AGB phase. It is thus possible that
this star has processed its accreted O
into N through the ON cycle. This phenomenon would also be a good
explanation for the peculiar C-depleted, O-normal and N-super-rich
star CS~30322-023, located at the tip of the AGB \citep{Masseron2006}.

\subsubsection{Statistics}
Finally, we stress that our hypothesis that CEMP-rs stars are formed solely by 
pollution from an AGB companion contradicts most of the 
scenarios suggested so far (see \citet{Jonsell2006} for
a thorough discussion of these). Using our extensive database of CEMP
stars, we wish to discuss more thoroughly
here one of  these scenarios, which invokes a double-pollution episode
(typically from AGB stars and Type-II supernovae), one contributing to the s-process enhancement
and the other to the r-process enhancement.
\citet{Jonsell2006} have a statistical argument against this 
most popular scenario.  The probability of finding
an AGB star and a Type-II supernova polluting a main-sequence star should be lower than
finding a Type-II supernova alone polluting a main-sequence
star. Therefore, the probability of finding CEMP-rs stars should be
lower than  finding rII stars (which were only polluted by a type-II supernova). According to
\citet{Jonsell2006}, this
is not supported by the available statistics,  since CEMP-rs stars are more
numerous than rII stars. According to the same argument, we should
find fewer CEMP-rs stars than CEMP-s stars. But in
Fig.~\ref{fig:BaFevsEuFe}, there are almost equal
numbers instead. 
In our picture, the statistics has a straightforward explanation in
terms of the IMF. If we consider that companions to CEMP-s stars
had an initial mass in the range 1 -- 3~M$_\odot$ whereas CEMP-rs stars
have 3 -- 8~M$_\odot$ companions, the resulting CEMP-rs to CEMP-s
frequency ratio is 0.62 adopting \citet{Miller1979} IMF and 0.95
adopting \citet{Lucatello2005IMF} IMF. With the
same IMFs, we would find frequency ratios of 0.68 and 1.01 when taking
2.4~M$_\odot$ instead of 3~M$_\odot$ as threshold between the two classes. Despite
the fact that this rough estimate crucially needs more stringent
constraints on the masses from theoretical models and more accurate
observed statistics, it appears natural that CEMP-rs are
approximately as numerous as CEMP-s stars, and still more numerous than
rII stars.

\newpage
\section{CEMP-no stars}\label{sec:CEMP-no}
\subsection{Absence of neutron-capture signature} 
In the case of CEMP-s and CEMP-rs stars, a strong argument in favour of mass transfer from an AGB companion is the enhancement in s-process elements. For those stars, the analysis was made much simpler by assuming that the AGB contribution to the observed heavy-element abundances overwhelms the primordial abundances in the accreting star. Unfortunately, for CEMP-no stars, this assumption does not hold true because CEMP-no stars have, by definition, low abundances of the neutron-capture elements. Moreover, because the neutron-capture enrichment for field stars is quite variable, we cannot reliably separate contributions from the AGB star from the pristine (unknown) abundances (Fig.~\ref{fig:BaFevsEuFe_diluCEMP}). 
The C and N enhancements in CEMP-no stars resemble those in CEMP-s and CEMP-rs stars. But models for both low-metallicity massive stars and low-metallicity AGB stars predict C and N enhancements \citep[e.g.][]{Hirschi2007,Siess2004}. As remarked in Sect.~\ref{sec:intro}, there is a lack of radial-velocity measurements to constrain the binary rate of CEMP-no stars. 

\begin{figure}[!h]
\begin{center}
\includegraphics[width=5cm,angle=-90]{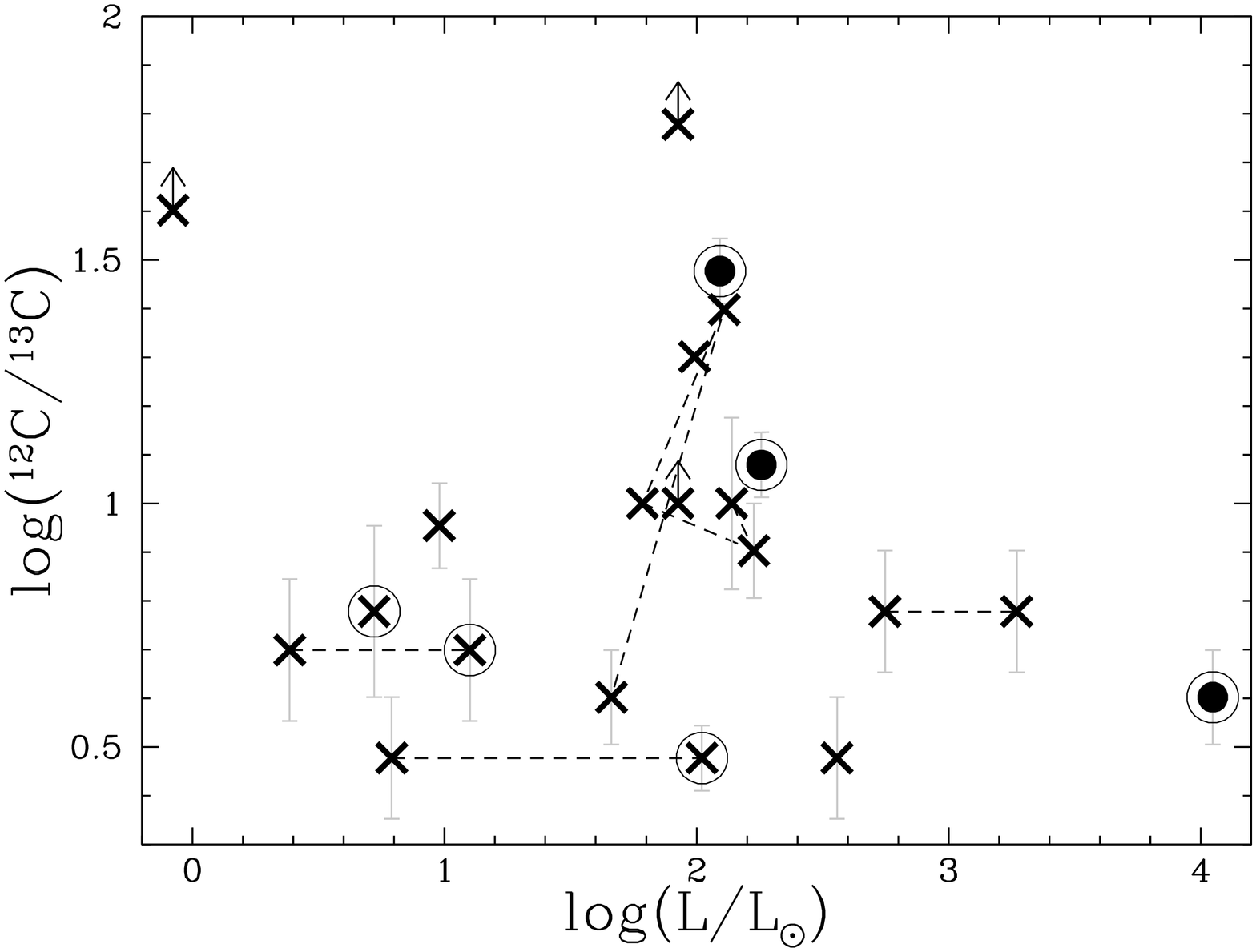}
\includegraphics[width=5cm,angle=-90]{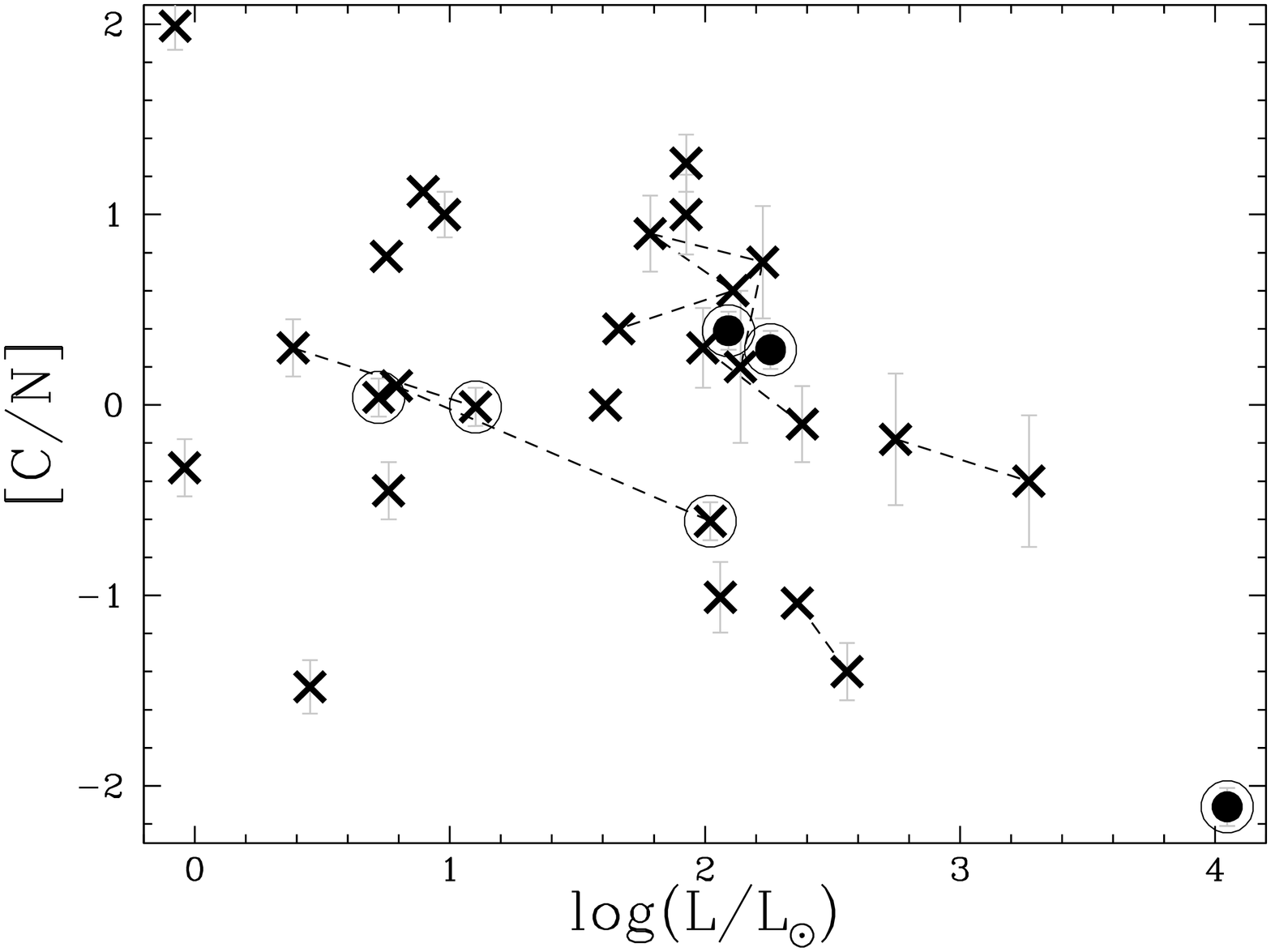}
  \caption{$\log$$\rm(^{12}C/^{13}C)$ and [C/N]  as a function of luminosity in CEMP-no stars. Some main sequence CEMP-no stars with enhanced N and and low $\rm ^{12}C/^{13}C$ ratios exist. }\label{fig:CNvsL_no}
\end{center}
\end{figure}
\citet{Ryan2005} observed that all CEMP-no stars in their
compilation were post-main-sequence stars, leading
these authors to suggest that CEMP-no stars have
undergone first dredge-up and have
processed some pristine C into N themselves. In their
scenario, CEMP-no stars were born of gas with high
C content from pollution by (possibly) low-energy supernovae
or winds from massive stars, C that was then processed
to show the high N and low $\rm ^{12}C/^{13}C$ in the CEMP-no stars
in their sample.
However, as already noted by \citet{Aoki2007}, some main sequence CEMP-no stars exist with similar N enhancements and similar $\rm ^{12}C/^{13}C$ ratios as CEMP-no giants (Fig.~\ref{fig:CNvsL_no}), thus ruling out the hypothesis of \citet{Ryan2005}. Therefore, if supernova and/or massive winds are responsible for
the C, N and isotopic ratios, the yields from these objects must
already bear the signature of CN-processed material at the time it leaves the massive star. This may be possible with rotating stars \citep[e.g. ][]{Meynet2002,Hirschi2007}
so we cannot rule out SN/massive stars based on this argument
alone, though AGB stars easily make CN-processed material as well.

\subsection{CEMP-no stars: the extremely metal-poor counterparts of CEMP s-process-rich (CEMP-s+CEMP-rs) stars} \label{sec:CEMP_relation}
Without clear diagnostics like mass (from an orbital solution), or 
s- or r-process abundance patterns, the nature
of the companion which polluted the CEMP-no star is very difficult to assess on a star by star
basis. However, thanks to our holistic approach, we may invoke several arguments collectively pointing towards the scenario of a mass transfer from a former AGB companion:
\begin{itemize}
\item [{\bf +}] It is remarkable that CEMP-low-s stars share many properties with CEMP-no stars: not only [Ba/Fe]$ < 1$ (their defining property), but also mild C-enrichments and similar [C+N/H] and [C+N+O/H] ratios \citep[][Figs.~\ref{fig:BaFevsFe} and \ref{fig:C+NvsFe}]{Aoki2002c}. It has been established in Sect.~\ref{sec:CEMP-low-s} that CEMP-low-s stars show Ba and Eu abundances compatible with the s-process, and hence, owe their peculiarities to AGB mass transfer (excluding the  CEMP-low-s star CS~30322-023, which is an intrinsic AGB star). To these three CEMP-low-s stars (black dots in Figs.~\ref{fig:BaFevsFe} and \ref{fig:C+NvsFe}), we can also add the CEMP-no star  CS~22956-28 (a blue straggler) which has been shown to be a binary with evidence for mass transfer from a former AGB companion \citep{Sneden2003}. There are thus hints for an AGB mass transfer in at least 4 CEMP-no or CEMP-low-s stars. Given the many similarities between CEMP-no and CEMP-low-s stars (the former could even turn into CEMP-low-s stars when their Eu abundance becomes available), it may thus be suspected that AGB mass transfer plays a role in many CEMP-no stars as well.

\begin{figure}[!h]
\begin{center}
\includegraphics[width=8cm,angle=-90]{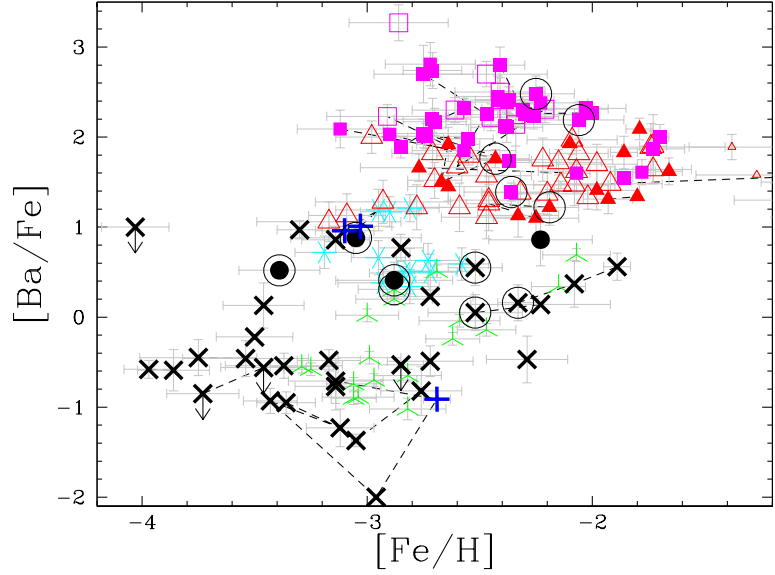}
  \caption{ [Ba/Fe] as a function of metallicity for all CEMP stars (see Fig.~\ref{fig:legend} for a description of the symbols). }\label{fig:BaFevsFe}
\end{center}
\end{figure}
\begin{figure}[!h]
\begin{center}
\includegraphics[width=5cm,angle=-90]{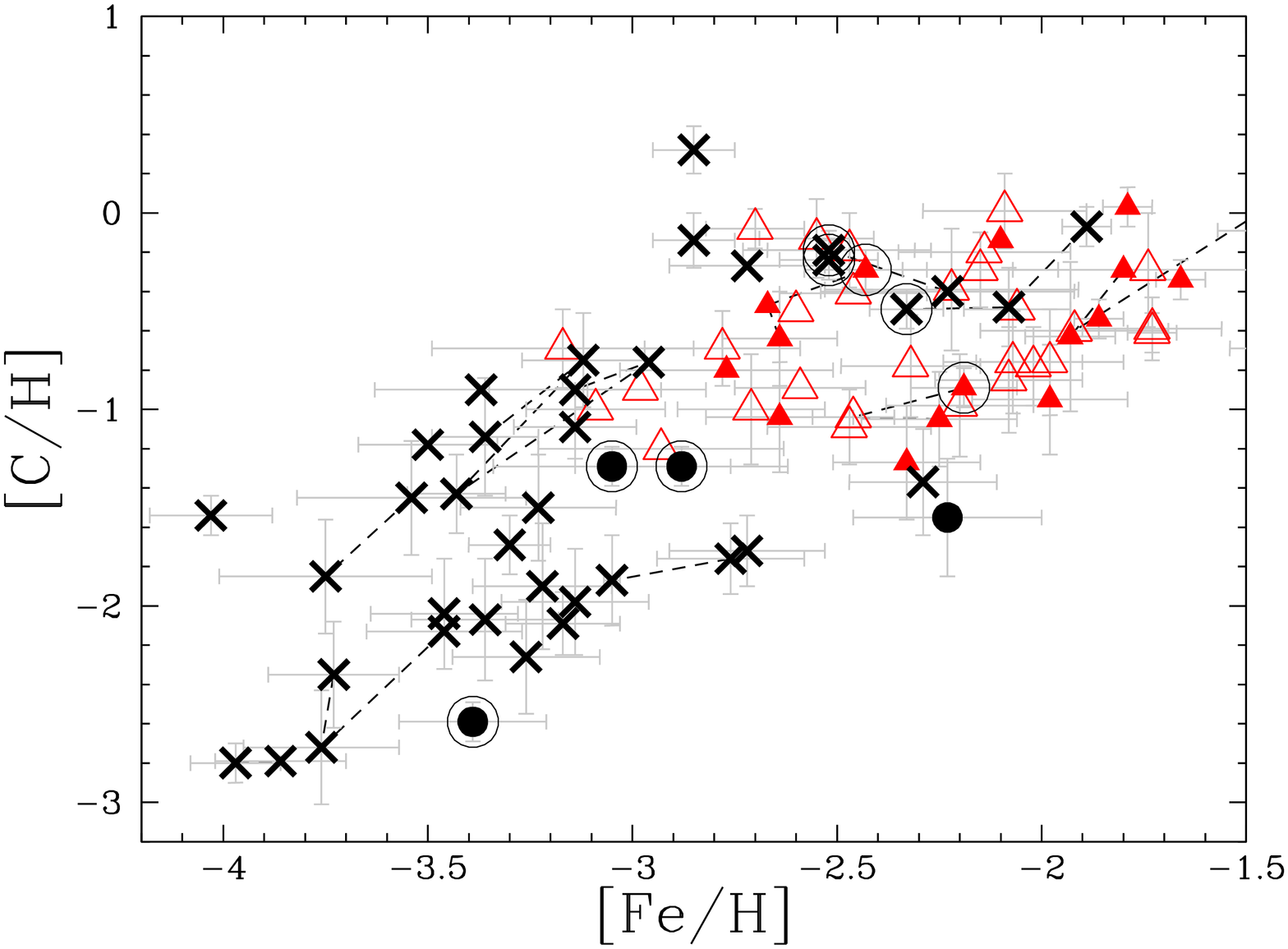}
\includegraphics[width=5cm,angle=-90]{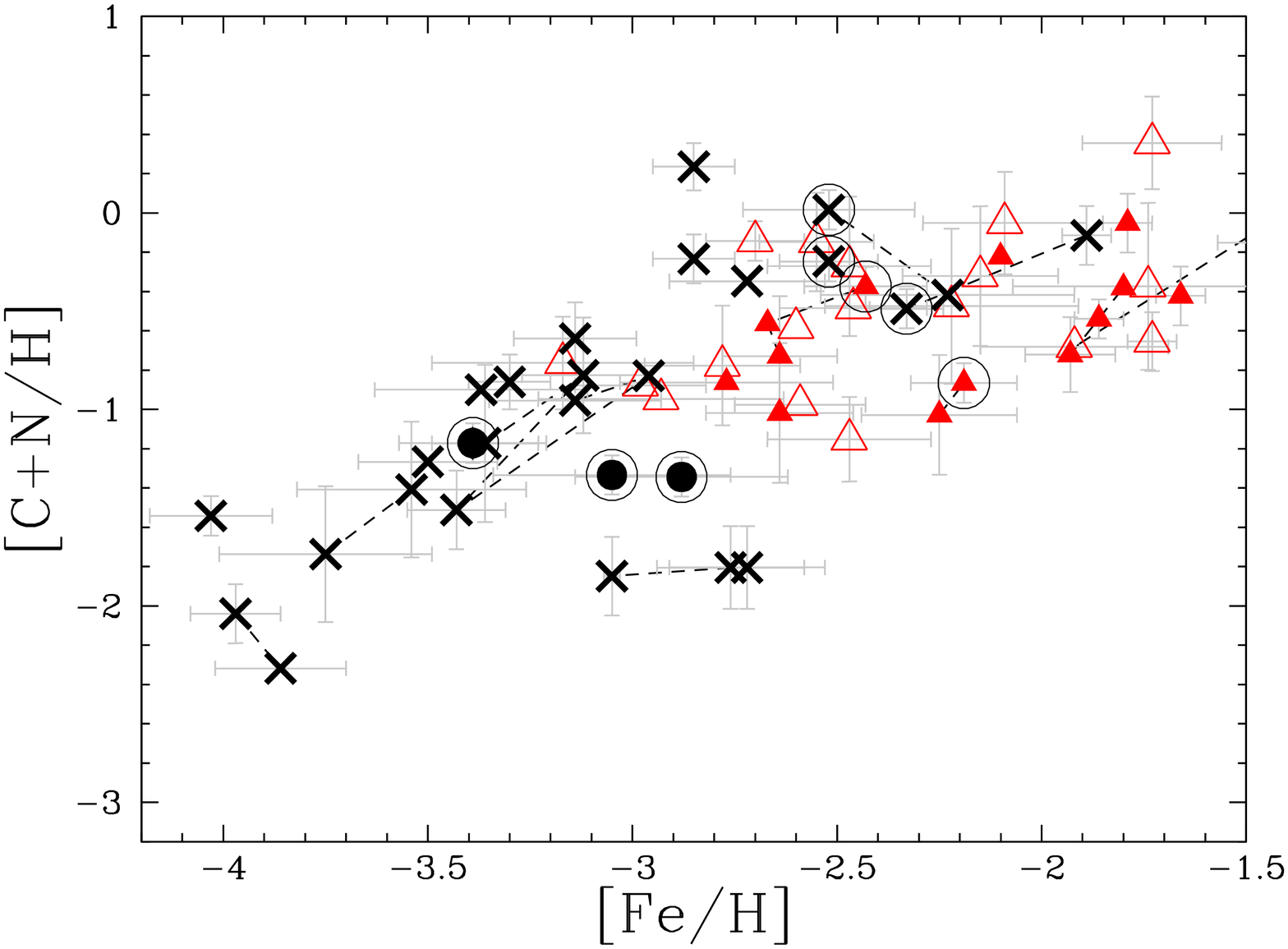}
\includegraphics[width=5cm,angle=-90]{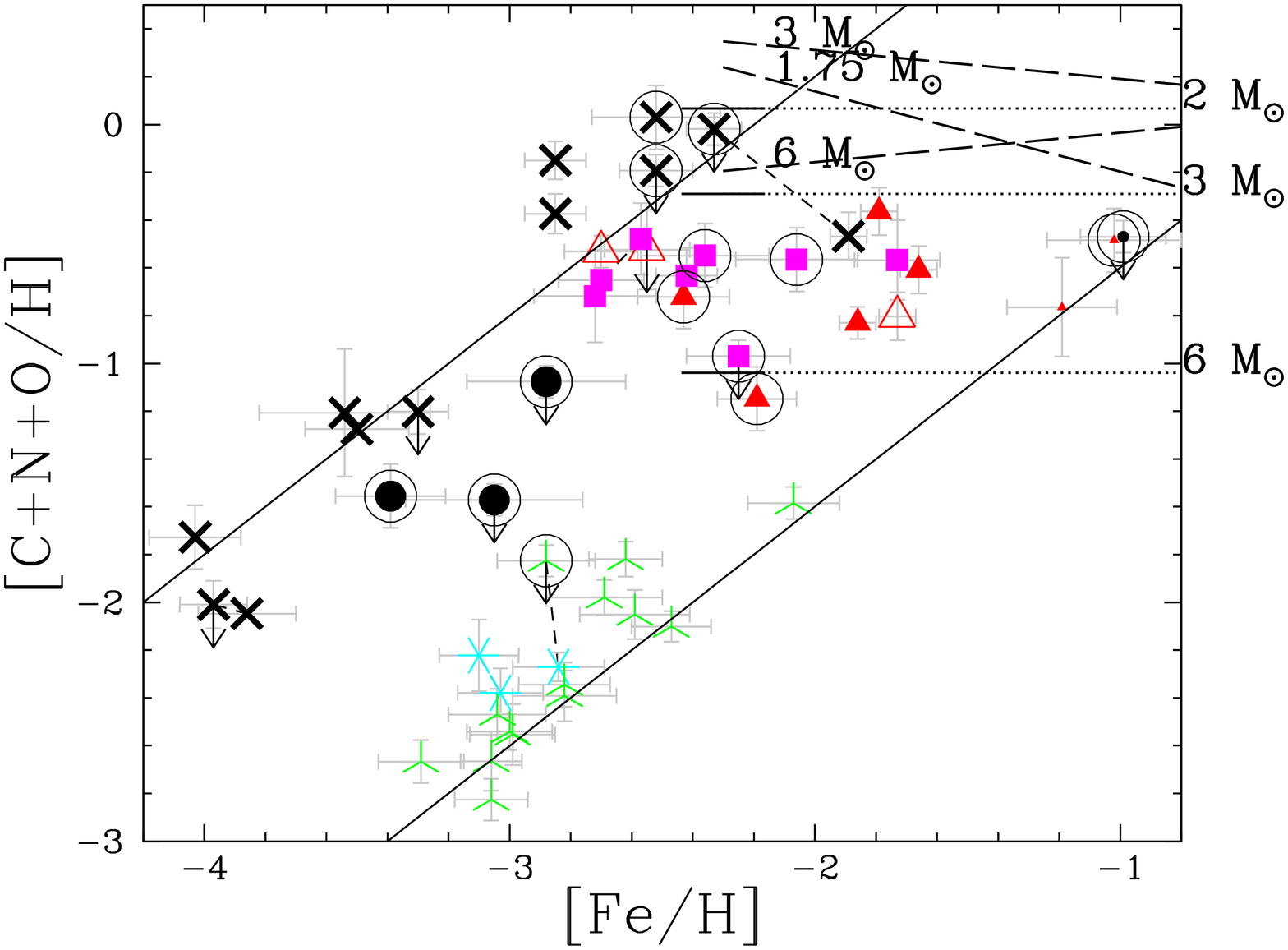}
  \caption{Top and middle panels: [C+N/H] and [C/H] vs metallicity in
  CEMP-no and CEMP-s stars (see Fig.~\ref{fig:legend} for a description of
  symbols). C and C+N abundances decline towards low
  metallicities in CEMP-no stars. Bottom panel: C+N+O vs metallicity in CEMP stars. The
  long-dashed  line corresponds to predictions from
  \citet{Karakas2007}, the dotted
  line to predictions from \citet{Herwig2004} for an AGB star with
  [Fe/H]~$=-2.3$. 
The thin solid lines correspond to [C+N+O/Fe]~=~0.4 and 2.2. 
C+N+O stays constant in CEMP-s and
  CEMP-rs stars (red triangles and magenta squares);
on the contrary,
C+N+O is proportional to metallicity in CEMP-no stars (black
  crosses). 
}\label{fig:C+NvsFe}
\end{center}
\end{figure}

\item [{\bf +}] There is a clear continuity in the abundance trends for CEMP-s, CEMP-rs and CEMP-no stars as a
function of metallicity, especially for O (Fig.~\ref{fig:OFevsFe}),
C+N (Fig.~\ref{fig:C+NvsFe}) and Mg
(Fig.~\ref{fig:MgFevsFe}). Moreover, Fig.~\ref{fig:C13vsCN} reveals
that CEMP-no stars have $\rm^{12}C/^{13}C$ and C/N ratios close to CEMP-s stars and CEMP-rs stars. In fact,
the CEMP-no stars are apparently divided in two subcategories, the O-
and Mg-enhanced and the O- and Mg-normal, as already noticed by
\citet{Aoki2002a}. By looking at Figs.~\ref{fig:OFevsFe} and \ref{fig:MgFevsFe}, these two subcategories may naturally be related to CEMP-rs stars
and CEMP-s stars, respectively. We do not
consider in the current discussion the 2 most metal-poor stars known
to date (HE~0107-5240 and HE~1327-2326), both being C-rich and
Ba-poor. We previously demonstrated for CEMP-s and CEMP-rs stars that
metallicity plays an important role in the outcome of nucleosynthesis,
and we therefore consider that
the metallicity of these two record-holders is too different from the
bulk of the sample to be safely included in the
comparison. \\

\begin{figure}[!h]
\begin{center}
\includegraphics[width=8cm,angle=-90]{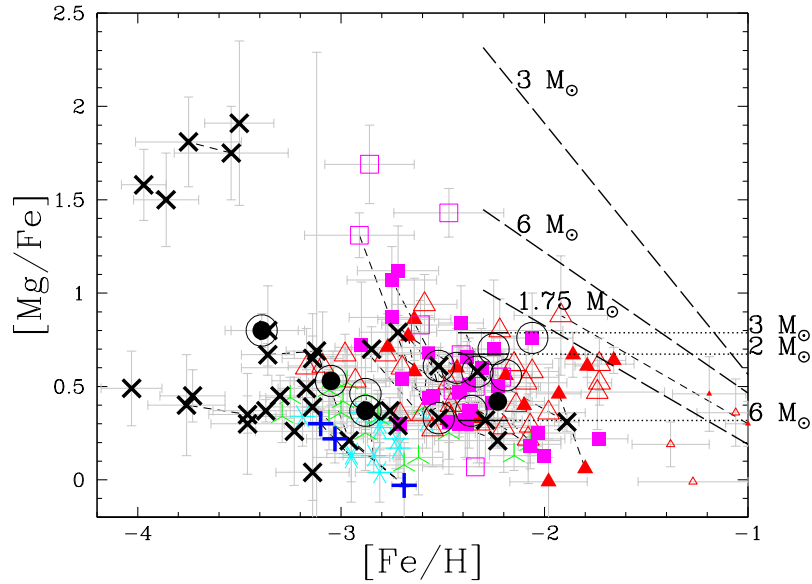}
  \caption{Mg enhancement as a function of metallicity for metal-poor
    stars (see Fig.~\ref{fig:legend} for a description of
    symbols). At extremely low metallicities,  the Mg enhancement
    seems to split CEMP-no stars into 2 categories, some being very
    Mg-enhanced as some CEMP-rs do, while most of the others do not show any Mg
    enhancement.
    }\label{fig:MgFevsFe}
\end{center}
\end{figure}

\item[{\bf +}] As already noticed by \citet{Aoki2007} and as shown in
Fig.~\ref{fig:numbervsFe}, the CEMP-no stars are more numerous at
low metallicity compared to CEMP-s and CEMP-rs stars. It is even more
puzzling that neither CEMP-rs stars nor CEMP-s stars have been
discovered below [Fe/H]~$< -3.2$. It is unlikely that binaries involving
a low- or intermediate-mass AGB star did not form at these very
low metallicities. Since the
lines from neutron-capture elements have a negligible impact on the stellar colours, it is very unlikely that the absence
of CEMP-s stars at low metallicities results from a selection effect acting against their detection when using broad-band colours as done in the HE or HK surveys. Thus, CEMP-no stars seem to be good candidates for 
being the more metal-poor counterparts of CEMP-s and/or CEMP-rs stars.
\begin{figure}[!h]
\begin{center}
\includegraphics[width=8cm,angle=-90]{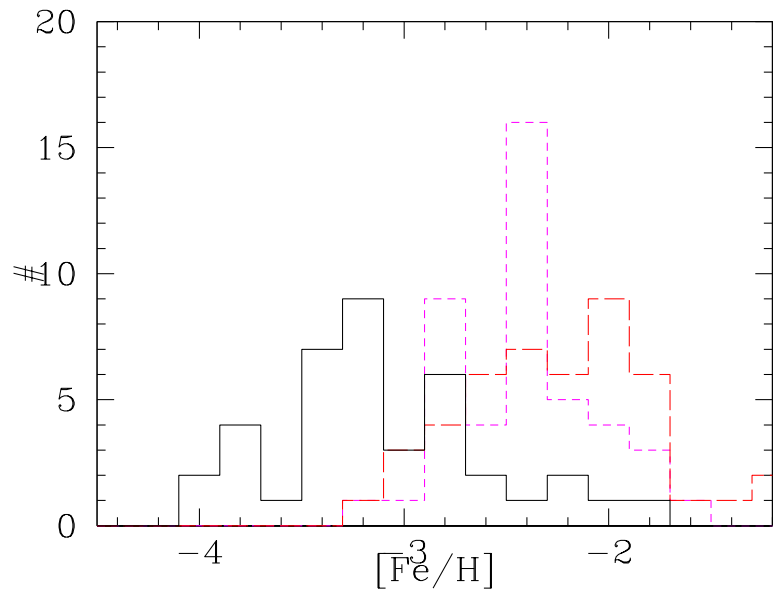}
  \caption{ Number of stars in the different CEMP subclasses in our
    sample as a function of metallicity: the black
    solid line represents CEMP-no stars, the red long-dashed line
    CEMP-s stars and the short-dashed magenta line CEMP-rs stars. CEMP-no stars generally show up at lower metallicities than the other CEMP stars.}\label{fig:numbervsFe}
\end{center}
\end{figure}
\end{itemize}
Although several observational facts seem to indicate that most of the CEMP-no stars had an AGB companion, it remains impossible to determine the origin of CEMP-no stars on a individual basis. Indeed, it is still possible that some among CEMP-no stars reflect the yields of the early massive stars with high C and N enhancements.  \\

Beyond the debate relating or not CEMP-no stars to AGB companions, the observed lack of CEMP-s stars for metallicities [Fe/H]$\lesssim$-3.0 might reveal a fundamental difference in the properties of extremely-low AGB stars compared to more metal-rich ones. 
Under the natural assumption that any low- and intermediate-mass star undergoes an AGB phase, irrespective of its metallicity, the changes observed in the abundances of CEMP stars between the metal-poor regime (-3.0$\lesssim$[Fe/H]$\lesssim$-1.5) and the extremely metal-poor regime ([Fe/H]$\lesssim$-3.0) must reflect similar changes in the properties of the s-process nucleosynthesis occurring in AGB stars in these two metallicity regimes. We now review possible causes for these differences.\\

\begin{figure}[!h]
\begin{center}
\includegraphics[width=5cm,angle=-90]{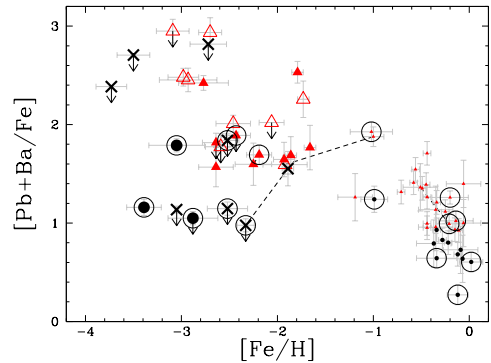}
  \caption{Pb+Ba enhancement as a function of metallicity in Ba, CEMP-s and CEMP-no stars (see Fig.~\ref{fig:legend} for a description of symbols). [Pb+Ba/Fe] is lower in CEMP-no stars than in CEMP-s stars.}\label{fig:Pb+BavsFe_sno}
\end{center}
\end{figure}

Is the s-process pushed toward the third peak at the very low metallicities?
When $\rm^{13}C(\alpha,n)^{16}O$ is the neutron source in AGB stars, models predict  that the efficiency of the s-process, expressed in terms of the number of neutrons captured per neutron seed nuclei, increases with decreasing metallicity \citep{Clayton1988,Goriely2001,Busso2001}. The s-process enrichment may not be obvious to detect then, as it may be restricted to a large overabundance involving the sole Pb.  The Pb measurements available in CEMP-low-s stars and the few upper limits in CEMP-no stars do not reveal strong overabundances, so that this possibility is ruled out (Fig.~ \ref{fig:Pb+BavsFe_sno}). Note that, because Pb has a condensation temperature similar that of Na and S \citep{Lodders2003}, the lack of Pb enhancement also confirms the statement of \citet{Venn2008} that peculiar abundances in CEMP-no stars with [Fe/H]$>$-4.0 cannot be explained by re-accretion of dust-to-gas segregated material (as in $\lambda$~Boo stars).   \\

If CEMP-no stars are the metal-poor counterparts of CEMP stars with s-process enhancements, what are the few metal-rich CEMP-no stars (with [Fe/]$>$-2.5 on Fig.~\ref{fig:numbervsFe})? 
It is useful to investigate in more details the properties of the relatively high-metallicity ([Fe/H]~$>-2.5$) CEMP-no stars, such as HKII~17435-00532, HE~1330-0354,  CS~22956-028, CS~22945-017 and HE~1410+0213, and compare them to the CEMP-s stars with the same metallicity.
  The abundance pattern of the first star is compatible with a strong dilution of the accreted material in a r-process-rich gas (see Sect.~\ref{sec:CEMP_relation}). The second one also shows a mild enrichment in both C and Ba. Therefore, the dilution scenario is also plausible for this star. Nevertheless, the abundance pattern of the last three ones is more puzzling. While they all show a C abundance comparable to CEMP-s stars, CS~22956-028 is recognized as a blue straggler
\citep[][ Paper~I]{Sneden2003} and HE~1410+0213 is extremely O-enhanced in the same manner as in more metal-poor CEMP-no stars. Note that the fact that  CS~22956-028 is a blue straggler is well in line with the fact that the large amount of material accreted modified its evolution, similarly to what is observed in CEMP-rs stars (Sect.~\ref{sec:CEMP-rs_dilu}). Thus, the various abundance patterns of these relatively metal-rich CEMP-no stars may come from various progenitors: accretion of matter from either an intermediate-mass AGB star (with oxygen coming from $\rm^{12}C(\alpha,\gamma)^{16}O$ operating in warm thermal pulses), or even from a massive star having exploded as a type~II supernova. Therefore, their connection with more metal-poor CEMP-no stars is not straightforward, and a more detailed study of these objects needs to be done.

It is likely that the complex interplay
between mass and metallicity has a decisive impact on the AGB nucleosynthesis at extremely low metallicity. Besides increasing the temperature at which the nucleosynthesis operates, the low metallicity regime might also enhance the effects of rotation on the abundances \citep{Herwig2003,Siess2004,Decressin2006} or might induce the occurrence of the hydrogen injection flash during the thermal pulse \citep[aka He-FDDM, ][]{Fujimoto2000}. Rotation is predicted to enhance N in agreement with what is observed for most CEMP-no stars and contrasting with CEMP-s and CEMP-rs stars \citep[][and Fig.~\ref{fig:C13vsCN}]{Ryan2005}. When rotation increases the N production, it inhibits the s-process, but is not predicted to enhance O in AGB stars (as observed for CEMP-no stars in Fig.~\ref{fig:OFevsFe}). He-FDDM has a drastic impact on the AGB structure, and is consequently expected to modify the AGB nucleosynthesis. Although the metallicity threshold below which this latter mechanism operates is still debated, this effect is predicted for all AGB models at low metallicity \citep{Fujimoto2000,Siess2002,Herwig2005,Campbell2008}, and may explain the rise of CEMP-no stars at extremely low metallicity. Actually, more extensive theoretical studies for AGB stars with metallicities in the range -5.0$<$[Fe/H]$<$-3.0 are needed to evaluate the importance of these effects in CEMP-no stars.\\

Finally, there is one striking fact that is difficult to reconcile with the binary scenario.
Fig.~\ref{fig:C+NvsFe} shows that C declines proportionally to metallicity in CEMP-no stars, as do [C+N/H] and [C+N+O/H]. This is very puzzling as the C yields from both AGB and massive stars yields is of primary nature (i.e. independent of metallicity), thus its observed abundance should similarly be independent of metallicity. Actually, when considering the effect of rotation in massive stars, the expected C and O abundances should even follow the opposite trend \citep{Hirschi2007}.

\newpage
\section{Summary}\label{sec:Summary}

\begin{table}
\begin{tabular}{l|lllll}
      & CEMP-no & CEMP-low-s & CEMP-s &CEMP-rs& comment\\
      \hline
C       & medium, $\nearrow$ with Fe    &  medium   & high  &  high   & from  AGB or low-Fe SNII  \\
$\rm ^{12}C/^{13}C$ &  low to high    & low to high   & low to high & low to high    & signature of CN cycling\\
N       & high     & high    & high  &  high   & not from HBB, from cool bottom processing, rotation...?\\
O       & high in most   &   std  & std  &  high in some   & SNII or high temperature pulses ? \\
Mg      & high     &  high in some   & std  &     high in some & $^{22}Ne(\alpha,n)^{25}Mg$ or $^{20}Ne(\alpha,\gamma)^{24}Mg$ ?\\
Ba      & low     & low    & high   &  very high   & correlated with C for CEMP-s and low-s\\
La      & low     & low    & high   &  very high   & from  $^{13}C(\alpha,n)^{16}O$ in CEMP-s and $^{22}Ne(\alpha,n)^{25}Mg$ in CEMP-rs \tabularnewline
Ce      & low                & low    &  high   & high and cst &   ''''   \tabularnewline
Eu      & not measurable     &  low   & medium  & high  &  '''' \tabularnewline
Pb      & not measured       &  low   & medium  & high  &  ''''\tabularnewline
\hline

\end{tabular}
\end{table}

The analysis of abundances in CEMP stars leads  to the following conclusions:

\begin{itemize}
\item Ba stars are the metal-rich counterparts of CEMP-s stars.

\item We demonstrate that CEMP-low-s stars are very likely the result of mass transfer from an AGB companion because of their C and s-process signatures. Although their light element-abundance pattern is identical to that of CEMP-no stars, it is not clear yet whether or not all CEMP-no stars are CEMP-low-s stars.

\item The neutron-capture element ratios in CEMP-s stars are consistent with low-mass AGB models where the $^{13}$C neutron source has operated.

\item The observed relation between the Ba/C ratio and metallicity in CEMP-s stars gives strong constraints on the mechanism responsible for the formation of $\rm^{13}$C in the He-rich shell during the interpulse. 

\item The low dispersion in Ce/Fe as a function of the luminosity in CEMP-rs stars suggests that a large amount of material (with a fixed Ce abundance) has been dumped into the CEMP-rs star envelope, thus erasing all variations caused by the dispersion of all the initial abundances by the so-called thermohaline mixing.

\item The correlation between [vhs/hs] and [N/H] ratios in CEMP-rs stars as well as the [Mg/Fe] in some of them suggest the operation of the $\rm^{22}$Ne neutron source

\item All CEMP stars have low $^{12}$C/$^{13}$C ratios, confirming that CN processing
occurring in low-metallicity AGB stars is responsible for the observed N, but the high [C/N] ratios suggest that the conditions offered by HBB in the convective envelope are not adequate to produce the observed N. In contrast, these ratios points toward characteristics compatible with cool bottom processing.

\item CEMP-rs stars have O enhancement as compared to CEMP-s stars;
this hints at temperatures larger in the He-burning shell of the companion of CEMP-rs companion than in the CEMP-s companion, in agreement with the operation of $\rm^{22}Ne(\alpha,n)^{25}Mg$.

\item We highlight the possible relation of CEMP-no stars and AGB stars.  Hence, most of CEMP-no stars are likely to be the metal-poor counterparts of CEMP-s and CEMP-rs stars.

\item The decline of C and s-process in CEMP-no stars points to lower C production and inefficient s-process in low-metallicity AGB stars. These two observations do not have any explanations in the current AGB models.

\end{itemize}

Nevertheless, one may attempt to use these constraints to sort CEMP subclasses according to the mass of the progenitor. The hotter conditions achieved in He-burning shell of the CEMP-rs stars companions compared to CEMP-s stars suggest that CEMP-rs stars had a more massive AGB companion than CEMP-s did. Furthermore, there is now a consensus for attributing the peculiar abundances of CEMP-s stars to low-mass AGB stars. Since the intermediate-mass counterparts of CEMP-s stars have not been identified yet \citep{Johnson2007}, CEMP-rs stars remain the best candidates.  This interpretation is also very satisfying regarding the high frequency of CEMP-rs stars \citep{Jonsell2006}. Concerning CEMP-no stars, they are likely to encompass both intermediate- and low-mass AGB stars at extremely low metallicity ([Fe/H]$\lesssim$-3.0). Indeed, the low metallicity regime seems to have drastic effect on nucleosynthesis.\\
However, the conditions for making N in the H-rich layer of the AGB companion to the CEMP stars are not well identified so far, since the [C/N] ratios are very similar in all categories of CEMP stars. Furthermore, it is expected that the more massive the AGB star, the more the synthesized material is diluted in the AGB envelope, thus the lower are the yields. This is in contradiction with the fact that the overabundances of neutron-capture elements are larger in CEMP-rs stars than in CEMP-s stars (Fig.~\ref{fig:numbervsFe}). Nevertheless, we also emphasize that $\rm^{22}Ne(\alpha,n)^{25}Mg$ can be very efficient at making s-process elements.  
AGB models also suffer from uncertainties, in particular the impact of HBB, the mechanism of CBP, the parametrization of 3$\rm^{rd}$ dredge-ups, the number of thermal pulses, the occurrence of the Hydrogen Injection Flash \citep{Fujimoto2000} and possible effects of rotation \citep{Siess2004} and thermohaline mixing \citep{Cantiello2008}, and finally, the impact of the uncertainty on the $\rm^{22}Ne(\alpha,n)^{25}Mg$ reaction rate. Therefore, it is not at all straightforward to deduce the mass of the former AGB companion of CEMP stars from the analysis of their abundances.

\subsection{Open questions}
\begin{itemize}
\item What is the mechanism for N production in AGB stars?  Is HBB really active at low metallicity? 

\item Since N/C and O/Fe seem to increase with decreasing metallicity, should we not look for NEMP or OEMP stars? But only a couple of N-rich metal-poor stars have been found so far \citep{Johnson2007}.
  
\item Where are the CEMP-s and CEMP-rs stars at extremely low metallicity?

\item Are there  metal-rich counterparts to CEMP-rs stars? The
  N-rich and Rb-rich stars from \citet{GarciaHernandez2006} might be possible candidates.

\item Are all CEMP-no stars the results of AGB mass transfer?

\item How low in metallicity does the C and s-process decline go? Do
  HE~1327-2326 and HE~0107-5240 fit in this scenario ?

\item What is the impact of the observed C decrease at very low metallicity on CEMP statistics? Do the strongest enhancements of elements in these stars (in particular N, O, Mg, s-process) have any implication on the chemical evolution of the Galaxy?  
\end{itemize}


\begin{acknowledgements}
T.M. deeply thanks P. Denissenkov, M. Pinsonneault and M. Lugaro for fruitful discussions. 
T.M. gratefully acknowledge support from ESO studentship program and from CNRS/INSU.
T.M. and J.A.J. also acknowledge support from AST-0707948. 
\end{acknowledgements}
\bibliographystyle{hapj}
\bibliography{thomas_AA}

\begin{thebibliography}{121}
\expandafter\ifx\csname natexlab\endcsname\relax\def\natexlab#1{#1}\fi

\bibitem[{{Abia} {et~al.}(2001){Abia}, {Busso}, {Gallino}, {Dom{\'{\i}}nguez},
  {Straniero}, \& {Isern}}]{Abia2001}
{Abia}, C., {Busso}, M., {Gallino}, R., {Dom{\'{\i}}nguez}, I., {Straniero},
  O., \& {Isern}, J. 2001, \apj, 559, 1117, arXiv:astro-ph/0105486

\bibitem[{{Allen} \& {Barbuy}(2006{\natexlab{a}})}]{Allen2006a}
{Allen}, D.~M., \& {Barbuy}, B. 2006{\natexlab{a}}, \aap, 454, 895,
  arXiv:astro-ph/0604036

\bibitem[{{Allen} \& {Barbuy}(2006{\natexlab{b}})}]{Allen2006b}
------. 2006{\natexlab{b}}, \aap, 454, 917, arXiv:astro-ph/0604064

\bibitem[{{Aoki} {et~al.}(2007){Aoki}, {Beers}, {Christlieb}, {Norris}, {Ryan},
  \& {Tsangarides}}]{Aoki2007}
{Aoki}, W., {Beers}, T.~C., {Christlieb}, N., {Norris}, J.~E., {Ryan}, S.~G.,
  \& {Tsangarides}, S. 2007, \apj, 655, 492, arXiv:astro-ph/0609702

\bibitem[{{Aoki} {et~al.}(2008){Aoki}, {Beers}, {Sivarani}, {Marsteller},
  {Lee}, {Honda}, {Norris}, {Ryan}, \& {Carollo}}]{Aoki2008}
{Aoki}, W. {et~al.} 2008, \apj, 678, 1351, 0801.4187

\bibitem[{{Aoki} {et~al.}(2006){Aoki}, {Bisterzo}, {Gallino}, {Beers},
  {Norris}, {Ryan}, \& {Tsangarides}}]{Aoki2006}
{Aoki}, W., {Bisterzo}, S., {Gallino}, R., {Beers}, T.~C., {Norris}, J.~E.,
  {Ryan}, S.~G., \& {Tsangarides}, S. 2006, \apjl, 650, L127,
  arXiv:astro-ph/0609138

\bibitem[{{Aoki} {et~al.}(2002{\natexlab{a}}){Aoki}, {Norris}, {Ryan}, {Beers},
  \& {Ando}}]{Aoki2002a}
{Aoki}, W., {Norris}, J.~E., {Ryan}, S.~G., {Beers}, T.~C., \& {Ando}, H.
  2002{\natexlab{a}}, \apjl, 576, L141

\bibitem[{{Aoki} {et~al.}(2002{\natexlab{b}}){Aoki}, {Norris}, {Ryan}, {Beers},
  \& {Ando}}]{Aoki2002d}
------. 2002{\natexlab{b}}, \pasj, 54, 933

\bibitem[{{Aoki} {et~al.}(2002{\natexlab{c}}){Aoki}, {Norris}, {Ryan}, {Beers},
  \& {Ando}}]{Aoki2002c}
------. 2002{\natexlab{c}}, \apj, 567, 1166, arXiv:astro-ph/0111297

\bibitem[{{Aoki} {et~al.}(2004){Aoki}, {Norris}, {Ryan}, {Beers}, {Christlieb},
  {Tsangarides}, \& {Ando}}]{Aoki2004}
{Aoki}, W., {Norris}, J.~E., {Ryan}, S.~G., {Beers}, T.~C., {Christlieb}, N.,
  {Tsangarides}, S., \& {Ando}, H. 2004, \apj, 608, 971

\bibitem[{{Aoki} {et~al.}(2001){Aoki}, {Ryan}, {Norris}, {Beers}, {Ando},
  {Iwamoto}, {Kajino}, {Mathews}, \& {Fujimoto}}]{Aoki2001}
{Aoki}, W. {et~al.} 2001, \apj, 561, 346, arXiv:astro-ph/0107040

\bibitem[{{Aoki} {et~al.}(2002{\natexlab{d}}){Aoki}, {Ryan}, {Norris}, {Beers},
  {Ando}, \& {Tsangarides}}]{Aoki2002b}
{Aoki}, W., {Ryan}, S.~G., {Norris}, J.~E., {Beers}, T.~C., {Ando}, H., \&
  {Tsangarides}, S. 2002{\natexlab{d}}, \apj, 580, 1149

\bibitem[{{Asplund}(2005)}]{Asplund2005}
{Asplund}, M. 2005, \araa, 43, 481

\bibitem[{{Barbuy} {et~al.}(1997){Barbuy}, {Cayrel}, {Spite}, {Beers}, {Spite},
  {Nordstroem}, \& {Nissen}}]{Barbuy1997}
{Barbuy}, B., {Cayrel}, R., {Spite}, M., {Beers}, T.~C., {Spite}, F.,
  {Nordstroem}, B., \& {Nissen}, P.~E. 1997, \aap, 317, L63

\bibitem[{{Barbuy} {et~al.}(1992){Barbuy}, {Jorissen}, {Rossi}, \&
  {Arnould}}]{Barbuy1992}
{Barbuy}, B., {Jorissen}, A., {Rossi}, S.~C.~F., \& {Arnould}, M. 1992, \aap,
  262, 216

\bibitem[{{Barbuy} {et~al.}(2005){Barbuy}, {Spite}, {Spite}, {Hill}, {Cayrel},
  {Plez}, \& {Petitjean}}]{Barbuy2005}
{Barbuy}, B., {Spite}, M., {Spite}, F., {Hill}, V., {Cayrel}, R., {Plez}, B.,
  \& {Petitjean}, P. 2005, \aap, 429, 1031

\bibitem[{{Barklem} {et~al.}(2005){Barklem}, {Christlieb}, {Beers}, {Hill},
  {Bessell}, {Holmberg}, {Marsteller}, {Rossi}, {Zickgraf}, \&
  {Reimers}}]{Barklem2005}
{Barklem}, P.~S. {et~al.} 2005, \aap, 439, 129

\bibitem[{{Beers} \& {Christlieb}(2005)}]{Beers2005}
{Beers}, T.~C., \& {Christlieb}, N. 2005, \araa, 43, 531

\bibitem[{{Beers} {et~al.}(1992){Beers}, {Preston}, \& {Shectman}}]{Beers1992}
{Beers}, T.~C., {Preston}, G.~W., \& {Shectman}, S.~A. 1992, \aj, 103, 1987

\bibitem[{{Bisterzo} \& {Gallino}(2008)}]{Bisterzo2008}
{Bisterzo}, S., \& {Gallino}, R. 2008, in American Institute of Physics
  Conference Series, Vol. 1001, Evolution and Nucleosynthesis in AGB Stars, ed.
  R.~{Guandalini}, S.~{Palmerini}, \& M.~{Busso}, 131--138

\bibitem[{{Bisterzo} {et~al.}(2006){Bisterzo}, {Gallino}, {Straniero}, {Ivans},
  {K{\"a}ppeler}, \& {Aoki}}]{Bisterzo2006}
{Bisterzo}, S., {Gallino}, R., {Straniero}, O., {Ivans}, I.~I., {K{\"a}ppeler},
  F., \& {Aoki}, W. 2006, Memorie della Societa Astronomica Italiana, 77, 985

\bibitem[{{Bonifacio} {et~al.}(1998){Bonifacio}, {Molaro}, {Beers}, \&
  {Vladilo}}]{Bonifacio1998}
{Bonifacio}, P., {Molaro}, P., {Beers}, T.~C., \& {Vladilo}, G. 1998, \aap,
  332, 672

\bibitem[{{Boothroyd} \& {Sackmann}(1999)}]{Boothroyd1999}
{Boothroyd}, A.~I., \& {Sackmann}, I.-J. 1999, \apj, 510, 232

\bibitem[{{Busso} {et~al.}(2001){Busso}, {Gallino}, {Lambert}, {Travaglio}, \&
  {Smith}}]{Busso2001}
{Busso}, M., {Gallino}, R., {Lambert}, D.~L., {Travaglio}, C., \& {Smith},
  V.~V. 2001, \apj, 557, 802

\bibitem[{{Campbell} \& {Lattanzio}(2008)}]{Campbell2008}
{Campbell}, S.~W., \& {Lattanzio}, J.~C. 2008, \aap, 490, 769, 0901.0799

\bibitem[{{Cantiello} \& {Langer}(2008)}]{Cantiello2008}
{Cantiello}, M., \& {Langer}, N. 2008, in IAU Symposium, Vol. 252, IAU
  Symposium, ed. L.~{Deng} \& K.~L. {Chan}, 103--109

\bibitem[{{Cayrel} {et~al.}(2004){Cayrel}, {Depagne}, {Spite}, {Hill}, {Spite},
  {Fran{\c c}ois}, {Plez}, {Beers}, {Primas}, {Andersen}, {Barbuy},
  {Bonifacio}, {Molaro}, \& {Nordstr{\" o}m}}]{Cayrel2004}
{Cayrel}, R. {et~al.} 2004, \aap, 416, 1117

\bibitem[{{Christlieb} {et~al.}(2002){Christlieb}, {Bessell}, {Beers},
  {Gustafsson}, {Korn}, {Barklem}, {Karlsson}, {Mizuno-Wiedner}, \&
  {Rossi}}]{Christlieb2002}
{Christlieb}, N. {et~al.} 2002, \nat, 419, 904

\bibitem[{{Christlieb} {et~al.}(2001){Christlieb}, {Green}, {Wisotzki}, \&
  {Reimers}}]{Christlieb2001}
{Christlieb}, N., {Green}, P.~J., {Wisotzki}, L., \& {Reimers}, D. 2001, \aap,
  375, 366

\bibitem[{{Christlieb} {et~al.}(2004){Christlieb}, {Gustafsson}, {Korn},
  {Barklem}, {Beers}, {Bessell}, {Karlsson}, \&
  {Mizuno-Wiedner}}]{Christlieb2004}
{Christlieb}, N., {Gustafsson}, B., {Korn}, A.~J., {Barklem}, P.~S., {Beers},
  T.~C., {Bessell}, M.~S., {Karlsson}, T., \& {Mizuno-Wiedner}, M. 2004, \apj,
  603, 708

\bibitem[{{Clayton}(1988)}]{Clayton1988}
{Clayton}, D.~D. 1988, \mnras, 234, 1

\bibitem[{{Cohen} {et~al.}(2008){Cohen}, {Christlieb}, {McWilliam}, {Shectman},
  {Thompson}, {Melendez}, {Wisotzki}, \& {Reimers}}]{Cohen2008}
{Cohen}, J.~G., {Christlieb}, N., {McWilliam}, A., {Shectman}, S., {Thompson},
  I., {Melendez}, J., {Wisotzki}, L., \& {Reimers}, D. 2008, \apj, 672, 320,
  arXiv:0709.0029

\bibitem[{{Cohen} {et~al.}(2004){Cohen}, {Christlieb}, {McWilliam}, {Shectman},
  {Thompson}, {Wasserburg}, {Ivans}, {Dehn}, {Karlsson}, \&
  {Melendez}}]{Cohen2004}
{Cohen}, J.~G. {et~al.} 2004, \apj, 612, 1107

\bibitem[{{Cohen} {et~al.}(2003){Cohen}, {Christlieb}, {Qian}, \&
  {Wasserburg}}]{Cohen2003}
{Cohen}, J.~G., {Christlieb}, N., {Qian}, Y.-Z., \& {Wasserburg}, G.~J. 2003,
  \apj, 588, 1082

\bibitem[{{Cohen} {et~al.}(2006){Cohen}, {McWilliam}, {Shectman}, {Thompson},
  {Christlieb}, {Melendez}, {Ramirez}, {Swensson}, \& {Zickgraf}}]{Cohen2006}
{Cohen}, J.~G. {et~al.} 2006, \aj, 132, 137, arXiv:astro-ph/0603582

\bibitem[{{Cristallo} {et~al.}(2009){Cristallo}, {Straniero}, {Gallino},
  {Piersanti}, {Dom{\'{\i}}nguez}, \& {Lederer}}]{Cristallo2009}
{Cristallo}, S., {Straniero}, O., {Gallino}, R., {Piersanti}, L.,
  {Dom{\'{\i}}nguez}, I., \& {Lederer}, M.~T. 2009, \apj, 696, 797

\bibitem[{{Decressin} \& {Charbonnel}(2006)}]{Decressin2006}
{Decressin}, T., \& {Charbonnel}, C. 2006, in International Symposium on
  Nuclear Astrophysics - Nuclei in the Cosmos, PoS(NIC--IX)086

\bibitem[{{Denissenkov} \& {Pinsonneault}(2008)}]{Denissenkov2008}
{Denissenkov}, P.~A., \& {Pinsonneault}, M. 2008, \apj, 679, 1541,
  arXiv:0709.4240

\bibitem[{{Depagne} {et~al.}(2002){Depagne}, {Hill}, {Spite}, {Spite}, {Plez},
  {Beers}, {Barbuy}, {Cayrel}, {Andersen}, {Bonifacio}, {Fran{\c c}ois},
  {Nordstr{\"o}m}, \& {Primas}}]{Depagne2002}
{Depagne}, E. {et~al.} 2002, \aap, 390, 187

\bibitem[{{Deroo} {et~al.}(2005){Deroo}, {Reyniers}, {van Winckel}, {Goriely},
  \& {Siess}}]{Deroo2005}
{Deroo}, P., {Reyniers}, M., {van Winckel}, H., {Goriely}, S., \& {Siess}, L.
  2005, \aap, 438, 987, arXiv:astro-ph/0504596

\bibitem[{{Fran{\c c}ois} {et~al.}(2007){Fran{\c c}ois}, {Depagne}, {Hill},
  {Spite}, {Spite}, {Plez}, {Beers}, {Andersen}, {James}, {Barbuy}, {Cayrel},
  {Bonifacio}, {Molaro}, {Nordstr{\"o}m}, \& {Primas}}]{Francois2007}
{Fran{\c c}ois}, P. {et~al.} 2007, \aap, 476, 935, arXiv:0709.3454

\bibitem[{{Frebel} {et~al.}(2005){Frebel}, {Aoki}, {Christlieb}, {Ando},
  {Asplund}, {Barklem}, {Beers}, {Eriksson}, {Fechner}, {Fujimoto}, {Honda},
  {Kajino}, {Minezaki}, {Nomoto}, {Norris}, {Ryan}, {Takada-Hidai},
  {Tsangarides}, \& {Yoshii}}]{Frebel2005}
{Frebel}, A. {et~al.} 2005, \nat, 434, 871

\bibitem[{{Frebel} {et~al.}(2006){Frebel}, {Christlieb}, {Norris}, {Aoki}, \&
  {Asplund}}]{Frebel2006}
{Frebel}, A., {Christlieb}, N., {Norris}, J.~E., {Aoki}, W., \& {Asplund}, M.
  2006, \apjl, 638, L17, arXiv:astro-ph/0512543

\bibitem[{{Frebel} {et~al.}(2007){Frebel}, {Norris}, {Aoki}, {Honda},
  {Bessell}, {Takada-Hidai}, {Beers}, \& {Christlieb}}]{Frebel2007}
{Frebel}, A., {Norris}, J.~E., {Aoki}, W., {Honda}, S., {Bessell}, M.~S.,
  {Takada-Hidai}, M., {Beers}, T.~C., \& {Christlieb}, N. 2007, \apj, 658, 534,
  arXiv:astro-ph/0612160

\bibitem[{{Fujimoto} {et~al.}(2000){Fujimoto}, {Ikeda}, \&
  {Iben}}]{Fujimoto2000}
{Fujimoto}, M.~Y., {Ikeda}, Y., \& {Iben}, I.~J. 2000, \apjl, 529, L25

\bibitem[{{Gallino} {et~al.}(1998){Gallino}, {Arlandini}, {Busso}, {Lugaro},
  {Travaglio}, {Straniero}, {Chieffi}, \& {Limongi}}]{Gallino1998}
{Gallino}, R., {Arlandini}, C., {Busso}, M., {Lugaro}, M., {Travaglio}, C.,
  {Straniero}, O., {Chieffi}, A., \& {Limongi}, M. 1998, \apj, 497, 388

\bibitem[{{Garc{\'{\i}}a-Hern{\'a}ndez}
  {et~al.}(2006){Garc{\'{\i}}a-Hern{\'a}ndez}, {Garc{\'{\i}}a-Lario}, {Plez},
  {D\'Antona}, {Manchado}, \& {Trigo-Rodr{\'{\i}}guez}}]{GarciaHernandez2006}
{Garc{\'{\i}}a-Hern{\'a}ndez}, D.~A., {Garc{\'{\i}}a-Lario}, P., {Plez}, B.,
  {D\'Antona}, F., {Manchado}, A., \& {Trigo-Rodr{\'{\i}}guez}, J.~M. 2006,
  Science, 314, 1751, arXiv:astro-ph/0611319

\bibitem[{{Garc{\'{\i}}a P{\'e}rez} {et~al.}(2006){Garc{\'{\i}}a P{\'e}rez},
  {Asplund}, {Primas}, {Nissen}, \& {Gustafsson}}]{GarciaPerez2006}
{Garc{\'{\i}}a P{\'e}rez}, A.~E., {Asplund}, M., {Primas}, F., {Nissen}, P.~E.,
  \& {Gustafsson}, B. 2006, \aap, 451, 621, arXiv:astro-ph/0512290

\bibitem[{{Gilroy}(1989)}]{Gilroy1989}
{Gilroy}, K.~K. 1989, \apj, 347, 835

\bibitem[{{Goriely}(1999)}]{Goriely1999}
{Goriely}, S. 1999, \aap, 342, 881

\bibitem[{{Goriely} \& {Mowlavi}(2000)}]{Goriely2000}
{Goriely}, S., \& {Mowlavi}, N. 2000, \aap, 362, 599

\bibitem[{{Goriely} \& {Siess}(2001)}]{Goriely2001}
{Goriely}, S., \& {Siess}, L. 2001, \aap, 378, L25

\bibitem[{{Goriely} \& {Siess}(2004)}]{Goriely2004}
------. 2004, \aap, 421, L25

\bibitem[{{Goriely} \& {Siess}(2005)}]{Goriely2005}
{Goriely}, S., \& {Siess}, L. 2005, in From Lithium to Uranium: Elemental
  Tracers of Early Cosmic Evolution, ed. V.~{Hill}, P.~{Fran{\c c}ois}, \&
  F.~{Primas}, IAU Symposium, 451--460

\bibitem[{{Goswami} {et~al.}(2006){Goswami}, {Aoki}, {Beers}, {Christlieb},
  {Norris}, {Ryan}, \& {Tsangarides}}]{Goswami2006}
{Goswami}, A., {Aoki}, W., {Beers}, T.~C., {Christlieb}, N., {Norris}, J.~E.,
  {Ryan}, S.~G., \& {Tsangarides}, S. 2006, \mnras, 372, 343,
  arXiv:astro-ph/0608106

\bibitem[{{Heger} \& {Woosley}(2002)}]{Heger2002}
{Heger}, A., \& {Woosley}, S.~E. 2002, \apj, 567, 532

\bibitem[{{Herwig}(2004)}]{Herwig2004}
{Herwig}, F. 2004, \apjs, 155, 651

\bibitem[{{Herwig}(2005)}]{Herwig2005}
------. 2005, \araa, 43, 435

\bibitem[{{Herwig} {et~al.}(2003){Herwig}, {Langer}, \& {Lugaro}}]{Herwig2003}
{Herwig}, F., {Langer}, N., \& {Lugaro}, M. 2003, \apj, 593, 1056,
  arXiv:astro-ph/0305491

\bibitem[{{Hill} {et~al.}(2000){Hill}, {Barbuy}, {Spite}, {Spite}, {Cayrel},
  {Plez}, {Beers}, {Nordstr{\"o}m}, \& {Nissen}}]{Hill2000}
{Hill}, V. {et~al.} 2000, \aap, 353, 557

\bibitem[{{Hill} {et~al.}(2002){Hill}, {Plez}, {Cayrel}, {Beers}, {Nordstr{\"
  o}m}, {Andersen}, {Spite}, {Spite}, {Barbuy}, {Bonifacio}, {Depagne},
  {Fran{\c c}ois}, \& {Primas}}]{Hill2002}
------. 2002, \aap, 387, 560

\bibitem[{{Hirschi}(2007)}]{Hirschi2007}
{Hirschi}, R. 2007, \aap, 461, 571, arXiv:astro-ph/0608170

\bibitem[{{Honda} {et~al.}(2006){Honda}, {Aoki}, {Ishimaru}, {Wanajo}, \&
  {Ryan}}]{Honda2006}
{Honda}, S., {Aoki}, W., {Ishimaru}, Y., {Wanajo}, S., \& {Ryan}, S.~G. 2006,
  \apj, 643, 1180, arXiv:astro-ph/0602107

\bibitem[{{Honda} {et~al.}(2004){Honda}, {Aoki}, {Kajino}, {Ando}, {Beers},
  {Izumiura}, {Sadakane}, \& {Takada-Hidai}}]{Honda2004}
{Honda}, S., {Aoki}, W., {Kajino}, T., {Ando}, H., {Beers}, T.~C., {Izumiura},
  H., {Sadakane}, K., \& {Takada-Hidai}, M. 2004, \apj, 607, 474

\bibitem[{{Ivans} {et~al.}(2005){Ivans}, {Sneden}, {Gallino}, {Cowan}, \&
  {Preston}}]{Ivans2005}
{Ivans}, I.~I., {Sneden}, C., {Gallino}, R., {Cowan}, J.~J., \& {Preston},
  G.~W. 2005, \apjl, 627, L145, arXiv:astro-ph/0505002

\bibitem[{{Johnson} \& {Bolte}(2002)}]{Johnson2002}
{Johnson}, J.~A., \& {Bolte}, M. 2002, \apjl, 579, L87

\bibitem[{{Johnson} \& {Bolte}(2004)}]{Johnson2004}
------. 2004, \apj, 605, 462

\bibitem[{{Johnson} {et~al.}(2007){Johnson}, {Herwig}, {Beers}, \&
  {Christlieb}}]{Johnson2007}
{Johnson}, J.~A., {Herwig}, F., {Beers}, T.~C., \& {Christlieb}, N. 2007, \apj,
  658, 1203

\bibitem[{{Jonsell} {et~al.}(2006){Jonsell}, {Barklem}, {Gustafsson},
  {Christlieb}, {Hill}, {Beers}, \& {Holmberg}}]{Jonsell2006}
{Jonsell}, K., {Barklem}, P.~S., {Gustafsson}, B., {Christlieb}, N., {Hill},
  V., {Beers}, T.~C., \& {Holmberg}, J. 2006, \aap, 451, 651,
  arXiv:astro-ph/0601476

\bibitem[{{Jorissen} \& {Van Eck}(2000)}]{Jorissen2000}
{Jorissen}, A., \& {Van Eck}, S. 2000, in IAU Symposium, Vol. 177, The Carbon
  Star Phenomenon, ed. R.~F. {Wing}, 259

\bibitem[{{Justham} {et~al.}(2009){Justham}, {Wolf}, {Podsiadlowski}, \&
  {Han}}]{Justham2009}
{Justham}, S., {Wolf}, C., {Podsiadlowski}, P., \& {Han}, Z. 2009, \aap, 493,
  1081, 0811.2633

\bibitem[{{Karakas} \& {Lattanzio}(2007)}]{Karakas2007}
{Karakas}, A., \& {Lattanzio}, J.~C. 2007, Publications of the Astronomical
  Society of Australia, 24, 103, arXiv:0708.4385

\bibitem[{{Karakas} \& {Lattanzio}(2003)}]{Karakas2003}
{Karakas}, A.~I., \& {Lattanzio}, J.~C. 2003, Publications of the Astronomical
  Society of Australia, 20, 279

\bibitem[{{Komiya} {et~al.}(2007){Komiya}, {Suda}, {Minaguchi}, {Shigeyama},
  {Aoki}, \& {Fujimoto}}]{Komiya2007}
{Komiya}, Y., {Suda}, T., {Minaguchi}, H., {Shigeyama}, T., {Aoki}, W., \&
  {Fujimoto}, M.~Y. 2007, \apj, 658, 367, arXiv:astro-ph/0610670

\bibitem[{{Limongi} \& {Chieffi}(2003)}]{Limongi2003}
{Limongi}, M., \& {Chieffi}, A. 2003, \apj, 592, 404

\bibitem[{{Lodders}(2003)}]{Lodders2003}
{Lodders}, K. 2003, \apj, 591, 1220

\bibitem[{{Lucatello}(2003)}]{LucatelloPhD}
{Lucatello}, S. 2003, PhD thesis, Universit\`a degli studi di Padova, Italy

\bibitem[{{Lucatello} {et~al.}(2006){Lucatello}, {Beers}, {Christlieb},
  {Barklem}, {Rossi}, {Marsteller}, {Sivarani}, \& {Lee}}]{Lucatello2006}
{Lucatello}, S., {Beers}, T.~C., {Christlieb}, N., {Barklem}, P.~S., {Rossi},
  S., {Marsteller}, B., {Sivarani}, T., \& {Lee}, Y.~S. 2006, \apjl, 652, L37,
  arXiv:astro-ph/0609730

\bibitem[{{Lucatello} {et~al.}(2003){Lucatello}, {Gratton}, {Cohen}, {Beers},
  {Christlieb}, {Carretta}, \& {Ram{\'{\i}}rez}}]{Lucatello2003}
{Lucatello}, S., {Gratton}, R., {Cohen}, J.~G., {Beers}, T.~C., {Christlieb},
  N., {Carretta}, E., \& {Ram{\'{\i}}rez}, S. 2003, \aj, 125, 875

\bibitem[{{Lucatello} {et~al.}(2005{\natexlab{a}}){Lucatello}, {Gratton},
  {Beers}, \& {Carretta}}]{Lucatello2005IMF}
{Lucatello}, S., {Gratton}, R.~G., {Beers}, T.~C., \& {Carretta}, E.
  2005{\natexlab{a}}, \apj, 625, 833

\bibitem[{{Lucatello} {et~al.}(2005{\natexlab{b}}){Lucatello}, {Tsangarides},
  {Beers}, {Carretta}, {Gratton}, \& {Ryan}}]{Lucatello2005bin}
{Lucatello}, S., {Tsangarides}, S., {Beers}, T.~C., {Carretta}, E., {Gratton},
  R.~G., \& {Ryan}, S.~G. 2005{\natexlab{b}}, \apj, 625, 825

\bibitem[{{Masseron}(2006)}]{MasseronPhD}
{Masseron}, T. 2006, PhD thesis, Observatoire de Paris, France

\bibitem[{{Masseron} {et~al.}(in prep., Paper~I){Masseron}, {Plez}, {Primas},
  F.~{van Eck}, \& {Jorissen}}]{masseron2009I}
{Masseron}, T., {Plez}, {Primas}, F.~{van Eck}, S., \& {Jorissen}, A. in prep.,
  Paper~I

\bibitem[{{Masseron} {et~al.}(2006){Masseron}, {van Eck}, {Famaey}, {Goriely},
  {Plez}, {Siess}, {Beers}, {Primas}, \& {Jorissen}}]{Masseron2006}
{Masseron}, T. {et~al.} 2006, \aap, 455, 1059, astro-ph/0605658

\bibitem[{{McClure} \& {Woodsworth}(1990)}]{McClure1990}
{McClure}, R.~D., \& {Woodsworth}, A.~W. 1990, \apj, 352, 709

\bibitem[{{Meynet} \& {Maeder}(2002)}]{Meynet2002}
{Meynet}, G., \& {Maeder}, A. 2002, \aap, 381, L25

\bibitem[{{Miller} \& {Scalo}(1979)}]{Miller1979}
{Miller}, G.~E., \& {Scalo}, J.~M. 1979, \apjs, 41, 513

\bibitem[{{Nollett} {et~al.}(2003){Nollett}, {Busso}, \&
  {Wasserburg}}]{Nollett2003}
{Nollett}, K.~M., {Busso}, M., \& {Wasserburg}, G.~J. 2003, \apj, 582, 1036,
  arXiv:astro-ph/0211271

\bibitem[{{Nomoto} \& {Kondo}(1991)}]{Nomoto1991}
{Nomoto}, K., \& {Kondo}, Y. 1991, \apjl, 367, L19

\bibitem[{{Norris} {et~al.}(2007){Norris}, {Christlieb}, {Korn}, {Eriksson},
  {Bessell}, {Beers}, {Wisotzki}, \& {Reimers}}]{Norris2007}
{Norris}, J.~E., {Christlieb}, N., {Korn}, A.~J., {Eriksson}, K., {Bessell},
  M.~S., {Beers}, T.~C., {Wisotzki}, L., \& {Reimers}, D. 2007, \apj, 670, 774,
  arXiv:0707.2657

\bibitem[{{Norris} {et~al.}(1997){Norris}, {Ryan}, \& {Beers}}]{Norris1997b}
{Norris}, J.~E., {Ryan}, S.~G., \& {Beers}, T.~C. 1997, \apjl, 489, L169

\bibitem[{{Pignatari} {et~al.}(2008){Pignatari}, {Gallino}, {Meynet},
  {Hirschi}, {Herwig}, \& {Wiescher}}]{Pignatari2008}
{Pignatari}, M., {Gallino}, R., {Meynet}, G., {Hirschi}, R., {Herwig}, F., \&
  {Wiescher}, M. 2008, \apjl, 687, L95, 0810.0182

\bibitem[{{Plez} \& {Cohen}(2005)}]{Plez2005}
{Plez}, B., \& {Cohen}, J.~G. 2005, \aap, 434, 1117

\bibitem[{{Plez} {et~al.}(2004){Plez}, {Hill}, {Cayrel}, {Spite}, {Barbuy},
  {Beers}, {Bonifacio}, {Primas}, \& {Nordstr{\"o}m}}]{Plez2004}
{Plez}, B. {et~al.} 2004, \aap, 428, L9, arXiv:astro-ph/0410628

\bibitem[{{Plez} {et~al.}(2008){Plez}, {Masseron}, {Van Eck}, {Godefroid},
  {Coheur}, \& {Jorissen}}]{Plez2008}
{Plez}, B., {Masseron}, T., {Van Eck}, S., {Godefroid}, M., {Coheur}, P.~F., \&
  {Jorissen}, A. 2008, in Astronomical Society of the Pacific Conference
  Series, Vol. 384, Astronomical Society of the Pacific Conference Series, ed.
  G.~{Van Belle}, 326

\bibitem[{{Preston} \& {Sneden}(2000)}]{Preston2000}
{Preston}, G.~W., \& {Sneden}, C. 2000, \aj, 120, 1014

\bibitem[{{Preston} \& {Sneden}(2001)}]{Preston2001}
------. 2001, \aj, 122, 1545

\bibitem[{{Proffitt} \& {Michaud}(1989)}]{Proffitt1989}
{Proffitt}, C.~R., \& {Michaud}, G. 1989, \apj, 345, 998

\bibitem[{{Roederer} {et~al.}(2008){Roederer}, {Frebel}, {Shetrone}, {Allende
  Prieto}, {Rhee}, {Gallino}, {Bisterzo}, {Sneden}, {Beers}, \&
  {Cowan}}]{Roederer2008}
{Roederer}, I.~U. {et~al.} 2008, \apj, 679, 1549, arXiv:0802.3701

\bibitem[{{Rossi} {et~al.}(1999){Rossi}, {Beers}, \& {Sneden}}]{Rossi1999}
{Rossi}, S., {Beers}, T.~C., \& {Sneden}, C. 1999, in ASP Conf. Ser. 165: The
  Third Stromlo Symposium: The Galactic Halo, 264

\bibitem[{{Ryan} {et~al.}(2005){Ryan}, {Aoki}, {Norris}, \& {Beers}}]{Ryan2005}
{Ryan}, S.~G., {Aoki}, W., {Norris}, J.~E., \& {Beers}, T.~C. 2005, \apj, 635,
  349, arXiv:astro-ph/0508475

\bibitem[{{Ryan} \& {Norris}(1991)}]{Ryan1991}
{Ryan}, S.~G., \& {Norris}, J.~E. 1991, \aj, 101, 1835

\bibitem[{{Siess} {et~al.}(2004){Siess}, {Goriely}, \& {Langer}}]{Siess2004}
{Siess}, L., {Goriely}, S., \& {Langer}, N. 2004, \aap, 415, 1089

\bibitem[{{Siess} {et~al.}(2002){Siess}, {Livio}, \& {Lattanzio}}]{Siess2002}
{Siess}, L., {Livio}, M., \& {Lattanzio}, J. 2002, \apj, 570, 329

\bibitem[{{Sivarani} {et~al.}(2006){Sivarani}, {Beers}, {Bonifacio}, {Molaro},
  {Cayrel}, {Herwig}, {Spite}, {Spite}, {Plez}, {Andersen}, {Barbuy},
  {Depagne}, {Hill}, {Fran{\c c}ois}, {Nordstr{\"o}m}, \&
  {Primas}}]{Sivarani2006}
{Sivarani}, T. {et~al.} 2006, \aap, 459, 125, arXiv:astro-ph/0608112

\bibitem[{{Sivarani} {et~al.}(2004){Sivarani}, {Bonifacio}, {Molaro}, {Cayrel},
  {Spite}, {Spite}, {Plez}, {Andersen}, {Barbuy}, {Beers}, {Depagne}, {Hill},
  {Fran{\c c}ois}, {Nordstr{\"o}m}, \& {Primas}}]{Sivarani2004}
------. 2004, \aap, 413, 1073

\bibitem[{{Sneden} {et~al.}(2003{\natexlab{a}}){Sneden}, {Cowan}, {Lawler},
  {Ivans}, {Burles}, {Beers}, {Primas}, {Hill}, {Truran}, {Fuller}, {Pfeiffer},
  \& {Kratz}}]{Sneden2003_CS22892}
{Sneden}, C. {et~al.} 2003{\natexlab{a}}, \apj, 591, 936,
  arXiv:astro-ph/0303542

\bibitem[{{Sneden} {et~al.}(2003{\natexlab{b}}){Sneden}, {Preston}, \&
  {Cowan}}]{Sneden2003}
{Sneden}, C., {Preston}, G.~W., \& {Cowan}, J.~J. 2003{\natexlab{b}}, \apj,
  592, 504

\bibitem[{{Spite} {et~al.}(2006){Spite}, {Cayrel}, {Hill}, {Spite}, {Fran{\c
  c}ois}, {Plez}, {Bonifacio}, {Molaro}, {Depagne}, {Andersen}, {Barbuy},
  {Beers}, {Nordstr{\"o}m}, \& {Primas}}]{Spite2006}
{Spite}, M. {et~al.} 2006, \aap, 455, 291, arXiv:astro-ph/0605056

\bibitem[{{Spite} {et~al.}(2005){Spite}, {Cayrel}, {Plez}, {Hill}, {Spite},
  {Depagne}, {Fran{\c c}ois}, {Bonifacio}, {Barbuy}, {Beers}, {Andersen},
  {Molaro}, {Nordstr{\" o}m}, \& {Primas}}]{Spite2005}
------. 2005, \aap, 430, 655

\bibitem[{{Stancliffe} {et~al.}(2007){Stancliffe}, {Glebbeek}, {Izzard}, \&
  {Pols}}]{Stancliffe2007}
{Stancliffe}, R.~J., {Glebbeek}, E., {Izzard}, R.~G., \& {Pols}, O.~R. 2007,
  \aap, 464, L57, arXiv:astro-ph/0702138

\bibitem[{{Straniero} {et~al.}(1995){Straniero}, {Gallino}, {Busso}, {Chiefei},
  {Raiteri}, {Limongi}, \& {Salaris}}]{Straniero1995}
{Straniero}, O., {Gallino}, R., {Busso}, M., {Chiefei}, A., {Raiteri}, C.~M.,
  {Limongi}, M., \& {Salaris}, M. 1995, \apjl, 440, L85

\bibitem[{{Suda} {et~al.}(2004){Suda}, {Aikawa}, {Machida}, {Fujimoto}, \&
  {Iben}}]{Suda2004}
{Suda}, T., {Aikawa}, M., {Machida}, M.~N., {Fujimoto}, M.~Y., \& {Iben}, I.~J.
  2004, \apj, 611, 476

\bibitem[{{Takeda}(2003)}]{Takeda2003}
{Takeda}, Y. 2003, \aap, 402, 343, arXiv:astro-ph/0105215

\bibitem[{{Tsangarides}(2005)}]{TsangaridesPhD}
{Tsangarides}, S.~A. 2005, PhD thesis, Open University (United Kingdom),
  England

\bibitem[{{Tumlinson}(2007)}]{Tumlinson2007}
{Tumlinson}, J. 2007, \apjl, 664, L63, arXiv:0706.2903

\bibitem[{{VandenBerg} {et~al.}(2006){VandenBerg}, {Bergbusch}, \&
  {Dowler}}]{VandenBerg2006}
{VandenBerg}, D.~A., {Bergbusch}, P.~A., \& {Dowler}, P.~D. 2006, \apjs, 162,
  375, arXiv:astro-ph/0510784

\bibitem[{{Vanhala} \& {Cameron}(1998)}]{Vanhala1998}
{Vanhala}, H.~A.~T., \& {Cameron}, A.~G.~W. 1998, \apj, 508, 291

\bibitem[{{Venn} \& {Lambert}(2008)}]{Venn2008}
{Venn}, K.~A., \& {Lambert}, D.~L. 2008, \apj, 677, 572, arXiv:0801.0752

\bibitem[{{Woosley} \& {Weaver}(1995)}]{Woosley1995}
{Woosley}, S.~E., \& {Weaver}, T.~A. 1995, \apjs, 101, 181

\bibitem[{{Zacs} {et~al.}(1998){Zacs}, {Nissen}, \& {Schuster}}]{Zacs1998}
{Zacs}, L., {Nissen}, P.~E., \& {Schuster}, W.~J. 1998, \aap, 337, 216

\end{thebibliography}

\end{document}